\documentclass[
 reprint,
 amsmath,amssymb,
 aps
]{revtex4-2}

\usepackage{graphicx} 
\usepackage{dcolumn}
\usepackage{bm} 
\usepackage[colorlinks=true, allcolors=blue]{hyperref}

\usepackage{amsmath}
\usepackage{mathtools}
\usepackage{lipsum}
\usepackage{dblfloatfix}
\usepackage{placeins}
\usepackage{nicefrac}
\usepackage{minitoc}
\usepackage{xcolor}
\usepackage{soul}
\usepackage{mhchem}
\usepackage{xfrac}

\usepackage{newtxtext,newtxmath}

\makeatletter
\newcommand{\subalign}[1]{
  \vcenter{
    \Let@ \restore@math@cr \default@tag
    \baselineskip\fontdimen10 \scriptfont\tw@
    \advance\baselineskip\fontdimen12 \scriptfont\tw@
    \lineskip\thr@@\fontdimen8 \scriptfont\thr@@
    \lineskiplimit\lineskip
    \ialign{\hfil$\m@th\scriptstyle##$&$\m@th\scriptstyle{}##$\hfil\crcr
      #1\crcr
    }
  }
}
\makeatother

\newcommand{\includepdfmt}[3]{\raisebox{-#2\height}{\includegraphics[scale=#1]{#3.pdf}}} 
\renewcommand{\vec}[1]{\boldsymbol{#1}} 
\newcommand{\diagsymb}[1]{\; #1 \;} 

\newcommand{\Tr}{\operatorname{Tr}}

\newcommand{\dlmassc}{\gamma_{c}}
\newcommand{\dlmassv}{\gamma_{v}}

\newcommand{\Xmodeindex}{\mu}
\newcommand{\XTM}{\vec{K}}

\newcommand{\XRM}{\vec{k}}
\newcommand{\XRMM}{k}
\newcommand{\BXTM}{\vec{Q}}

\newcommand{\BXRM}{\vec{q}}
\newcommand{\BXRMM}{q}
\newcommand{\genericXLabel}{X}
\newcommand{\aBohr}{a_{0}}
\newcommand{\XWF}[4]{\Phi^{#1#2}_{#3#4}}
\newcommand{\XWFt}[4]{\tilde{\Phi}^{#1#2}_{#3#4}}
\newcommand{\BXWF}[4]{\Psi^{#1#2}_{#3#4}}
\newcommand{\NormalizationFunction}{R}
\newcommand{\TdepX}{X} 
\newcommand{\elecEnergy}{\xi}
\newcommand{\NormalizationFunctionVeff}{R}

\DeclareSymbolFontAlphabet{\mathcal}{symbols}
\DeclareMathAlphabet{\mathcal}{OMS}{cmsy}{m}{n}
\SetMathAlphabet{\mathcal}{bold}{OMS}{cmsy}{b}{n}
\allowdisplaybreaks 

\makeatletter
\g@addto@macro\bfseries{\boldmath}
\makeatother

\usepackage{totcount} 
\newcounter{commentcounter}
\regtotcounter{commentcounter}

\begin{document}


\title[Variational and field-theoretical approach to exciton--exciton interactions and biexcitons in semiconductors]{Variational and field-theoretical approach to exciton--exciton interactions and biexcitons in semiconductors}

\author{P. A. Noordman}
\affiliation{Institute for Theoretical Physics and Center for Extreme Matter and Emergent Phenomena, Utrecht University, Princetonplein 5, 3584 CC Utrecht, The Netherlands}
 
\author{L. Maisel Licerán}
\email{l.maiselliceran@uu.nl}
\affiliation{Institute for Theoretical Physics and Center for Extreme Matter and Emergent Phenomena, Utrecht University, Princetonplein 5, 3584 CC Utrecht, The Netherlands}
 
\author{H. T. C. Stoof}
\affiliation{Institute for Theoretical Physics and Center for Extreme Matter and Emergent Phenomena, Utrecht University, Princetonplein 5, 3584 CC Utrecht, The Netherlands}

\date{\today}

\begin{abstract}
	

Bound electron--hole pairs in semiconductors known as excitons are the subject of intense research because of their potential for optoelectronic devices and applications, especially in the realm of two-dimensional materials. 
While the properties of free excitons in these systems are well understood, a general description of the interactions between these quasiparticles is complicated owing to their composite nature, which leads to important exchange processes that can take place between the identical fermions of different excitons.
In this work, we employ a variational approach to study interactions between Wannier excitons and obtain an effective interaction potential between two ground-state excitons in a system of spin-degenerate electrons and holes.
This potential is in general nonlocal in position space and depends on the combined spin configurations of the electrons and holes.
When particularized to the case of hydrogen-like excitons with a heavy hole, this potential becomes local and exactly reproduces the Heitler--London result for two interacting hydrogen atoms.
Thus, our result can be interpreted as a generalization of the Heitler--London potential to the case of arbitrary masses.
We also show how including corrections arising from excited states into the theory results in a van der Waals potential at large distances, which is expected given the the induced dipole--dipole nature of the interactions.
Our approach is also applicable to more complicated systems with nonhydrogenic exciton series, such as layered semiconductors with Rytova--Keldysh interactions.
Additionally, we use a path-integral formalism to develop a many-body theory for a dilute gas of excitons, resulting in an excitonic action that formally includes many-body interactions between excitons.
While in this approach the field representing the excitons is exactly bosonic, we clarify how the internal exchange processes arise in the field-theoretical treatment, and show that the diagrams corresponding to the interactions between excitons align with our variational calculation when evaluated on shell.
Our methods and results lay the groundwork for a generalized theory of exciton--exciton interactions and their application to the study of biexciton spectra and correlated excitonic matter.
\end{abstract}

\maketitle
\section{Introduction}
\label{sec: Introduction}

Since their original prediction by Frenkel \cite{Frenkel_1931} and Wannier \cite{Wannier_1937} almost a century ago, bound electron--hole pairs in semiconductors known as excitons have sparked extensive theoretical and experimental research. 
These photoexcited quasiparticles play an important role in a multitude of systems, out of which two-dimensional (2D) materials are of special interest given their potential for optoelectronic applications. 
Examples are transition-metal dichalcogenide monolayers \cite{mak2016photonics,wang2018colloquium,zheng2018light,ugeda2014giant,hanbicki2015measurement,zhao2015electronic,shang2015observation,palummo2015exciton,moody2016exciton,ceballos2016exciton,robert2016exciton,robert2017fine,krasnok2018nanophotonics,guo2019exchange,Jiang_2021} and heterostructures \cite{rivera2018interlayer,miller2017long,jin2018ultrafast,calman2018indirect,kunstmann2018momentum,alexeev2019resonantly,tran2019evidence,Jiang_2021,ciarrocchi2022excitonic,Regan_2022}, phosphorene \cite{Carvalho_2016,yang2015optical,lu2016light,xu2016extraordinarily,akhtar2017recent}, and even topological insulators \cite{kung2019observation,syperek2022observation,mori2023spin}. 
In these materials, the most prevalent type of exciton is the Wannier exciton, whose size is much larger than the underlying lattice spacing.
The properties of single excitons in these two-dimensional materials have been studied in detail \cite{Chernikov_2014,Selig_2016,Haining_2016,Quan_2016,Zhang_2017,Xiao_2017,Mueller_2018,Rodin_2020,Lee_2023,MaiselLiceran_2023,Glazov_2024}.

A particularly interesting topic beyond the free exciton picture concerns exciton--exciton interactions.
These have been studied experimentally in a variety of systems, and it has been shown that they can influence the dynamics and photoluminescence spectrum of the system and contribute to phenomena such as exciton--exciton annihilation and valley depolarization \cite{peyghambarian1984blue,kossacki2005exciton,stone2009exciton,sim2013exciton,mouri2014nonlinear,sun2014observation,kumar2014exciton,soavi2016exciton,Dostal_2018,mahmood2018observation,purz2021coherent,Birkmeier_2022,steinhoff2024exciton}.
One important effect of exciton--exciton interactions concerns the formation of biexcitons, quasiparticles of two bound excitons.
Their formation and dynamics have been studied in different kinds of systems such as quantum dots and quantum wires \cite{banyai1987excitons,bacher1999biexciton,vonk2021biexciton,sun2022biexciton,diroll2023long,huang2023nonlocal}, transition-metal dichalcogenides \cite{you2015observation,wang2019studying,pei2017excited,steinhoff2018biexciton,conway2022direct,sharma2022engineering}, excited semiconductor nanoplatelets \cite{kunneman2014nature,peng2021observation}, and perovskite nanocrystals \cite{thouin2018stable,cho2024size}.
 
While it is clear that the signatures of exciton--exciton interactions and biexcitonic physics are experimentally accessible, theoretical modeling of interactions between Wannier excitons is challenging due to their composite nature.
Understanding these interactions could provide significant insights into the dynamics and collective behavior of excitons in these systems and aid in the prediction of biexciton spectra.
In this work, we develop a general approach to the study of interactions between Wannier excitons.
Specifically, we derive an effective pair potential between ground-state excitons, as well as an effective many-body field-theoretical description of a dilute exciton gas.
While our work is motivated by the advent of 2D systems, our methods are not restricted to any particular spatial dimension.
Therefore, the theory can also be straightforwardly applied to three-dimensional (3D) systems.

In prior descriptions of exciton--exciton interactions, the simplest and often-made assumption is to take the exciton operators to be exactly bosonic \cite{Inoue_1998,Elistratov_2016,Berman_2017}. 
This neglects the composite nature of the excitons, meaning that two excitons can no longer exchange their identical constituents with other excitons.
In order to obtain an effective exciton--exciton interaction, some works \cite{deLeon_2001,Thilagam_2001,Shahnazaryan_2017,Katsch_2018,Gribakin_2021} calculate the two-exciton scattering matrix elements before making the bosonic approximation.
This matrix is then taken to be the interaction between two exactly bosonic excitons and is included in a second-quantized Hamiltonian in terms of elementary bosonic exciton operators.
However, this approach does not take into account the particle exchanges that can occur between excitons in the absence of interaction, i.e., those associated purely with the Fermi--Dirac statistics of the constituent electrons and holes.

Alternatively, because of the apparent similarities between an exciton and the hydrogen atom, the Heitler--London approach \cite{Heitler_1927} has been used as a means to study exciton--exciton interactions \cite{Zimmermann_2007,Zimmermann_2008}.
Originally, this method was employed to obtain an effective potential between two hydrogen atoms responsible for the formation of the hydrogen molecule.
However, the simplicity and physical transparency of this approach largely stem from the important assumption that one of the constituents is much heavier than the other.
While this is true for the proton and electron that make up the hydrogen atom, in the majority of systems of interest the electrons and holes have very similar masses.
As such, the usual Heitler--London method cannot provide an accurate description of interacting excitons.

Another more sophisticated approach is to perform a bosonization procedure where the exciton operator is mapped onto a space where it is represented by a bosonic operator \cite{Hanamura_1974,Haug_1984,Rochat_2001,Okumura_2001}. 
A bosonic second-quantized Hamiltonian for excitons is then obtained by mapping the standard electronic Hamiltonian to this bosonic space.
Nevertheless, as pointed out in Ref.\ \cite{Combescot_2007}, this approach also does not exactly take into account the particle exchanges between excitons.
Furthermore, as discussed in Ref.\ \cite{Keldysh_1968}, effects associated with the deviation from exact Bose statistics are expected to be of the same order as those connected to the nonideal nature of a Bose gas.
Thus, they must be considered in any approach aimed at describing interactions between two or more excitons.

The goal of this work is two-fold.
Firstly, we obtain a two-body exciton potential by exactly taking into account the nonbosonic nature of these quasiparticles.
This is done via a variational approach resulting in an effective two-exciton Schrödinger equation in the standard two-body form, from where an effective exciton--exciton potential can be read off.
While the latter is in general nonlocal in position space, it exactly reduces to the local Heitler--London potential in the limit of the hole being much more massive than the electron (or vice-versa).
This justifies the need to account for all possible exchange processes between constituents.
At low exciton densities, where it should be possible to approximate the interactions by a sum of interactions between exciton pairs, the obtained potential precisely gives the best approximation (in a variational sense) to the corresponding two-body term.
Secondly, we use the finite-temperature path-integral formalism to effectively bosonize the excitons by introducing an auxiliary bosonic field whose excitations correspond precisely to the excitons.
We obtain an effective action that incorporates the many-body effects of interacting excitons up to arbitrary order.
We clarify how the exchange processes related to the Fermi statistics of the underlying constituents are recovered in the effective theory even though the exciton field is exactly bosonic in nature.

Our article is organized as follows.
In Sec.\ \ref{ref: theoretical framework} we introduce the general framework used to describe excitons in this article, and discuss the systems where our theory applies.
In Sec.\ \ref{sec: Variational Approach} we perform the aforementioned variational calculation.
We start by introducing biexciton states (Sec.\ \ref{sec: biexciton states}) and use them to write the biexciton problem as a (generalized) eigenvalue equation (Sec.\ \ref{sec: general biexciton ev problem}).
We then derive an effective potential between two ground-state excitons and discuss its applicability and limitations (Sec.\ \ref{sec: ground state excitons}).
We also analyze the long-distance behavior of the exciton--exciton interaction, obtaining the expected van der Waals coupling between induced dipoles (Sec.\ \ref{sec: corrections excited}).
Finally, we discuss some aspects to be taken into account from the point of view of numerical implementations of our method and provide a summary guide of the steps to follow to achieve this (Sec.\ \ref{sec: numeric notes}).
In Sec.\ \ref{sec: Hydrogenic Example} we compute the potential explicitly for hydrogenic excitons in the heavy-hole limit to show that it exactly reduces to that obtained via the Heitler--London approach, and also comment on the opposite regime of similar electron and hole masses.
In Sec.\ \ref{sec: Field Theoretic Approach} we perform the path-integral calculation to derive a formal action for the bosonic exciton field. 
Specifically, after setting up the electronic action (Sec.\ \ref{sec:FTElectronicAction}), we obtain an effective action for the interband polarization via a Hubbard--Stratonovich transformation (Sec.\ \ref{sec:FTPolarizationAction}), which we use to derive the free exciton propagator (Sec.\ \ref{sec: Free exciton propagator}) as well as the effective excitonic action whose interaction term reproduces the previous variational calculation (Sec. \ref{sec: Effective exciton action}).
Finally, in Sec.\ \ref{sec: Conclusion} we give our conclusions and outlook for further research.

\section{Theoretical framework}
\label{ref: theoretical framework}


In this section we introduce the general framework used to describe single excitons.
We first introduce the system Hamiltonian along with a discussion on the applicability of our theory to follow.
We then review the composite nature of the excitons and their theoretical treatment via the Bethe--Salpeter equation.

\subsection{System Hamiltonian}

We consider a system with conduction and valence bands labeled by well-defined (pseudo)spins, respectively, which from now on we refer to as just spins.
For simplicity, we assume that the repulsive electrostatic interaction between electrons is local in position space and spin-independent.
The (grand-canonical) second-quantized Hamiltonian describing this system reads
\begin{align}
\begin{split}
    \hat{\mathcal{H}} = \sum_{a \sigma \XRM} \elecEnergy^{a}_{\sigma \XRM} \hat{\psi}^{\dagger}_{a \sigma \XRM} \hat{\psi}_{a \sigma \XRM} + \frac{1}{2\mathcal{V}} \sum_{aa'} \sum_{\sigma \sigma'} \sum_{\XTM  \XRM  \XRM '} V(\XRM - \XRM')& \\
    \times \, \hat{\psi}^{\dagger}_{a \sigma,\XTM /2+\XRM }\hat{\psi}^{\dagger}_{a' \sigma',\XTM /2-\XRM} \hat{\psi}_{a' \sigma',\XTM /2-\XRM '}\hat{\psi}_{a \sigma,\XTM /2+\XRM '}&,
\end{split}
\end{align}
where $\hat{\psi}^{\dagger}$ and $\hat{\psi}$ are electronic creation and annihilation operators, respectively, and $V$ is the repulsive interaction between the electrons.
The indices $a, a' \in \{c,v\}$ label the type of band, and we will often write $\hat{c} \equiv \hat{\psi}_{c}$ and $\hat{v} \equiv \hat{\psi}_{v}$.
While we use $\sigma, \sigma'$ to compactly label the spin degrees of freedom of either type of band, the spins associated specifically to the conduction and valence bands will be labeled as $\alpha, \alpha', \dots$ and $\beta, \beta', \dots$, respectively.
Note that in principle $\alpha$ and $\beta$ are most appropriately understood as collective indices including the total-spin and spin-projection quantum numbers.
However, in practice it is often the case that the former has a single, fixed value for each kind of particle (in III--V semiconductor compounds, this is $\nicefrac{1}{2}$ for electrons, and $\nicefrac{1}{2}$ or $\nicefrac{3}{2}$ for holes).
We thus take $\alpha$ and $\beta$ to simply be the spin projections and omit the total spin, in the understanding that the latter is known.
Furthermore, $\elecEnergy^{a}_{\sigma \XRM} \equiv \epsilon^{a}_{\sigma \XRM} - \mu$, where $\epsilon^{a}_{\sigma \XRM}$ and $\mu$ are the single-particle dispersions and the chemical potential, respectively.
Our only assumption is that the system is gapped and $\mu$ lies within the gap.

Because the conduction or valence character of the bands is conserved at the interaction vertex, our theory applies to systems where the band-geometrical, band-topological, Berry-curvature, and moiré effects are not important.
Thus, our theory is broadly applicable to conventional semiconductors, but modifications are expected especially in systems with topological bands whose Wannier functions cannot be exponentially localized, which results in $\mathrm{U}(1)$-symmetry-breaking components in the interaction \cite{liceran2025unconventional}.
In a forthcoming paper, we will generalize the theory developed here to be also applicable to these systems, thus providing a general treatment of exciton--exciton interactions in systems with arbitrary underlying electronic Hamiltonians.

\subsection{Exciton states}

Having defined the creation and annihilation operators of single electrons, we introduce the exciton creation operator as
\begin{align}
\label{eq: exciton creation operator}
    \hat{X}^{\dagger}_{\Xmodeindex \XTM} = \frac{1}{\sqrt{\mathcal{V}}} \sum_{\alpha\beta\XRM}
    \Phi^{\alpha \beta}_{\mu \XTM}(\XRM )
    \hat{c}^{\dagger}_{\alpha, \XRM +\dlmassc \XTM}
    \hat{v}_{\beta, \XRM -\dlmassv \XTM} .
\end{align}
The exciton is labeled by its total momentum $\XTM $ and a set of quantum numbers collectively denoted by $\Xmodeindex$.
For instance, for hydrogen-like excitons in 2D we can write $\Xmodeindex = (n, m, S, m_{S})$, where $n$ is the principal quantum number, $m$ the azimuthal quantum number, $S$ the total exciton spin, and $m_{S}$ the associated magnetic quantum number.
The product of conduction and valence operators in Eq.\ \eqref{eq: exciton creation operator} is sometimes referred to as the polarization operator.
We note that annihilating a valence electron is analogous to creating a hole with opposite momentum and spin, so that one could equally work with the hole creation operator $\hat{h}^{\dagger}_{\beta \boldsymbol{p}} \equiv \hat{v}_{{-}\beta, {-}\boldsymbol{p}}$.
Furthermore, $\XRM$ is the relative exciton momentum, $\mathcal{V}$ stands for the volume of the system in the dimensionality of interest, and $\XWF{\alpha}{\beta}{\Xmodeindex}{\XTM}(\XRM)$ is the relative exciton wave function satisfying the normalization condition 
\begin{align}
    \frac{1}{\mathcal{V}} \sum_{\alpha \beta \XRM} | \XWF{\alpha}{\beta}{\Xmodeindex}{\XTM}(\XRM) |^{2} = 1.
\end{align}

In writing the momenta of the single particles in Eq.\ \eqref{eq: exciton creation operator} we have defined two numbers $\dlmassc $ and $\dlmassv $ such that $\dlmassc  + \dlmassv  = 1$, so that the exciton indeed has total momentum $\XTM$.
Generally speaking, the exciton binding energy and the relative wave function depend on $\XTM$ \cite{Wu_2015,MaiselLiceran_2023}.
However, an exception to this is when the effective-mass approximation for the underlying electrons is valid.
In this case, one can most conveniently choose $\dlmassc  = m_c / M_{X}$ and $\dlmassv  = m_v / M_{X}$, with $m_{c}$ and $m_{v}$ the masses of the electron and the hole, respectively, and $M_{X} = m_c + m_v$ that of the exciton.
Note that $m_{v}$ is thus defined to be a positive number.
In this case, the total momentum $\XTM$ can also be called the center-of-mass (CoM) momentum of the exciton.
When the effective-mass approximation holds, the CoM and relative motions completely decouple, the dispersion of the excitons is also parabolic, and the wave functions and binding energies become independent of $\XTM$. 
Consequently, the CoM-momentum label on the exciton wave function becomes redundant and can be omitted.
By contrast, when the effective-mass approximation does not hold, there is no preferred convention and we may choose $\dlmassc$ and $\dlmassv$ freely as long as $\dlmassc  + \dlmassv  = 1$.
A practical solution in this situation is to choose $\dlmassc  = \dlmassv  = \nicefrac{1}{2}$ or to choose one of them to be unity and the other zero.
Keeping $\dlmassc$ and $\dlmassv$ general, as we do here, is thus able to account for both situations.

The exciton creation and annihilation operators commute amongst themselves, i.e.,
\begin{align} \label{eq: exciton commutation}
    [\hat{X}_{\Xmodeindex\XTM},\hat{X}_{\Xmodeindex'\XTM'}] = [\hat{X}^{\dagger}_{\Xmodeindex\XTM},\hat{X}^{\dagger}_{\Xmodeindex'\XTM'}] = 0.
\end{align}
By contrast, the commutator of a creation and an annihilation operator reads \cite{Keldysh_1968}
\begin{widetext}
    \begin{equation}
        \label{eq: exciton operator commutator}
        \begin{split}
            [\hat{X}_{\Xmodeindex\XTM},\hat{X}^{\dagger}_{\Xmodeindex'\XTM'}] = 
            \delta_{\XTM \XTM '}\delta_{\Xmodeindex\Xmodeindex'} 
            -\frac{1}{\mathcal{V}} \sum_{\alpha\beta\XRM} \bigg\{
            &\sum_{\alpha'} \big[\XWF{\alpha}{\beta}{\Xmodeindex}{\XTM}(\XRM  + \dlmassv \XTM )\big]^* 
            \XWF{\alpha'}{\beta}{\Xmodeindex'}{\XTM'}(\XRM + \dlmassv \XTM')
            \hat{c}^\dagger_{\alpha', \XRM  + \XTM '} \hat{c}_{\alpha, \XRM  + \XTM} 
            \\
            + \, &\sum_{\beta'}
            \big[\XWF{\alpha}{\beta}{\Xmodeindex}{\XTM}(\XRM  - \dlmassc \XTM )\big]^* 
            \XWF{\alpha}{\beta'}{\Xmodeindex'}{\XTM'}(\XRM - \dlmassc \XTM')
            \hat{v}_{\beta', \XRM  - \XTM'} \hat{v}^\dagger_{\beta, \XRM  - \XTM}
            \biggr\} .
        \end{split}
    \end{equation}
\end{widetext}
The matrix elements of the operator on the right-hand side are of order $n_{X} a^{d}_{X}$ in $d$ dimensions, where $n_{X}$ is the exciton number density and $a_{X}$ the typical size of the exciton.
Eq.\ \eqref{eq: exciton operator commutator} shows that an exciton is not an exact boson, for if this were the case only the very first term would appear.
The aforementioned bosonic approximation is then obtained by neglecting the latter two terms on the right-hand side of this equation. 
To reiterate, we will not adopt this assumption, and instead we will always consider excitons as composite particles throughout this article.

\subsection{The exciton Bethe--Salpeter equation}

We calculate the expectation of $\hat{\mathcal{H}}$ in the state $\hat{X}^{\dagger}_{\Xmodeindex \XTM} \vert G \rangle$, where $\vert G \rangle$ is the neutral semiconductor ground state, and minimize the resulting energy functional with respect to the variational wave function with the usual normalization constraint.
In this way we obtain the eigenvalue equation satisfied by $\XWF{\alpha}{\beta}{\Xmodeindex}{\XTM}(\XRM)$, namely
\begin{equation}
    \label{eq: exciton BSE}
    \Delta^{\alpha\beta}_{\XTM \XRM }\XWF{\alpha}{\beta}{\Xmodeindex}{\XTM}(\XRM )
    -\frac{1}{\mathcal{V}} \sum_{\XRM '} 
    V(\XRM  - \XRM ')
    \XWF{\alpha}{\beta}{\Xmodeindex}{\XTM}(\XRM ') = \varepsilon^{\Xmodeindex}_{\XTM }\XWF{\alpha}{\beta}{\Xmodeindex}{\XTM}(\XRM ).
\end{equation}
Here, $ \varepsilon^{\Xmodeindex}_{\XTM }$ is the total exciton eigenenergy and we have defined
\begin{align} \label{eq: Delta definition}
     \Delta^{\alpha\beta}_{\XTM \XRM } \equiv \elecEnergy^c_{\alpha, \XRM  + \dlmassc \XTM} - \elecEnergy^v_{\beta, \XRM  - \dlmassv \XTM}.
\end{align}
In the literature, Eq.\ \eqref{eq: exciton BSE} is sometimes referred to as the Bethe--Salpeter equation (BSE).
It reduces to the well-known Wannier equation for excitons in conventional semiconductors in the case of parabolic bands.
As stated before, the interaction between the conduction and valence electrons is repulsive, such that the minus sign in Eq.\ \eqref{eq: exciton BSE} results in an attractive electron--hole interaction that allows for the formation of the bound state.
Moreover, the minus sign in front of the valence energy in Eq.\ \eqref{eq: Delta definition} is due to the fact that the energy of a hole is minus that of a valence electron.

Lastly, the excitonic envelope wave functions satisfy the completeness relation
\begin{equation}
\label{eq: wavefunction idendity eigenvalues}
    \sum_{\Xmodeindex} 
    \big[\XWF{\alpha}{\beta}{\Xmodeindex}{\XTM}(\XRM ) \big]^*
    \XWF{\alpha'}{\beta'}{\Xmodeindex}{\XTM}(\XRM' ) = \mathcal{V} \delta_{\XRM \XRM '}\delta_{\alpha\alpha'}\delta_{\beta\beta'},
\end{equation}
and the normalization condition
\begin{equation}
\label{eq: wavefunction idendity momenta and spin}
    \frac{1}{\mathcal{V}}\sum_{\alpha\beta\XRM} 
    \big[\XWF{\alpha}{\beta}{\Xmodeindex}{\XTM}(\XRM ) \big]^*
    \XWF{\alpha}{\beta}{\Xmodeindex'}{\XTM}(\XRM ) = \delta_{\Xmodeindex\Xmodeindex'}.
\end{equation}
An important thing to note is that here $\Xmodeindex$ jointly denotes \emph{all} particle--hole states, which include both bound states and scattering states.
The latter refer to solutions of Eq.\ \eqref{eq: exciton BSE} that asymptotically do not decay to zero, and for which the label $\mu$ contains a wave number $\vec{p}$ which becomes continuous in the thermodynamic limit \cite{elliott1957intensity,michel2008direct,mukhamedzhanov2008completeness}.
The energy of these states lies above the bound-state dissociation threshold.
We will then talk about ``excitons'' to refer to the bound states only.
We also note that a sum over $\alpha$ and $\beta$ has been included in Eq.\ \eqref{eq: wavefunction idendity momenta and spin}, despite the fact that they can be chosen as good quantum numbers in view of the fact that the electrostatic potential is spin-independent.
In this case $\XWF{\alpha}{\beta}{\Xmodeindex}{\XTM}(\XRM) \propto \delta_{\alpha \alpha_{\mu}} \delta_{\beta \beta_{\mu}}$, where $\alpha_{\mu}$ and $\beta_{\mu}$ are the spin quantum numbers contained in $\mu$.
In writing Eq.\ \eqref{eq: wavefunction idendity momenta and spin} as it stands we reserve the freedom to not necessarily label the exciton states by the individual spins of the electron and the hole, but possibly by the total exciton spin in the coupled basis as mentioned in the previous section.

\section{Variational approach}
\label{sec: Variational Approach}


In this section, we first introduce two-exciton states known as biexcitons, and then derive an eigenvalue equation for biexcitons via a variational principle.
This equation is in fact an exact rewriting of the four-particle Schrödinger equation and expresses how a bound exciton--exciton state arises as a superposition of all bound and scattering electron--hole states.
However, as discussed below, it is impractical in analyzing the interaction between two ground-state excitons.
To remedy this, we reduce the variational freedom to ground-state excitons only and obtain the equivalent equation in this case, allowing us to identify the effective potential between two such quasiparticles.
Readers mainly interested in the final result may skip ahead to Sec.\ \ref{sec: ground state excitons}, in particular to Eq.\ \eqref{eq: exciton--exciton potential effective} and the surrounding explanation.

\subsection{Biexciton states}
\label{sec: biexciton states}

There are two equivalent ways to study the formation of a biexciton.
One is to consider the simultaneous binding of two electrons and two holes, the other to consider the formation of a bound state between two preexisting excitons.
In either case, the result is a four-particle bound state.
Since we are looking for an effective potential between two excitons, it will be convenient to adopt the latter perspective.
We accordingly define a biexciton creation operator as
\begin{align}
\label{eq: biexciton creation operator}
    \hat{B}^\dagger_{\BXTM }=\frac{1}{2\sqrt{\mathcal{V}}}\sum_{\Xmodeindex_1\Xmodeindex_2\BXRM}
    \BXWF{\Xmodeindex_{1}}{\Xmodeindex_{2}}{}{\BXTM}(\BXRM )
    \hat{X}^\dagger_{\Xmodeindex_1, \BXTM /2+\BXRM}
    \hat{X}^\dagger_{\Xmodeindex_2, \BXTM /2-\BXRM} .
\end{align}
In this equation, the sums over the labels $\Xmodeindex_{1}$ and $\Xmodeindex_{2}$ can run over the entirety of the particle--hole space (including both bound states, i.e., the excitons, as well as scattering states) or only over a preferred variational subspace.
We will study the two cases in Secs. \ref{sec: general biexciton ev problem} and \ref{sec: ground state excitons}, respectively.
The prefactor of $1/2$ in Eq.\ \eqref{eq: biexciton creation operator} ensures that the condition $(1 / \mathcal{V}) \sum_{\Xmodeindex_{1} \Xmodeindex_{2} \BXRM} |\BXWF{\Xmodeindex_{1}}{\Xmodeindex_{2}}{}{\BXTM}(\BXRM)|^{2} = 1$ leads to a normalized state when the sums over $\mu_{1}$ and $\mu_{2}$ run over all states.
This has a similar form to the exciton creation operator defined in Eq.\ \eqref{eq: exciton creation operator}, with $\BXTM$ and $\BXRM$ the biexciton CoM and relative momenta, respectively.
These are defined in terms of the individual exciton momenta $\XTM_1$ and $\XTM_2$ as
\begin{equation} \label{eq: CoM momentum coords (biexcitons)}
    \BXTM  = \XTM _1+\XTM _2,
    \qquad
    \BXRM  = \frac{1}{2}(\XTM _1-\XTM _2).
\end{equation}
For simplicity, we have assumed both excitons share the same mass.
If this were not the case, it would be convenient to rewrite these expressions in terms of excitonic mass ratios, in analogy with the approach outlined for the single-exciton case.
Furthermore, we note that one should also include an additional set of quantum numbers labeling the particular biexciton state under consideration, which would play the same role as the collective index $\Xmodeindex$ in Eq.\ \eqref{eq: exciton creation operator}.
However, in this work we restrict ourselves to single-biexciton states and omit this label in what follows.
For generality, we let the biexciton wave function depend on the total momentum $\BXTM$ of the biexciton, which becomes relevant when the exciton dispersion relation is not quadratic.

As a result of the commuting nature of the exciton creation operators, it follows from Eq.\ \eqref{eq: biexciton creation operator} that the biexciton wave function is symmetric under exciton exchange, i.e.,
\begin{align}
\label{eq: biexciton wavefunction exciton exchange symmetry}
	\BXWF{\Xmodeindex_{1}}{\Xmodeindex_{2}}{}{\BXTM}(\BXRM ) = \BXWF{\Xmodeindex_{2}}{\Xmodeindex_{1}}{}{\BXTM}({-}\BXRM ).
\end{align}
This symmetry is equivalent to the simultaneous exchange of both electrons and holes within the composite state and reflects the partially bosonic nature of the excitons.
The internal structure of the exciton creation operators in Eq.\ \eqref{eq: biexciton creation operator} actually results in additional symmetry constraints on the biexciton wave function.
These stem from the fact that such a state is a two-electron and two-hole state which must be antisymmetric under the separate exchanges of both types of identical particles.
These ``hidden'' constraints are detailed in Appendix\ \ref{app: A relations}.
They involve an operator $\mathcal{A}$, which is an antisymmetrizer encompassing all possible exchanges of identical particles whose matrix elements are given by
\begin{widetext}
	\begin{equation}
		\label{eq: biexciton projector}
	    \mathcal{A}^{\Xmodeindex_1'\Xmodeindex_2'}_{\Xmodeindex_1\Xmodeindex_2}(\BXTM ,\BXRM ,\BXRM ') = 
	    \frac{1}{4}\bigg(\delta_{\BXRM \BXRM '} \delta_{\Xmodeindex_1\Xmodeindex_1'}\delta_{\Xmodeindex_2\Xmodeindex_2'} 
	    - \frac{1}{\mathcal{V}}[\mathcal{K}^{\mathrm{c}}]^{\Xmodeindex_1'\Xmodeindex_2'}_{\Xmodeindex_1\Xmodeindex_2}(\BXTM ,\BXRM ,\BXRM ') 
        - \frac{1}{\mathcal{V}}[\mathcal{K}^{\mathrm{v}}]^{\Xmodeindex_1'\Xmodeindex_2'}_{\Xmodeindex_1\Xmodeindex_2}(\BXTM ,\BXRM ,\BXRM ')
	    + \delta_{\BXRM ,-\BXRM '} \delta_{\Xmodeindex_1\Xmodeindex_2'}\delta_{\Xmodeindex_2\Xmodeindex_1'} \bigg).
	\end{equation}
	The first and fourth terms represent the identity and the exciton-exchange processes, respectively.
	The second component describes the electron exchange and is represented by the overlap integral
	\begin{equation}
	\label{eq: conduction electron exchange element}
		\begin{split}
			[\mathcal{K}^{\mathrm{c}}]^{\Xmodeindex_1'\Xmodeindex_2'}_{\Xmodeindex_1\Xmodeindex_2}(\BXTM ,\BXRM ,\BXRM ') =
		    \frac{1}{\mathcal{V}} \sum_{\alpha \beta \XRM} \sum_{\alpha' \beta'}
		    \big[ \XWF{\alpha}{\beta}{\Xmodeindex_1}{,\BXTM /2+\BXRM}(\XRM ) \big]^*
		    \big[ \XWF{\alpha'}{\beta'}{\Xmodeindex_2}{,\BXTM /2-\BXRM }(\XRM  + \dlmassc (\BXRM  + \BXRM ') - \dlmassv (\BXRM  - \BXRM ')) \big]^* \\
		    \times \, \XWF{\alpha'}{\beta}{\Xmodeindex_1'}{,\BXTM /2+\BXRM '}(\XRM  - \dlmassv (\BXRM  - \BXRM '))
		    \XWF{\alpha}{\beta'}{\Xmodeindex_2'}{,\BXTM /2-\BXRM '}(\XRM  + \dlmassc (\BXRM  + \BXRM ')).
		\end{split}
	\end{equation}
\end{widetext}
Finally, the third term describes hole exchange, which can be understood as a combined electron and exciton exchange via $[\mathcal{K}^{\mathrm{v}}]^{\Xmodeindex_1'\Xmodeindex_2'}_{\Xmodeindex_1\Xmodeindex_2}(\BXTM ,\BXRM ,\BXRM ') = [\mathcal{K}^{\mathrm{c}}]^{\Xmodeindex_2'\Xmodeindex_1'}_{\Xmodeindex_1\Xmodeindex_2}(\BXTM ,\BXRM ,-\BXRM ')$.
An important aspect is that $\mathcal{A}$ acts as a projector ($\mathcal{A}^{2} = \mathcal{A}$) when the sums over the $\Xmodeindex$ indices run over all two-particle states.

\subsection{General biexciton eigenvalue problem}
\label{sec: general biexciton ev problem}

In this section we assume that the sums over $\Xmodeindex_{1}$ and $\Xmodeindex_{2}$ in Eq.\ \eqref{eq: biexciton creation operator} indeed run over the entire set of states obtained from the single-exciton BSE.
The eigenvalue equation for the biexciton wave function can be derived by minimizing the energy functional
\begin{align}
\label{eq: biexciton energy functional}
    \mathcal{F}[\Psi^*,\Psi] = \langle G| \hat{B}_{\BXTM }\hat{\mathcal{H}}\hat{B}^{\dagger}_{\BXTM } |G\rangle 
    - \mathcal{E}_{\BXTM }\langle G| \hat{B}_{\BXTM }\hat{B}^{\dagger}_{\BXTM }|G\rangle,
\end{align}
where $\mathcal{E}_{\BXTM }$ is a Lagrange multiplier taking into account the normalization condition for the biexciton state, where again $\vert G \rangle$ is the neutral ground state of the semiconductor.
Minimizing the energy functional with respect to $\Psi^{*}$ leads to the biexciton eigenvalue problem
\begin{equation}
\label{eq: biexciton eigenvalue equation} 
\begin{split}
	\sum_{\Xmodeindex_1' \Xmodeindex_2' \BXRM'} \!
	[H - \mathcal{E}_{\BXTM } \mathcal{A}]_{\Xmodeindex_1 \Xmodeindex_2}^{\Xmodeindex_1' \Xmodeindex_2'} (\BXTM, \BXRM, \BXRM') \BXWF{\Xmodeindex_1'}{\Xmodeindex_2'}{}{\BXTM}(\BXRM ') = 0 .
\end{split}
\end{equation}
Here, $\mathcal{E}_{\BXTM }$ is interpreted as the biexciton eigenenergy and the Hamiltonian reads
\begin{align}
\label{eq: Hamiltonian exciton--exciton}
	&H_{\Xmodeindex_1 \Xmodeindex_2}^{\Xmodeindex_1' \Xmodeindex_2'}(\BXTM, \BXRM, \BXRM')
		\\ 
	&= \frac{1}{2} E_{\Xmodeindex_1 \Xmodeindex_2}^{\Xmodeindex_1' \Xmodeindex_2'}(\BXTM, \BXRM, \BXRM') \mathcal{A}_{\Xmodeindex_1 \Xmodeindex_2}^{\Xmodeindex_1' \Xmodeindex_2'}(\BXTM, \BXRM, \BXRM')
    +\frac{1}{2 \mathcal{V}} \mathcal{U}_{\Xmodeindex_1 \Xmodeindex_2}^{\Xmodeindex_1' \Xmodeindex_2'}(\BXTM, \BXRM, \BXRM') . \nonumber
\end{align}
This contains the sum of a kinetic term with
\begin{equation}
	\label{eq: E factor eigenenergies}
	E_{\Xmodeindex_1 \Xmodeindex_2}^{\Xmodeindex_1' \Xmodeindex_2'} (\BXTM, \BXRM, \BXRM') = \varepsilon^{\Xmodeindex_1}_{\BXTM /2+\BXRM }+\varepsilon^{\Xmodeindex_2}_{\BXTM /2-\BXRM } +	\varepsilon^{\Xmodeindex_1'}_{\BXTM /2+\BXRM '} + \varepsilon^{\Xmodeindex_2'}_{\BXTM /2-\BXRM '} ,
\end{equation}
and a potential term written as a sum of four different matrix elements,
\begin{equation}
    \label{eq: exciton--exciton interaction}
    	\mathcal{U}_{\Xmodeindex_1 \Xmodeindex_2}^{\Xmodeindex_1' \Xmodeindex_2'} (\BXTM, \BXRM, \BXRM') = \frac{1}{2} \big[ \mathcal{U}^{0} + \mathcal{U}^{\mathrm{c}} + \mathcal{U}^{\mathrm{v}} + \mathcal{U}^{\mathrm{X}} \big]_{\Xmodeindex_1 \Xmodeindex_2}^{\Xmodeindex_1' \Xmodeindex_2'} (\BXTM, \BXRM, \BXRM') .
\end{equation}
Their explicit expressions are given in Appendix \ref{app: Second Quantization Interaction Components}.
Additional details on the derivation of Eq.\ \eqref{eq: biexciton eigenvalue equation} can be found in Appendix \ref{app: biexciton eigenvalue problem derivation}.

The interaction of Eq.\ \eqref{eq: exciton--exciton interaction} is invariant under the exchange of the in- and outgoing excitons and contains all possible scatterings and exchange processes between two excitons in the initial state $(\Xmodeindex_1', \Xmodeindex_2', \BXRM ')$ and final state $(\Xmodeindex_1, \Xmodeindex_2, \BXRM )$.
All components of the interaction $\mathcal{U}$ are nonlocal quantities, i.e., upon Fourier transformation they will separately depend on two position-space coordinates.

We can now exploit the exciton-exchange symmetry Eq.\ \eqref{eq: biexciton wavefunction exciton exchange symmetry} of the biexciton wave function in order to remove the hole-exchange and exciton-exchange components of Eq.\ \eqref{eq: biexciton eigenvalue equation}.
This property is solely due to the commutativity of the exciton creation operators [cf. Eq.\ \eqref{eq: exciton commutation}] and is independent of the chosen subspace of exciton states, i.e., it does not depend on the range of the summation over $\Xmodeindex_1$, $\Xmodeindex_2$ in Eq.\ \eqref{eq: biexciton creation operator}.
This reduces the eigenvalue equation to
\begin{equation}
\label{eq: biexciton eigenvalue equation reduced}
	\frac{1}{\mathcal{V}} \! \sum_{\Xmodeindex_1' \Xmodeindex_2' \BXRM'} \! 
	[h - \mathcal{E}_{\BXTM } \NormalizationFunction]_{\Xmodeindex_1 \Xmodeindex_2}^{\Xmodeindex_1' \Xmodeindex_2'} (\BXTM, \BXRM, \BXRM') \BXWF{\Xmodeindex_1'}{\Xmodeindex_2'}{}{\BXTM}(\BXRM ') = 0 .
\end{equation}
The ``reduced'' Hamiltonian appearing in this equation reads
\begin{equation}
\label{eq: Hamiltonian exciton--exciton reduced}
\begin{split}
	h_{\Xmodeindex_1 \Xmodeindex_2}^{\Xmodeindex_1' \Xmodeindex_2'}(\BXTM, \BXRM, \BXRM') 
	&= \frac{1}{2} E_{\Xmodeindex_1 \Xmodeindex_2}^{\Xmodeindex_1' \Xmodeindex_2'}(\BXTM, \BXRM, \BXRM') \NormalizationFunction_{\Xmodeindex_1 \Xmodeindex_2}^{\Xmodeindex_1' \Xmodeindex_2'}(\BXTM, \BXRM, \BXRM') 
	\\
	&+	[\mathcal{U}^{\mathrm{0}} + \mathcal{U}^{\mathrm{c}}]_{\Xmodeindex_1 \Xmodeindex_2}^{\Xmodeindex_1' \Xmodeindex_2'}(\BXTM, \BXRM, \BXRM') ,
\end{split}
\end{equation}
where 
\begin{equation}
\label{eq: biexciton projector reduced}
	\NormalizationFunction_{\Xmodeindex_1 \Xmodeindex_2}^{\Xmodeindex_1' \Xmodeindex_2'}(\BXTM, \BXRM, \BXRM') = \mathcal{V}\delta_{\BXRM \BXRM'} 
	\delta_{\Xmodeindex_1 \Xmodeindex_1'} \delta_{\Xmodeindex_2 \Xmodeindex_2'} - [\mathcal{K}^{\mathrm{c}}]_{\Xmodeindex_1 \Xmodeindex_2}^{\Xmodeindex_1' \Xmodeindex_2'}(\BXTM, \BXRM, \BXRM') .
\end{equation}
Note that we have defined the above quantities to allow for a straightforward implementation of the thermodynamic limit, namely $(1/\mathcal{V})\sum_{\BXRM} \rightarrow \int \mathrm{d}^{d} k / (2 \pi)^{d}$ and $\mathcal{V} \delta_{\BXRM \BXRM'} \rightarrow (2 \pi)^{d} \delta(\BXRM - \BXRM')$.
This formulation will be particularly convenient when applying the upcoming results explicitly for hydrogenic excitons in Sec.\ \ref{sec: Hydrogenic Example}.

Because of the presence of $R$ on the right-hand side of Eq.\ \eqref{eq: Hamiltonian exciton--exciton reduced} and within the kinetic term of $h$, the eigenvalue problem does not seem to have the typical form, and is instead a \emph{generalized} eigenvalue problem.
However, it turns out that Eq.\ \eqref{eq: Hamiltonian exciton--exciton reduced} can be further reduced by employing the aforementioned hidden constraints satisfied by $\BXWF{}{}{}{}$, which rely on the completeness of the particle--hole basis.
As explained in Appendix\ \ref{app: biexciton eigenvalue problem derivation}, this results in the familiar Schrödinger equation for two electrons and two holes, now projected on the coupled particle--hole basis [cf. in particular to Eq.\ \eqref{eq: biexciton BSE full variational freedom}].
While this is formally correct, the resulting equation is impractical and does not give much insight into the interactions between excitons in specific states, as it mixes all exciton states.
This is not surprising given that the exciton states are not true eigenstates of the microscopic semiconductor Hamiltonian and thus possess overlaps among each other.

To make progress, we must remember that we are interested in an approximate description of interacting excitons within some fixed subspace of states, which will be valid for timescales shorter than the typical exciton lifetimes within this subset.
This means that we can simply consider Eq.\ \eqref{eq: biexciton eigenvalue equation reduced} and restrict the sum over states to this subspace \emph{without} employing the completeness of the particle--hole basis, so that the exchange processes are explicitly kept.
In fact, this is completely equivalent to restricting the variational freedom in the ansatz of Eq.\ \eqref{eq: biexciton creation operator} by letting the $\Xmodeindex$ indices run over the desired subspace only.
In the next section, we use this approach by only considering ground-state excitons for a system with a spin-independent interaction.
This will allow us to derive an effective exciton--exciton interaction potential describing how two ground-state excitons bind into a biexciton.
Our procedure can be readily generalized and used to describe the binding of excitons in excited states in more complicated situations.

\subsection{Effective potential between ground-state excitons}
\label{sec: ground state excitons}

We are now in the position to derive the sought-after effective potential between ground-state excitons, which is presented in Eq.\ \eqref{eq: exciton--exciton potential effective} and the surrounding discussion.
For transparency, we consider a system where the electron--electron interaction is independent of the spin and assume that the ground state is spin degenerate, and also that no further degeneracy exists.
A particularly interesting example of this situation is that of hydrogen-like excitons in 2D and 3D.
The hydrogenic 2D case will be explored below in some detail, but we stress that our theory is much more general and in particular allows for an electron--hole interaction different from the $1/r$ Coulomb potential. 
To obtain an effective interaction potential between excitons, we now restrict the reduced eigenvalue problem of Eq.\ \eqref{eq: biexciton eigenvalue equation reduced} to the spin-degenerate subspace of ground-state exciton states.
In the particular case of hydrogen-like excitons, this corresponds to setting the principal and azimuthal quantum numbers to zero, i.e., $n = m = 0$ everywhere.
In what follows, we omit the ground-state quantum numbers from all expressions and only highlight the spin dependence, which is the only relevant degree of freedom.
Because the interaction is spin independent and the bands have well-defined spins, the orbital exciton wave function separates from its spin component as
\begin{align}
\label{eq: exciton wavefunction spin seperation}
	\XWF{\alpha}{\beta}{\vec{S}}{\XTM}(\XRM) = \XWF{}{}{}{\XTM}(\XRM) \langle \alpha \beta | \vec{S} \rangle.
\end{align}
Here, $\vec{S}$ is the spin state of the exciton, which is often written in the coupled conduction--valence spin basis, but can in principle also stand for the individual bases of the conduction and valence spins due to the spin degeneracy.
Since in the variational ansatz we sum over all possibilities for $\vec{S}$, both approaches are equivalent for our purposes, although the latter simplifies the calculations.
While the system considered here corresponds to the simplest possible scenario, it serves as a transparent example whose procedure below can be generalized to situations with additional pseudospin degrees of freedom, such as the valley index in transition-metal dichalcogenides.
Finally, we note that in the hydrogenic case the total momentum label $\XTM$ on the exciton wave function becomes redundant and can be omitted.

\begin{figure}[!t]
    \centering
    \includegraphics[width=0.75\linewidth]{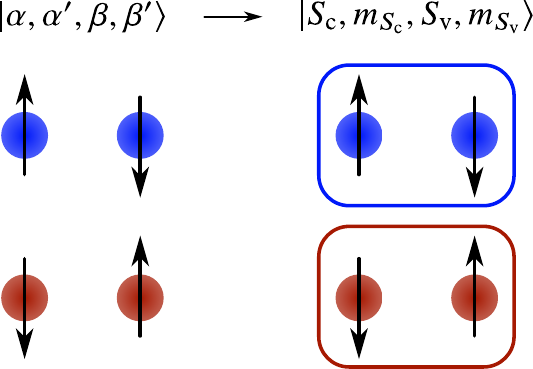}
    \caption{Schematic illustration of the spin-basis transformation explicitly performed in Appendix\ \ref{app: Spin-Basis Transformation}.
    The blue and red circles represent the conduction and valence electrons, respectively, and the biexciton eigenproblem simplifies when one considers the coupled electron spins on the one hand and the coupled valence spins on the other.
    }
    \label{fig: spin transformation}
\end{figure}

While in the system under consideration there is no preferred basis to label the spin state of individual excitons, this is not true for the exciton--exciton interaction.
The latter is not diagonal when expressed in terms of the total exciton spins, but becomes diagonal upon a change of basis to the states of pairwise coupled electrons and holes (in actuality, we use the spins of the valence bands where the holes reside, which are opposite to those of the holes themselves).
That is, instead of specifying the spin state of each exciton individually, we specify the spin state of the combined conduction electrons and that of the combined valence electrons, as schematically shown in Fig.\ \ref{fig: spin transformation}.
The spins and magnetic quantum numbers of the combined conduction and valence electrons are denoted by $S_{\mathrm{c}}, m_{S_{\mathrm{c}}}$ and $S_{\mathrm{v}}, m_{S_{\mathrm{v}}}$, respectively.
The transformation is performed explicitly in Appendix\ \ref{app: Spin-Basis Transformation}.
As a result of the exciton-exchange symmetry of Eq.\ \eqref{eq: biexciton wavefunction exciton exchange symmetry}, the orbital biexciton wave function in this basis satisfies
\begin{align}
\label{eq: biexciton effective wavefunction exciton exchange}
	\BXWF{S_{\mathrm{c}}}{S_{\mathrm{v}}}{}{\BXTM}(\BXRM ) = (-1)^{S_{\mathrm{c}} + S_{\mathrm{v}}} \hspace{0.3mm} \BXWF{S_{\mathrm{c}}}{S_{\mathrm{v}}}{}{\BXTM}(-\BXRM ) .
\end{align}
This immediately implies that the parity of the wave function under reflection must be the same as that of the combination $S_{\mathrm{c}} + S_{\mathrm{v}}$, which can be understood by writing the total wave function as the product state $|\Psi(\BXRM)\rangle \otimes |S_{\mathrm{c}}, m_{S_{\mathrm{c}}}\rangle \otimes |S_{\mathrm{v}}, m_{S_{\mathrm{v}}} \rangle$ and demanding symmetry under exciton exchange.
Note that the magnetic quantum numbers $m_{S_{\mathrm{c}}}$ and $m_{S_{\mathrm{v}}}$ have been omitted from the biexciton wave function of Eq.\ \eqref{eq: biexciton effective wavefunction exciton exchange} because the latter turns out to be independent of these variables, since the effective potential derived below does not depend on them.

The reduced two-exciton Hamiltonian of Eq.\ \eqref{eq: Hamiltonian exciton--exciton reduced} is expressed in the conduction--valence spin basis as
\begin{equation} 
\label{eq: Hamiltonian exciton--exciton ground state}
	\begin{split}
		h_{S_{\mathrm{c}}}(\BXTM, \BXRM, \BXRM') &= \frac{1}{2} E(\BXTM, \BXRM, \BXRM') \NormalizationFunction_{S_{\mathrm{c}}}(\BXTM, \BXRM, \BXRM')
		\\  
		&+ [ \mathcal{U}^{\mathrm{0}} - (-1)^{S_{\mathrm{c}}} \mathcal{U}^{\mathrm{c}} ](\BXTM, \BXRM, \BXRM'),
	\end{split}
\end{equation}
where $E$ is understood to be Eq.\ \eqref{eq: E factor eigenenergies} with all energies set to that of the excitonic ground state, and
\begin{equation}
\label{eq: biexciton projector ground state}
	\NormalizationFunction_{S_{\mathrm{c}}}(\BXTM ,\BXRM ,\BXRM ') = \mathcal{V} \delta_{\BXRM \BXRM '} 
	+ (-1)^{S_{\mathrm{c}}} \hspace{0.1mm} \mathcal{K}^{\mathrm{c}}(\BXTM ,\BXRM ,\BXRM ') .
\end{equation}
Since the variational freedom has been restricted to the ground state, $R_{S_{\mathrm{c}}}$ can be inverted, with $R^{-1}_{S_{\mathrm{c}}}$ being defined through
\begin{equation}
	\frac{1}{\mathcal{V}} \sum_{\vec{p}} \NormalizationFunction_{S_{\mathrm{c}}}(\BXTM ,\BXRM ,\vec{p}) 
	\NormalizationFunction^{-1}_{S_{\mathrm{c}}}(\BXTM ,\vec{p} ,\BXRM ')
	= \mathcal{V} \delta_{\BXRM \BXRM '} .
\end{equation}
It is straightforward to show that $R^{-1}_{S_{\mathrm{c}}}$ satisfies the implicit relation
\begin{equation} 
\label{eq: inverse of P}
	\begin{split}
		&\NormalizationFunction^{-1}_{S_{\mathrm{c}}}(\BXTM ,\BXRM ,\BXRM ') 
		\\
		&= \mathcal{V} \delta_{\BXRM \BXRM '}  - \frac{(-1)^{S_{\mathrm{c}}}}{\mathcal{V}} \sum_{\boldsymbol{p}} 
		\NormalizationFunction^{-1}_{S_{\mathrm{c}}}(\BXTM ,\BXRM ,\boldsymbol{p})\mathcal{K}^{\mathrm{c}}(\BXTM ,\boldsymbol{p},\BXRM ') .
	\end{split}
\end{equation}
The function $\mathcal{K}^{\mathrm{c}}$ remains that of Eq.\ \eqref{eq: conduction electron exchange element} but with all states restricted to the ground state and the spin dependencies removed.
Likewise, the interaction components $\mathcal{U}^{0}$ and $\mathcal{U}^{\mathrm{c}}$ are simply the matrix elements found in Appendix\ \ref{app: Second Quantization Interaction Components}, again with the spin variables removed and only considering the ground-state excitons. 
That is, one keeps only the orbital wave function $\XWF{}{}{}{\XTM}(\XRM)$ in the expressions and ignores the spin sums and indices.
Because of the form of Eq.\ \eqref{eq: exciton wavefunction spin seperation}, the spin sums factor out and simply yield the spin-dependent prefactors present in Eqs.\ \eqref{eq: Hamiltonian exciton--exciton ground state} and \eqref{eq: biexciton projector ground state}.
In fact, the full unreduced exciton--exciton interaction $\mathcal{U}$ in this coupled basis reads
\begin{align} \label{eq: exciton--exciton interaction spin-basis}
	&\mathcal{U}_{S_{\mathrm{c}} S_{\mathrm{v}}}(\BXTM ,\BXRM ,\BXRM ') 
	\\ \nonumber
    &= \frac{1}{2} [\mathcal{U}^0 -\!(-1)^{S_{\mathrm{c}}} \, \mathcal{U}^{\mathrm{c}} -\!(-1)^{S_{\mathrm{v}}} \, \mathcal{U}^{\mathrm{v}} +\! (-1)^{S_{\mathrm{c}} + S_{\mathrm{v}}} \, \mathcal{U}^{\mathrm{X}}]
    (\BXTM ,\BXRM, \BXRM ') ,
\end{align}
where the spin-dependent prefactor of each component indicates the type of exchange that it describes.
How the $\mathcal{U}^{\mathrm{v}}$ and $\mathcal{U}^{\mathrm{X}}$ components are connected to $\mathcal{U}^{0}$ and $\mathcal{U}^{\mathrm{c}}$ in this spin basis is described in Appendix\ \ref{app: Spin-Basis Transformation}.

Using the reduced Hamiltonian of Eq.\ \eqref{eq: Hamiltonian exciton--exciton reduced}, the biexciton eigenvalue problem restricted to the ground state ultimately reads
\begin{equation}
\label{eq: biexction effective BSE}
	\begin{split}
		&(\varepsilon_{\BXTM /2+\BXRM }+\varepsilon_{\BXTM /2-\BXRM }) \BXWF{S_{\mathrm{c}}}{S_{\mathrm{v}}}{}{\BXTM}(\BXRM ) \\
	    &+ \frac{1}{\mathcal{V}} \sum_{\BXRM '} V^{\mathrm{eff}}_{S_{\mathrm{c}}}(\BXTM ,\BXRM ,\BXRM ') \BXWF{S_{\mathrm{c}}}{S_{\mathrm{v}}}{}{\BXTM}(\BXRM') = \mathcal{E}_{\BXTM } \BXWF{S_{\mathrm{c}}}{S_{\mathrm{v}}}{}{\BXTM}(\BXRM ) ,
	\end{split}
\end{equation}
where
\begin{equation}
\label{eq: exciton--exciton potential effective}
	\begin{split}
		V^{\mathrm{eff}}_{S_{\mathrm{c}}}(\BXTM ,\BXRM ,\BXRM ') = \frac{(-1)^{S_{\mathrm{c}}}}{2 \mathcal{V}}\sum_{\boldsymbol{p}} \NormalizationFunction^{-1}_{S_{\mathrm{c}}}(\BXTM ,\BXRM ,\boldsymbol{p}) \mathcal{K}^{\mathrm{c}}(\BXTM ,\boldsymbol{p},\BXRM ')& \\
		\times \, (\varepsilon_{\BXTM /2+\boldsymbol{p}} + \varepsilon_{\BXTM /2-\boldsymbol{p}} - \varepsilon_{\BXTM /2+\BXRM '} - \varepsilon_{\BXTM /2-\BXRM '})&
		\\
		+ \, \frac{1}{\mathcal{V}}\sum_{\boldsymbol{p}} \NormalizationFunction^{-1}_{S_{\mathrm{c}}}(\BXTM ,\BXRM ,\boldsymbol{p}) 
		[\mathcal{U}^0 - (-1)^{S_{\mathrm{c}}} \hspace{0.3mm} \mathcal{U}^{\mathrm{c}}](\BXTM ,\boldsymbol{p},\BXRM ')& .
	\end{split}
\end{equation}
The result of Eq.\ \eqref{eq: biexction effective BSE} has the usual form of a two-particle Schrödinger equation, so that $V^{\mathrm{eff}}_{S_{\mathrm{c}}}$ can be interpreted as an effective potential responsible for the binding of two ground-state excitons into a biexciton.
Here, $\NormalizationFunction^{-1}_{S_{\mathrm{c}}}$ effectively acts as a normalization factor for the potential.
Even though the effective Hamiltonian matrix does not explicitly depend on $S_{\mathrm{v}}$, each biexciton state obtained from Eq.\ \eqref{eq: biexction effective BSE} for a given value of $S_{\mathrm{c}}$ will correspond to a state with a particular $S_{\mathrm{v}}$ depending on its parity according to Eq.\ \eqref{eq: biexciton effective wavefunction exciton exchange}.
Alternatively, as explained in Appendix\ \ref{app: effective int with e and h exchange}, the biexciton problem can be written in a completely analogous form where the potential depends on both $S_{\mathrm{c}}$ and $S_{\mathrm{v}}$, and which does not require to explicitly consider the parity of the resulting wave functions.
Furthermore, while $V^{\text{eff}}_{S_{\mathrm{c}}}$ is explicitly non-hermitian, the biexciton energies will be real if $R_{S_{\mathrm{c}}}$ is positive definite.
This follows from the fact that Eq.\ \eqref{eq: biexction effective BSE} arises from the generalized eigenvalue problem of Eq.\ \eqref{eq: biexciton eigenvalue equation reduced}, which has real eigenvalues for positive-definite $R$ because both $h$ and $R$ are hermitian \cite{parlett1998symmetric}.
The effective potential of Eq.\ \eqref{eq: exciton--exciton potential effective} is one of the main results of this work.

The spin-dependent effective potential of Eq.\ \eqref{eq: exciton--exciton potential effective} between two ground-state excitons is in general nonlocal.
This means that the potential term in the Schrödinger equation in position space does not have the usual product form $V(\vec{r}) \Psi(\vec{r})$; rather, it looks like $\int \mathrm{d}^{d} r' \, V(\vec{r}, \vec{r}') \Psi(\vec{r}')$.
Moreover, the potential consists of two qualitatively different terms whose interpretation is as follows.
The term in the second line contains the matrix elements $\mathcal{U}^{0}$ and $\mathcal{U}^{\mathrm{c}}$, which describe direct and electron-exchange interaction processes, respectively.
These arise simply from the electrostatic interaction between all possible combinations of the constituent particles of each exciton.
Their detailed expressions, along with explanations of the scattering processes that take place, are found in Appendix\ \ref{app: Second Quantization Interaction Components}.
Meanwhile, the term in the first line of Eq.\ \eqref{eq: exciton--exciton potential effective} does not depend on the electrostatic interaction.
Instead, it contains the exciton eigenenergies and arises purely from an exciton's ability to exchange its constituents with the other exciton.
Because of the presence of these kinetic operators, when expressed in coordinate space this term will lead to derivatives of the biexciton wave function.
It thus represents nonadiabatic corrections to the otherwise fully electrostatic potential.
In fact, when the effective-mass approximation holds, this term becomes directly proportional to the inverse exciton mass $M_{X}$.
In the limit of an infinitely heavy hole, to be analyzed in Sec.\ \ref{sec: Hydrogenic Example}, this represents corrections to the Born--Oppenheimer approximation.
Finally, the presence of the overlap integral $\mathcal{K}^{\mathrm{c}}$ (which is also present within $R$) is a direct consequence of the composite nature of the excitons.
Setting $\mathcal{K}^{\mathrm{c}}$ to zero (which in turn gives $R = R^{-1} = 1$) implies neglecting the wave function overlap of the two excitons.
It thus removes the nonadiabatic corrections and leads to a potential made up of purely electrostatic processes.
This is then precisely the interaction that appears in bosonized Hamiltonians, such as the one presented in Ref.\ \cite{Hanamura_1977}.

Eqs.\ \eqref{eq: biexction effective BSE} and \eqref{eq: exciton--exciton potential effective} are valid for excitons in any dimension and for an arbitrary electrostatic potential between electrons, as long as the underlying electrons and holes possess a two-state (pseudo)spin degree of freedom that separates from the relative exciton wave function.
While the latter assumption significantly simplifies the expressions, the variational approach employed here can be straightforwardly generalized to situations where this is not the case.
In Sec.\ \ref{sec: Hydrogenic Example} we will study the potential of Eq.\ \eqref{eq: exciton--exciton potential effective} in the case of 2D hydrogenic excitons in the heavy-hole limit.

\subsection{Corrections from excited states}
\label{sec: corrections excited}

In the previous section we assumed that the effects of excited states on the exciton--exciton interactions are negligible.
This was motivated by the fact that in parabolic-band semiconductors, the ground-state excitons lie much lower in energy than their excited counterparts, as their binding energies depend on the principal quantum number $n = 0, 1, 2, \dots$ as $(n + 1/2)^{-2}$ and $(n + 1)^{-2}$ in 2D and 3D, respectively.
However, the potential we have obtained does not always give the correct behavior at long distances.
In 3D systems and 2D monolayers, the latter is expected to be of van der Waals type, as excitons are polarizable quasiparticles which can give rise to induced electric dipoles via virtual transitions to excited states.
However, our procedure above neglects these dipole transitions, resulting in an effective potential which in these systems decays much faster than the expected behavior.

To remedy this, we consider the effect of excited states by starting from the variational problem in the form of Eq.\ \eqref{eq: biexciton eigenvalue equation reduced}.
At large distances, their inclusion leads to a correction term $h \rightarrow h + \delta h$ in the generalized eigenvalue problem, which in the case of the ground-state excitons reads
\begin{equation}
\label{eq: delta h 0000}
	\begin{split}
		&\delta h_{00}^{00}(\BXTM, \BXRM, \BXRM') \\
		&\approx {-} \frac{1}{\mathcal{V}} \sum_{\nu \nu' \vec{p}} \frac{[\mathcal{U}^{0}]_{00}^{\nu \nu'} \! (\BXTM, \BXRM, \vec{p}) [\mathcal{U}^{0}]_{\nu \nu'}^{00}(\BXTM, \vec{p}, \BXRM')}{\varepsilon^{\nu}_{\BXTM/2 + \vec{p}} + \varepsilon^{\nu'}_{\BXTM/2 - \vec{p}} - \mathcal{E}_{\BXTM}} .
	\end{split}
\end{equation}
Here, the index ``0'' labels the exciton ground state, while $\nu$ and $\nu'$ run over excited states only.
The detailed computation of Eq.\ \eqref{eq: delta h 0000} is found in Appendix\ \ref{app: corrections excited}.
While this expression is general, we can evaluate this further in the situation where the effective-mass approximation is valid.
In this case the relative exciton wave functions do not depend on the total exciton momentum, and thus neither does $\mathcal{U}^{0}$.
Then the matrix elements appearing in the above expression can be conveniently computed in the dipole approximation as
\begin{equation}
	[\mathcal{U}^{0}]_{00}^{\nu \nu'} \! (\BXRM, \BXRM') \approx V(\BXRM - \BXRM') [\vec{d}_{0 \nu} \vec{\cdot} (\BXRM - \BXRM')] [\vec{d}_{0 \nu'} \vec{\cdot} (\BXRM - \BXRM')] ,
\end{equation}
where $\vec{d}_{0 \nu} = \langle 0 \vert \vec{r} \vert \nu \rangle$ is the transition-dipole matrix element between the ground state and the excited state $\nu$.
Furthermore, the denominator of Eq.\ \eqref{eq: delta h 0000} can be well approximated by simply the difference in binding energies, so that it becomes independent of the kinetic energy, i.e.,
\begin{equation}
	\varepsilon^{\nu}_{\BXTM/2 + \vec{p}} + \varepsilon^{\nu'}_{\BXTM/2 - \vec{p}} - \mathcal{E}_{\BXTM} \approx 2 \varepsilon^{\mathrm{b}}_{0} - \varepsilon^{\mathrm{b}}_{\nu} - \varepsilon^{\mathrm{b}}_{\nu'} .
\end{equation}
With these approximations, the interaction energy shift of Eq.\ \eqref{eq: delta h 0000} gives rise to a local potential in position space which reads
\begin{equation}
\label{eq: Vld effective mass}
	V^{\text{ld}}(\vec{r}) = {-} \sum_{\nu \nu'} \frac{|(\vec{d}_{0 \nu} \vec{\cdot} \vec{\nabla}) (\vec{d}_{0 \nu'} \vec{\cdot} \vec{\nabla}) V(r)|^{2}}{2 \varepsilon^{\mathrm{b}}_{0} - \varepsilon^{\mathrm{b}}_{\nu} - \varepsilon^{\mathrm{b}}_{\nu'}} ,
\end{equation}
where the superscript ``ld'' emphasizes that this is valid at long distances.
In this regime it is $V^{\text{ld}}$, as opposed to the potential $V^{\text{eff}}$ of Eq.\ \eqref{eq: exciton--exciton potential effective}, that correctly approximates the interaction between two excitons.
As an example we consider hydrogenic excitons in 2D with the standard Coulomb interaction $V(r) = e^{2} / 4 \pi \epsilon r$, with $\epsilon$ the effective dielectric constant, and find
\begin{equation}
	V^{\text{ld}}(\vec{r}) = {-} \bigg(\frac{e^{2}}{4 \pi \epsilon r^{3}}\bigg)^{2} \sum_{\nu \nu'} \frac{|\vec{d}_{0 \nu} \vec{\cdot} \vec{d}_{0 \nu'} - 3 (\vec{d}_{0 \nu} \vec{\cdot} \hat{\vec{r}})(\vec{d}_{0 \nu'} \vec{\cdot} \hat{\vec{r}})|^{2}}{2 \varepsilon^{\mathrm{b}}_{0} - \varepsilon^{\mathrm{b}}_{\nu} - \varepsilon^{\mathrm{b}}_{\nu'}} ,
\end{equation}
where $\hat{\vec{r}}$ is a unit vector in the direction of $\vec{r}$.
In view of the $s$-wave nature of the exciton ground state, we expect the angular dependence to drop out from the above expression once the sum over states is explicitly carried out.
We then obtain a van der Waals interaction reflecting the induced dipole--dipole nature of the exciton--exciton interaction at large distances.
Indeed, keeping only the first set of degenerate excited states in the sums over $\nu$ and $\nu'$ yields.
\begin{equation}
	V^{\mathrm{ld}}(\vec{r}) \approx {-} \frac{45}{64} \bigg(\frac{e^{2}}{4 \pi \epsilon}\bigg)^{2} \frac{\mu_{X} \aBohr^{2} d_{01}^{4}}{\hbar^{2}} \frac{1}{r^{6}} ,
\end{equation}
where $\mu_{X}$ and $\aBohr$ are the reduced mass and the Bohr radius of the exciton, respectively, and $d_{01} = (27 / 32 \sqrt{3}) \aBohr$ is the magnitude of the dipole matrix element between the ground state and the $p$-wave excited states with azimuthal quantum number $m = \pm 1$.
The above result immediately provides the value of the van der Waals $C_{6}$ coefficient via $V^{\mathrm{ld}}(r) = {-} C_{6} / r^{6}$ \cite{weiner1999experiments,kohler2006production,chin2010feshbach}.

Finally, we note that the large-distance behavior of Eq.\ \eqref{eq: exciton--exciton potential effective} will not necessarily be incorrect in all systems.
For instance, in the case of electron--hole bilayers, the excitons form permanent dipoles and thus the interaction is expected to decay as $1 / r^{3}$ at long distances.
This behavior must already arise from the effective potential between ground-state excitons \emph{without} including excited-state corrections if the correct bilayer electron--hole potential is used in the computation.
In cases like this, the corrections presented in this section will be subleading.

\subsection{Notes on numerical implementation}
\label{sec: numeric notes}

In Sec.\ \ref{sec: ground state excitons} we have cast the biexciton problem in the form of a generalized eigenvalue equation and an equivalent Schrödinger equation with a kinetic term and a potential term.
From the point of view of parameter-free first-principles calculations of biexciton states, it is worth devoting some attention to the discussion of the computational cost of solving this problem.

In order to obtain the biexciton spectrum via our method, one must first solve the exciton BSE of Eq.\ \eqref{eq: exciton BSE} to obtain the ground-state exciton energies and wave functions as a function of the total exciton momentum $\XTM$.
Then, the interaction matrix elements $\mathcal{U}^{0}$ and $\mathcal{U}^{\mathrm{c}}$ of Eqs.\ \eqref{eqappx: direct term} and \eqref{eqappx: conduction exchange term} must be evaluated for a given (fixed) total biexciton momentum $\BXTM$.
Similarly, one must evaluate the overlap integral $\mathcal{K}^{\mathrm{c}}$ of Eq.\ \eqref{eq: conduction electron exchange element} in order to obtain $R$ for a given spin $S_{\mathrm{c}}$ according to Eq.\ \eqref{eq: biexciton projector ground state}.
Note that the $\BXRM$ and $\BXRM'$ labels in these matrix elements depend on the total exciton momentum $\XTM$, thus the exciton spectrum must be known on a dense enough mesh.
With these ingredients, one can compute the Hamiltonian of Eq.\ \eqref{eq: Hamiltonian exciton--exciton ground state} for the spin $S_{\mathrm{c}}$ and proceed to solve the generalized eigenvalue equation $(h - \mathcal{E} R) \Psi = 0$, or the equivalent Schrödinger equation of Eq.\ \eqref{eq: biexction effective BSE} to obtain the biexciton energies and wave functions.
While this discussion focuses on the specific case of spin degeneracy and a single exciton ground state, our method can be used in more complicated situations by considering the generalized eigenvalue equation of Eq.\ \eqref{eq: biexciton eigenvalue equation reduced}, which includes coupling between all excitonic states labeled by $\Xmodeindex$.

In order to obtain accurate results, close attention must be paid to the convergence of the interaction matrix elements and overlap integrals.
In this regard, there are two especially important factors to consider.
Firstly, if a continuum model is used for the exciton states (such as in the case of the Wannier model), then care must be taken to correctly handle the edge effects that arise from the momentum shifts within the wave functions appearing in the matrix elements of $\mathcal{U}$ and $\mathcal{K}^{\mathrm{c}}$.
This problem completely disappears if a lattice model is used, as the compact Brillouin zone enforces the periodicity of the wave functions and dispersions, or if the wave functions are known analytically or a good variational approximation is available.
Secondly, attention must be paid to the effects of a potential infrared divergence of the electron--hole interaction by either fixing a reasonable cutoff or extrapolating the results to the divergent case.

If the exciton BSE is solved on a momentum grid with $N_{\XRMM}$ points in each direction and taking into account $N_{\mathrm{c}}$ and $N_{\mathrm{v}}$ conduction and valence bands, respectively, then each exciton wave function at fixed total exciton momentum $\XTM$ consists of an array of $N_{k}^{d} \times N_{\mathrm{c}} \times N_{\mathrm{v}}$ points, with $d$ the number of spatial dimensions.
If these wave functions (and their associated energies) are known on a $\XTM$ mesh of $N_{q}$ points in each direction, then the biexciton Hamiltonian at a fixed total biexciton momentum $\BXTM$ will be an array of $(N_{q}^{d} \times N_{\mathrm{X}}^{2})^{2}$ points, where $N_{\mathrm{X}}$ is the number of exciton states to be taken into account for the biexciton calculation.
Because of the fast-growing nature of the problem with all variables involved, it is convenient to analyze the symmetries of the exciton problem under consideration in order to potentially reduce the computational cost of the calculation by splitting it into orthogonal blocks.
An example of this is precisely our rewriting of the problem in terms of the coupled conduction and valence spins $S_{\mathrm{c}}$ and $S_{\mathrm{v}}$, which effectively leads to solving two problems with $N_{\mathrm{X}} = 1$ instead of a single problem with $N_{\mathrm{X}} = 4$.

While we expect the computational cost of the theory to be significant in complicated systems, we have evaluated the effective exciton--exciton potential and used it to compute biexciton binding energies in tight-binding models with a contact electron--hole interaction with a single conduction and valence band.
In this simple case, convergence in most cases was excellent already for $N_{\XRMM}$ of the order of $20$.
The results will be presented in a forthcoming paper, where the theory we have developed will be generalized to include the effects of topology, Berry curvature, and band geometry, among others.
This shows that our approach can indeed be used to obtain physical results, and we hope it will inspire the development of high-performing algorithms and solvers to facilitate its application also in more complex situations.

\section{An example: Hydrogenic excitons in 2D}
\label{sec: Hydrogenic Example}

In general, the effective potential of Eq.\ \eqref{eq: exciton--exciton potential effective} is a nonlocal quantity in position space.
To gain some insight into its behavior, in this section we study hydrogen-like excitons in 2D.
We consider a 2D system with parabolic bands with masses $m_{c}$ and $m_{v}$ for electrons and holes, respectively, so that the energies of the conduction and valence bands are given by
\begin{subequations}
\label{eq: quadratic dispersion relations}
    \begin{align}
        \epsilon^c_{\vec{p}} &= \hphantom{{-}}\frac{E_{\mathrm{g}}}{2} + \frac{p^2}{2m_{c}}, \\
        \epsilon^v_{\vec{p}} &=  {-}\frac{E_{\mathrm{g}}}{2} - \frac{p^2}{2m_{v}} ,
    \end{align}
\end{subequations}
with $E_{\mathrm{g}}$ the band gap of the semiconductor.
Note that we work in units where $\hbar = 1$.
We model the electron--electron repulsion via the standard Coulomb potential
\begin{equation}
	 V(\vec{p}) = \frac{e^2}{2 \epsilon p} ,
\end{equation}
where ${-}e$ is the charge of the electron, and $\epsilon = \epsilon_{0} \epsilon_{\mathrm{r}}$ is the total dielectric constant, with $\epsilon_0$ the vacuum permittivity and $\epsilon_\mathrm{r}$ the (dimensionless) effective relative dielectric constant of the surrounding medium.
The eigenstates of the excitonic Wannier problem are completely analogous to those of the 2D hydrogen atom with a reduced mass $\mu_{X} = m_{c} m_{v} / (m_{c} + m_{v})$.
This problem admits well-known analytical solutions  \cite{chao1991analytical,yang1991analytic,Parfitt_2002,efimkin2021electron}.
The eigenstates are labeled by a principal quantum number $n$ and an azimuthal quantum number $m$, both of which are zero in the ground state.
The corresponding ground-state wave function in momentum space reads
\begin{align}
\label{eq: Hydrogen wavefunction}
    \Phi(\XRM)=\frac{2 \sqrt{2\pi} a_{X}}{(1 + a_{X}^{2} \XRMM^2)^{\nicefrac{3}{2}}} ,
\end{align}
where we have omitted the ground-state quantum numbers $n, m = 0$.
Here, $a_{X}$ is the average radius of an exciton in the hydrogenic ground state, i.e., $\langle \Phi \vert r \vert \Phi \rangle = a_{X}$.
In terms of the Bohr radius $\aBohr = 4 \pi \epsilon / \mu_{X} e^{2}$, one has $a_{X} = a_{0} / 2$ in 2D.
Meanwhile, the ground-state binding energy is $\varepsilon^{\mathrm{b}}_{X} = 1 / (2 \mu_{X} a_{X}^{2}) = e^{2} / 4 \pi \epsilon a_{X}$.
Thus, the total energy of a 2D hydrogenic ground-state exciton with CoM momentum $\XTM$ is given by
\begin{align}
	\varepsilon_{\XTM} = E_{\mathrm{g}} + \frac{\XTM^2}{2 M_{X}} - \varepsilon^{\mathrm{b}}_{X} ,
\end{align}
where we recall that $M_{X} = m_{c} + m_{v}$ is the total (effective) mass of the exciton.
The exciton dispersion is parabolic due to the decoupling of the CoM and relative motions that takes place in the case of parabolic bands.
Consequently, the relative wave function of Eq.\ \eqref{eq: Hydrogen wavefunction} does not depend on the total exciton momentum.
As a result, neither the potential of Eq.\ \eqref{eq: exciton--exciton potential effective} nor the relative biexciton wave function depend on the total biexciton momentum $\BXTM$.
Thus, we conclude that the biexciton dispersions in a system of parabolic bands are also parabolic, in the same way as those of the single excitons.

In the following section we study the infinite-mass limit for the holes, in which case the effective potential can be evaluated explicitly.
We then comment on the regime of similar masses.

\subsection{Heavy-hole limit}

\begin{figure}
    \centering
    \includegraphics[width=\linewidth]{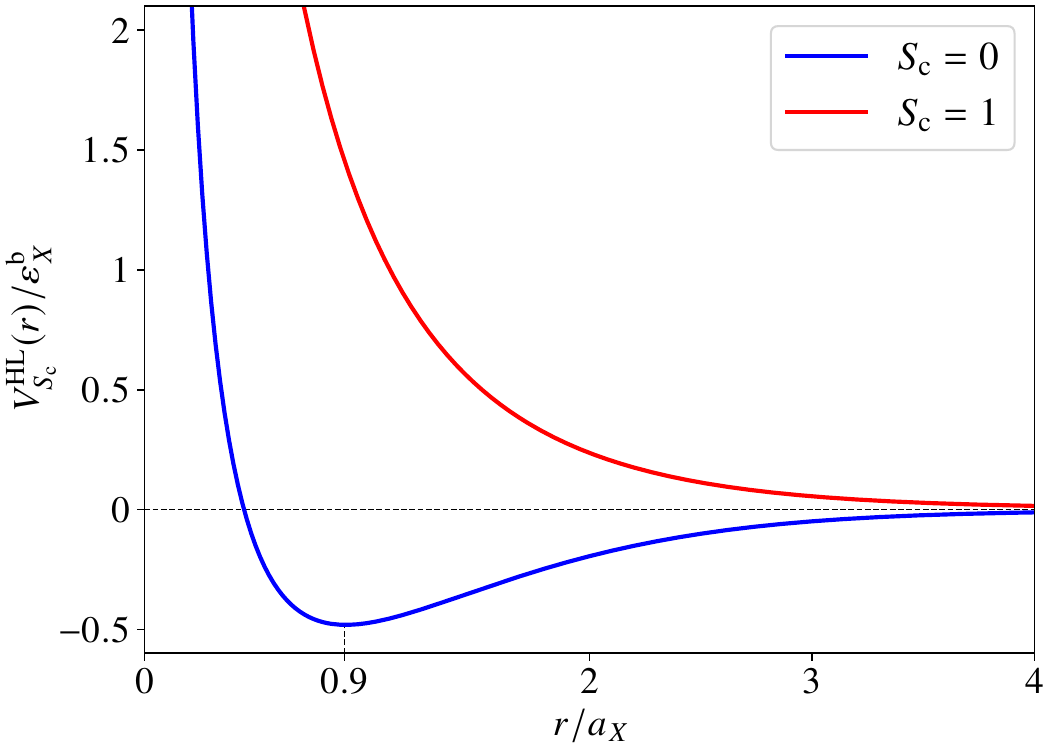}
    \caption{Exciton--exciton potential in the heavy-hole limit.
    The blue curve corresponds to the singlet state of the coupled conduction electrons $(S_{\mathrm{c}} = 0)$ and displays an attractive part with a minimum at around $r = 0.9 a_{X}$.
    The red curve corresponds to the triplet configuration $(S_{\mathrm{c}} = 1)$ and is repulsive.
    These functions are exactly the singlet and triplet Heitler--London potentials obtained in the Born--Oppenheimer approximation for the dihydrogen molecule.
    The radial coordinate and the effective momentum-space potential have been made dimensionless via the mean exciton radius $a_{X}$ and the exciton binding energy $\varepsilon^{\mathrm{b}}_{X}$, respectively, both defined in the main text.
    }
    \label{fig: Heitler--London potential}
\end{figure}

We now consider the heavy-hole limit by taking $m_{v} \rightarrow \infty$.
In this case we will see that the potential becomes a local quantity in position space.
The first thing to note in this case is that $\varepsilon_{\XTM}$ becomes a momentum-independent constant and thus the first term of Eq.\ \eqref{eq: exciton--exciton potential effective} vanishes.
Furthermore, the electron-exchange overlap integral can be computed analytically.
Its position-space expression reads $\mathcal{K}^{\mathrm{c}}(\boldsymbol{r},\boldsymbol{r}') = \mathcal{K}^{\mathrm{c}}(r) \delta(\boldsymbol{r}-\boldsymbol{r}')$ with
\begin{align}
\label{eq: hydrogen integral conduction exchange term}
	 \mathcal{K}^{\mathrm{c}}(r) = \frac{1}{4} \bigg(\frac{r}{a_{X}}\bigg)^{4} K_{2}\bigg(\frac{r}{a_{X}}\bigg)^2 ,
\end{align}
where $K_{n}$ is the modified Bessel function of the second kind of order $n$.
Because of the Dirac delta in $\mathcal{K}^{\mathrm{c}}(\boldsymbol{r},\boldsymbol{r}')$, both $\NormalizationFunction_{S_{\mathrm{c}}}$ and $\NormalizationFunction^{-1}_{S_{\mathrm{c}}}$ become local operators in position space.
The integrals contained in $\mathcal{U}^{0}$ and $\mathcal{U}^{\mathrm{c}}$ can all be performed analytically except for a single one, and we give their expressions in Appendix\ \ref{app: Heavy-Hole Integrals}.
Ultimately we obtain the local interaction potential
\begin{align}
\label{eq: Heitler--London potential real space}
    V^{\mathrm{HL}}_{S_{\mathrm{c}}}(r) =
    \frac{\mathcal{U}^0(r) - (-1)^{S_{\mathrm{c}}}\hspace{0.3mm}\mathcal{U}^{\mathrm{c}}(r)}{1 + (-1)^{S_{\mathrm{c}}}\hspace{0.3mm}\mathcal{K}^{\mathrm{c}}(r)} ,
\end{align}
which only depends on the magnitude of $\vec{r}$.
As mentioned, $S_{\mathrm{c}}$ is the coupled total spin of the two conduction electrons and takes values $0$ or $1$ for spin-$\nicefrac{1}{2}$ electrons, but we stress that this result is also valid when one or both of the particles have higher total spin.
The label ``HL'' indicates that the two possibilities in Eq.\ \eqref{eq: Heitler--London potential real space} are identical to the singlet and triplet potentials of the Heitler--London problem for the dihydrogen molecule in the Born--Oppenheimer approximation \cite{Heitler_1927}.
Hence, the heavy-hole limit of Eq.\ \eqref{eq: exciton--exciton potential effective} exactly reproduces the Heitler--London physics where the heavy holes play the role of the protons in the dihydrogen problem.
Thus, the general interaction potential of Eq.\ \eqref{eq: exciton--exciton potential effective} may be understood as a generalization of the Heitler--London result to the case of arbitrary masses.
In Fig.\ \ref{fig: Heitler--London potential} we have plotted $V^{\mathrm{HL}}_{S_{\mathrm{c}}}$ for $S_{\mathrm{c}} = 0, 1$, corresponding to spin-$\nicefrac{1}{2}$ electrons in the singlet and triplet configuration, respectively.
As is well known from the Heitler--London physics, in the limit of one species being much heavier than the other it is the spin state of the lighter particles that determines the nature of the potential \footnote{Note that we could have also eliminated $S_{\mathrm{c}}$ in favor of $S_{\mathrm{v}}$ in Eq.\ \eqref{eq: exciton--exciton potential effective} and performed the heavy-electron limit instead of the heavy-hole limit, which would result in Eq.\ \eqref{eq: Heitler--London potential real space} depending only on $S_{\mathrm{v}}$.}.
The potential is fully repulsive in the triplet configuration but has an attractive well in the singlet configuration.
In this limit the spin state of the heavy particles only plays an ancillary role leading to the excitonic analogs of the orthohydrogen and parahydrogen isomers.

\subsection{Similar masses}

We briefly comment on the opposite limit, namely that of equally massive electrons and holes.
Our considerations in this section are not restricted to the Coulomb potential, but are valid for an arbitrary central potential.
However, we do assume the validity of the effective-mass approximation, so that the wave functions do not depend on the total exciton momentum $\XTM$.
Under these conditions, the direct interaction $\mathcal{U}^{0}$ vanishes for excitons in an $s$-wave state.
Furthermore, we find that $\mathcal{U}^{\mathrm{c}}(\BXRM, \BXRM') = \mathcal{U}^{\mathrm{c}}(\BXRM, {-}\BXRM')$, and similarly for $\mathcal{K}^{\mathrm{c}}$, where we have omitted the dependence on $\BXTM$ as we consider the validity of the effective-mass approximation.
As a result, the effective potential for $m_{c} = m_{v}$ as a whole satisfies $V^{\text{eff}}_{S_{\mathrm{c}}}(\BXRM, \BXRM') = V^{\text{eff}}_{S_{\mathrm{c}}}(\BXRM, {-}\BXRM')$.
It is easy to show that then $\sum_{\BXRM'} V^{\text{eff}}_{S_{\mathrm{c}}}(\BXRM, \BXRM') \Psi^{S_{\mathrm{c}} S_{\mathrm{v}}}(\BXRM') = 0$ for wave functions with negative parity under reflection, i.e., $\Psi^{S_{\mathrm{c}} S_{\mathrm{v}}}({-}\BXRM) = {-} \Psi^{S_{\mathrm{c}} S_{\mathrm{v}}}(\BXRM)$.
Thus, in the case of equal masses, the potential we have derived yields no biexciton states with negative parity.
Recalling the property of Eq.\ \eqref{eq: biexciton effective wavefunction exciton exchange} for the biexciton wave function, we see that effectively this means that there are no solutions with $S_{\mathrm{c}} + S_{\mathrm{v}} = 1$.
Consequently, for such spin states, the excitons effectively do not interact at this level of approximation.
The leading corrections would have to be obtained by including in the effective potential the effect of states with principal number $n \ge 1$.
In any case, we expect biexcitons with $S_{\mathrm{c}} + S_{\mathrm{v}} = 1$ to be very lightly bound in comparison with their counterparts with $S_{\mathrm{c}} + S_{\mathrm{v}} = 0, 2$.
We stress that the validity of the effective-mass approximation is crucial for this argument, as otherwise the effective potential does generally not satisfy the aforementioned property.

On another note, away from the heavy-hole limit, the term depending on the exciton energies in the first line of Eq.\ \eqref{eq: exciton--exciton potential effective} is nonzero.
This term indicates that the effective potential between excitons is not only a result of the scattering processes that take place between the individual constituents, but has a part originating purely from the Pauli exchange principle incorporated by the presence of $\mathcal{K}^{\mathrm{c}}$.
In the regime of similar masses this term is of the same order as that of the third line, and thus we expect it to significantly influence the binding energies of biexcitons.
This contrasts with the study of Ref.\ \cite{cam2022symmetry} on exciton--exciton interactions in transition-metal dichalcogenides, where the biexciton energies are obtained by considering only the effect of our $\mathcal{U}^{\mathrm{c}}$ when $m_{c} = m_{v}$.


In summary, we have shown how the potential of Eq.\ \eqref{eq: exciton--exciton potential effective} reproduces the Heitler--London singlet and triplet potentials in the limit of heavy holes.
This equivalence is exact in the case of hydrogen-like excitons with a $1 / r$ electron--hole attraction and we expect similar results for other more realistic interactions.
In particular, it would be interesting to consider a potential of the Rytova-Keldysh type, which more accurately models the attraction between electrons and holes in many quasi-2D systems and semiconductor quantum wells.
Given the fact that for this interaction the wave functions cannot be obtained in an analytic closed form, in this work we have focused on the idealized Coulomb scenario and leave the investigation of more complicated potentials for future works with a more numerical focus.

\section{Field-theoretical approach}
\label{sec: Field Theoretic Approach}


In this section, we set up the many-body theory for excitons using the path-integral formalism.
Within this formalism, excitons will no longer be described by operators, but by bosonic fields.
To accomplish this, we will first perform appropriate Hubbard--Stratonovich transformations on the action describing interacting conduction and valence electrons.
This procedure yields a formal effective action for the so-called polarization field, whose fluctuations correspond to interband excitations, i.e., excitons.
The resulting action incorporates a two-body interaction term, which we will compare to the exciton--exciton interaction obtained in the previous section.
Additionally, this effective action will be used to calculate the two-exciton propagator.

\subsection{Electronic action} \label{sec:FTElectronicAction}
We consider two Grassmann-valued fields $\phi_{c}$ and $\phi_{v}$ describing conduction and valence electrons, respectively.
These fields depend on a position-space label $\boldsymbol{x}$, imaginary time $\tau$, and spin label $\alpha$.
By defining a combined spacetime index $x \equiv (\vec{x}, \tau)$, the Euclidean action describing a gas of interacting conduction and valence electrons reads
\begin{align}
	\nonumber
	&S[\phi^*_c,\phi^*_v,\phi_c,\phi_v] \\
	\label{eq: initial conduction valence action}
	&= \sum_{a \sigma}\int_{x}
	\phi^*_{a \sigma}(x) 
	\bigg( \frac{\partial}{\partial\tau} + \elecEnergy^{a}_{\sigma}({-} \mathrm{i} \vec{\nabla}) \bigg)
	\phi_{a \sigma}(x)
	\\
	&+ \frac{1}{2} \sum_{\substack{aa' \\ \sigma \sigma'}} \int_{xx'} \phi^*_{a \sigma}(x) \phi^*_{a' \sigma'}(x') 
	V(x - x') \phi_{a' \sigma'}(x') \phi_{a \sigma}(x) , \nonumber
\end{align}
where $a \in \{c,v\}$ and we consider an instantaneous interaction potential $V(x) = V(\vec{x}) \delta(\tau)$.
We also note that formally we should include in the quadratic part of the action an effective potential associated with the positively charged ionic background.
However, it is known that this cancels against the zero-momentum contribution of the Hartree self-energy, and thus we omit it for the sake of brevity.
The position-space integrals run over the system volume and the imaginary-time integrals run from $0$ to $\beta = 1/k_{\mathrm{B}}T$, and $\int_{x} \equiv \int \mathrm{d}^{d} x \, \mathrm{d} \tau$.
Note that the inverse thermal energy must not be confused with the valence-band spin index, which we also denote by $\beta$.
Using the functional inner-product notation of Ref.\ \cite{Stoof_2008} and reviewed also in Appendix\ \ref{app: inner product notation}, we rewrite this action as
\begin{equation}
\begin{split}
\label{eq: action electrons suggestive}
	&S[\phi^*_c, \phi^*_v,\phi_c,\phi_v] 
    \\
    &= -\sum_{a}\bigg\{ (\phi_a|G^{-1}_{0,a}|\phi_a) - \frac{1}{2}(\phi^*_a\phi_a \lvert V \rvert \phi^*_a\phi_a) \bigg\}
    \\
    &\hphantom{= \,\,} -(\phi^*_v\phi_c \lVert V \rVert \phi^*_v\phi_c).
\end{split}
\end{equation}
The inverse noninteracting Green's function of the system is $G^{-1}_{0, a; \sigma \sigma'}(x,x') = G^{-1}_{0, a \sigma}(x,x') \delta_{\sigma \sigma'}$ with
\begin{equation} \label{eq: Green's function non-interacting real space}
	{-}G^{-1}_{0, a \sigma}(x, x') = \bigg( \frac{\partial}{\partial\tau} + \elecEnergy^{a}_{\sigma}({-} \mathrm{i} \vec{\nabla}) \bigg) \delta(x - x') .
\end{equation}
Finally, the partition function associated with the above Euclidean action reads
\begin{equation} \label{eq: partition function conduction valence}
	\mathcal{Z} = \int\mathcal{D}\phi^*_{c} \, \mathcal{D}\phi_{c} \, \mathcal{D}\phi^*_{v} \, \mathcal{D}\phi_{v} \, \mathrm{e}^{-S[\phi^*_c,\phi^*_v,\phi_c,\phi_v] }.
\end{equation}

\subsection{Polarization-field action} \label{sec:FTPolarizationAction}
In this section we will perform multiple Hubbard--Stratonovich transformations to arrive at a formal action for the polarization field.
We first introduce a complex bosonic polarization field, whose expectation value we demand to be related to the electron fields via
\begin{equation}
	\langle \mathcal{P}_{\alpha\beta}(\boldsymbol{x},\boldsymbol{x}',\tau)\rangle = \langle \phi^*_{v\beta}(\boldsymbol{x}',\tau)\phi_{c\alpha}(\boldsymbol{x},\tau)\rangle.
\end{equation}
While this field can be used to decouple the interband electron--hole interaction term, we must also remove the purely repulsive couplings between electrons of the same species.
Therefore, we additionally introduce two real bosonic fields $\rho_{c}, \rho_{v}$ which satisfy
\begin{equation}
	\langle \rho_{a \sigma}(x) \rangle = 
	\langle \phi^*_{a \sigma}(x)\phi_{a \sigma}(x) \rangle .
\end{equation}
In practice, we consider a homogeneous system and absorb the Dirac-sea effects associated with the filled band into the single-particle propagator.
Consequently, the only contribution from the density fields effectively arises from their fluctuations.
With these definitions, we multiply the partition function of Eq.\ \eqref{eq: partition function conduction valence} by
\begin{subequations}
\begin{align} \label{eq: HST polarization field}
	1 &=\int \mathcal{D}\mathcal{P}^{*} \, \mathcal{D} \mathcal{P} \, \exp \bigg\{{-}(\mathcal{P} - \phi^*_v\phi_c \lVert V \rVert \mathcal{P} - \phi^*_v\phi_c)\bigg\},
	\\ \label{eq: HST density field}
	1 &= \prod_{a} \int \mathcal{D}\rho_a \, \exp \bigg\{\frac{1}{2}(\rho_a - \phi^*_a\phi_a |V| \rho_a - \phi^*_a\phi_a)\bigg\},
\end{align}
\end{subequations}
where the integral measures contain appropriate normalization factors of $\exp(\pm \Tr \ln V)$.
After integrating out the fermionic fields, the resulting action for the combined $\mathcal{P}$, $\rho_c$, and $\rho_v$ fields reads
\begin{align}
	S[\mathcal{P}^*,\mathcal{P},\rho_c,\rho_v] &= (\mathcal{P}|V|\mathcal{P}) - \frac{1}{2}(\rho_c|V|\rho_c) - \frac{1}{2}(\rho_v|V|\rho_v) \nonumber
	\\
	\label{eq: action transformed polarization and density}
	&- \Tr \ln \big[ {-} \textbf{G}^{-1}_{0} + \boldsymbol{\Sigma}^{\rho} + \boldsymbol{\Sigma}^{\mathcal{P}} \big] .
\end{align}
The boldface objects stand for matrices in a $2 \times 2$ space representing the conduction and valence degrees of freedom, henceforth referred to as the ``band space'', and carry additional spin and spacetime indices.
They read
\begin{subequations}
	\begin{align} \label{eq: Green's function non-interacting matrix}
		\textbf{G}^{-1}_{0;\alpha \beta}(x, x') &= 
			\begin{bmatrix}
				G^{-1}_{0, c \alpha}(x, x') & 0 
	        		\\
	        		0 & G^{-1}_{0, v \beta}(x, x')
			\end{bmatrix} \delta_{\alpha \beta} ,
		\\ \label{eq: selfenergy density matrix}
		\boldsymbol{\Sigma}^{\rho}_{\alpha\beta}(x, x') &= 
			 \begin{bmatrix}
				[\rho_c \cdot V]_{\alpha}(x) & 0 
				\\
				0 & [\rho_v \cdot V]_{\beta}(x)
			\end{bmatrix}
		\delta_{\alpha \beta} \delta(x - x'),
		\\
		\begin{split}
			\label{eq: selfenergy polarization matrix}
			\boldsymbol{\Sigma}^{\mathcal{P}}_{\alpha\beta}(x, x') &= {-}
				\begin{bmatrix}
					0 & \mathcal{P}_{\alpha\beta}(\boldsymbol{x},\boldsymbol{x}',\tau)
					\\
					\mathcal{P}^*_{\beta\alpha}(\boldsymbol{x}',\boldsymbol{x},\tau) & 0
				\end{bmatrix} \\
			&\hspace{0pt + \widthof{$= - \,\,$}} \times V(\boldsymbol{x} - \boldsymbol{x}')\delta(\tau - \tau').
		\end{split}
	\end{align}
\end{subequations}
These correspond to the inverse noninteracting Green's function, the Hartree-like self-energy associated with the density fields, and the Fock-like selfenergy associated with the polarization field, respectively.
Note that in writing Eqs.\ \eqref{eq: Green's function non-interacting matrix}--\eqref{eq: selfenergy polarization matrix} we have implicitly assumed that the total spins of the conduction and valence bands take on the same values.
In a more general situation, the theory can be developed along the same lines but will be somewhat more notationally inconvenient.

Since the Hubbard--Stratonovich transformation is an exact procedure, the action of Eq.\ \eqref{eq: action transformed polarization and density} is formally exact.
We have transformed the original electronic action to one describing a pair of density-fluctuation fields interacting with a polarization field.
However, we are interested in an effective action for $\mathcal{P}$ only, as we will shortly see that this field corresponds to the excitons.
Integrating out the density fluctuations at the Gaussian level formally results in the action for the polarization field
\begin{align}
\label{eq: action exact polarization}
	\begin{split}
		S[\mathcal{P}^*,\mathcal{P}] = (\mathcal{P}|V|\mathcal{P}) - \frac{1}{2} \Tr \ln [1 - (\mathbf{G}^{0}\boldsymbol{\Sigma}^{\mathcal{P}})^2 ]&
	    \\
		+ \, \frac{1}{2}(\boldsymbol{\eta} | V \cdot \mathbf{G}^{0, \rho'} \cdot V |\boldsymbol{\eta}) + 
		\frac{1}{2} \Tr \ln (1 - \boldsymbol{\pi}\cdot V)& .
	\end{split}
\end{align}
The quantities $\vec{\eta}$, $\vec{\pi}$, and $\mathbf{G}^{0, \rho'}$ are unimportant for the present discussion, and their definitions can be found in Appendix\ \ref{app: polarization field action notation}.
Here we only mention that the polarization-independent part of $V \cdot \mathbf{G}^{0, \rho'} \!\cdot V$ gives rise to the random-phase approximation (RPA) for each electron species separately.

In the next section, we show that the polarization field differs from the exciton field by a basis transformation. 
Therefore, we can assert that the above action provides a many-body description for a gas of Wannier excitons up to arbitrary order in the exciton field. This action, in principle, contains all many-body interactions between the excitons.
After deriving the exciton propagator, we will reduce this action to one that is quartic in the exciton fields, in order to obtain an effective description for a dilute gas of excitons with a two-body interaction.

Before moving on, we give two additional remarks on the results so far. 
Firstly, on physical grounds, one might object to the use of the partition function for the description of a gas of interacting excitons, since excitons would not necessarily be present if the system (i.e., a semiconductor) were in equilibrium. 
Nevertheless, thermalization of excitons after their formation can be much faster than the recombination of the electron--hole pair, so that they can exist in a quasi-equilibrium state \cite{Hanamura_1974,sun2014observation,Selig_2016,Steinhoff_2017,Ma_2021}.
In this transient regime our description applies. 
Secondly, when deriving the action for the polarization field, we performed a Hartree theory on the terms that were of the order $|\phi_c|^4$and $|\phi_v|^4$. 
However, it is just as well possible to do a Fock Hubbard--Stratonovich transformation for these terms.
Its derivation, which is essentially identical to the Hartree theory, is given in Sec.\ S.II of the Supplemental Material \cite{supp}.
Both approaches result in the same aforementioned effective exciton action up to the quartic order in the exciton fields, but there will be differences when truncating the full action beyond the quartic order. 
As is usual, after performing the Hubbard--Stratonovich transformation and introducing the polarization field, the most general result would be obtained by doing a Hartree--Fock theory on the remaining quartic conduction and valence-electron fields. 

\subsection{Free exciton propagator} \label{sec: Free exciton propagator}
Before introducing the effective exciton action we will derive the free inverse propagator of the polarization field.
We will use the latter to obtain the exciton BSE, which will allow us to introduce a proper exciton field.

We begin by expanding Eq.\ \eqref{eq: action exact polarization} up to quadratic order in $\mathcal{P}$.
We note that in the normal semiconductor state $\langle \mathcal{P}\rangle = 0$, so that the polarization field is equivalent to its fluctuations.
The electronic conduction (valence) propagators contained in these quadratic terms are dressed by their own RPA corrections, i.e., the conduction (valence) propagators are not dressed by the RPA bubbles of the valence (conduction) electrons.
We formally perform the resulting resummation by assuming that the underlying electronic bands already include the associated effects and do not explicitly consider such corrections any further.
In practice this is not a problem since the band structure is typically taken to reproduce $GW$ calculations which already include the effects associated with a filled Dirac sea.

In what follows it will be more convenient to work in momentum space, particularly in the exciton CoM coordinates used in Secs.\ \ref{sec: Introduction} and \ref{sec: Variational Approach}.
Accordingly, we perform the transformation
\begin{equation}
\label{eq: Fourier transform polarization field}
	\begin{split}
		\mathcal{P}_{\alpha\beta}(\XRM,\XTM, \mathrm{i}\Omega_{n}) &= \frac{1}{\sqrt{\beta} \mathcal{V}} \int_{\tau \vec{x} \vec{x}'} \mathcal{P}_{\alpha\beta}(\boldsymbol{x},\boldsymbol{y},\tau)
		\\
		&\times \,
		\mathrm{e}^{-\mathrm{i}\XRM \boldsymbol{\cdot} (\boldsymbol{x} - \boldsymbol{y})}
		\mathrm{e}^{-\mathrm{i}\XTM \boldsymbol{\cdot} (\dlmassc \boldsymbol{x} + \dlmassv \boldsymbol{y})}
		\mathrm{e}^{\mathrm{i} \Omega_{n}\tau} ,
	\end{split}
\end{equation}
with $\XRM$ and $\XTM$ the relative and total momenta, respectively, and $\Omega_{n} = 2\pi n/\beta$ a bosonic Matsubara frequency.
Furthermore, the noninteracting electron Green's functions are diagonal in momentum and frequency space, i.e.,
\begin{equation}
	G^{0}_{a \sigma}(\vec{p},\mathrm{i} \omega_n) = \frac{1}{\mathrm{i} \omega_n - \elecEnergy^{a}_{\sigma \vec{p}}} ,
\end{equation}
where $\omega_n = (2n + 1)\pi/\beta$ is a fermionic Matsubara frequency.
The free propagator of the polarization fields can now be identified from
\begin{equation}
\label{eq: PVP plus trace polarization propagator}
	(\mathcal{P} | V | \mathcal{P}) + \Tr \big[ G^{0}_{c} \cdot \Sigma^{\mathcal{P}}_{cv} \cdot G^{0}_{v} 
	\cdot \Sigma^{\mathcal{P}}_{vc} \big] \equiv {-}(\mathcal{P}|G^{-1}_{0,\mathcal{P}}|\mathcal{P}) ,
\end{equation}
where $\Sigma^{\mathcal{P}}_{cv}$ and $\Sigma^{\mathcal{P}}_{vc}$ refer to the off-diagonal components of the polarization-field self-energy of Eq.\ \eqref{eq: selfenergy polarization matrix}.
Because of conservation of total momentum and energy, and in our case also that of the spins, the inverse polarization propagator can be written as 
\begin{equation}
	\begin{split}
	    &G^{-1}_{0,\mathcal{P};\alpha\beta, \alpha' \beta'}(\XRM,\XTM,\mathrm{i}\Omega_{n}; \XRM',\XTM',\mathrm{i}\Omega_{n'})
	    \\
	    &= G^{-1}_{0,\mathcal{P};  \alpha \beta}(\XRM,\XRM';\XTM,\mathrm{i}\Omega_{n})
	    \delta_{\XTM\XTM'} \delta_{\alpha\alpha'} \delta_{\beta\beta'} \delta_{nn'} .
	\end{split}
\end{equation}
The same holds for the propagator itself and will also be true for the exciton propagator to be introduced shortly.
From Eq.\ \eqref{eq: PVP plus trace polarization propagator} it follows that
\begin{widetext}
	\begin{equation} \label{eq: polarization field non-interacting propagator}
		-G^{-1}_{0,\mathcal{P};\alpha\beta}(\XRM,\XRM';\XTM,\mathrm{i}\Omega_{n}) = \frac{1}{\mathcal{V}} \bigg[ V(\XRM-\XRM') + \frac{1}{\mathcal{V}}\sum_{\boldsymbol{p}}V(\XRM-\boldsymbol{p}) \Pi_{\alpha\beta}^{cv}(\boldsymbol{p}+\dlmassc \XTM,\boldsymbol{p}-\dlmassv \XTM,\mathrm{i}\Omega_n) 
		V(\boldsymbol{p}-\XRM')
		\bigg] .
	\end{equation}
\end{widetext}
Note that here we have performed an internal fermionic Matsubara summation, namely
\begin{align} \nonumber
	\Pi_{\alpha\beta}^{cv}(\boldsymbol{p}_c,&\boldsymbol{p}_v,\mathrm{i}\Omega_{n}) \\
	&= \frac{1}{\beta}\sum_{m} G^{0}_{c\alpha}(\boldsymbol{p}_c,\mathrm{i}\Omega_{n} + \mathrm{i}\omega_{m}) G^{0}_{v\beta}(\boldsymbol{p}_v,\mathrm{i}\omega_{m}) \nonumber
	\\ \label{eq: polarization bubble 2-loop}
	&= \frac{N_{ \mathrm{F}}(\elecEnergy^c_{\alpha \boldsymbol{p}_c}) - N_\mathrm{F}(\elecEnergy^v_{\beta \boldsymbol{p}_v}) }
	{\elecEnergy^c_{\alpha\boldsymbol{p}_c} - \elecEnergy^v_{\beta\boldsymbol{p}_v} - \mathrm{i}\Omega_{n}},
\end{align}
with $N_{\mathrm{F}}(x) = (\mathrm{e}^{\beta x} + 1)^{-1}$ the Fermi--Dirac distribution.
The exciton BSE may be derived by finding the zeros of the inverse polarization propagator Eq.\ \eqref{eq: polarization field non-interacting propagator} after the analytical continuation $\mathrm{i} \Omega_{n} \rightarrow \Omega + \mathrm{i} 0^{+}$.
In other words, we introduce a function $\Phi$ satisfying $G^{-1}_{0,\mathcal{P}} \cdot \Phi = 0$. 
This can be solved as a function of the good quantum number $\XTM$ to obtain a set of eigenmodes labeled by $\Xmodeindex$ with frequencies $\Omega = \varepsilon^{\Xmodeindex}_{\XTM}$.
This leads to the eigenvalue equation
\begin{equation}
\label{eq: temperature-dependent exciton BSE}
	\begin{split}
		\Delta^{\alpha\beta}_{\XTM\XRM} \XWF{\alpha}{\beta}{\Xmodeindex}{\XTM}(\XRM)
		- \big[\mathcal{N}^{\mathrm{F}}_{\alpha \beta}(\XRM,\XTM)\big]^{2} \frac{1}{\mathcal{V}}\sum_{\XRM'}V(\XRM-\XRM') \XWF{\alpha}{\beta}{\Xmodeindex}{\XTM}(\XRM')& \\
		= \varepsilon^{\Xmodeindex}_{\XTM} \XWF{\alpha}{\beta}{\Xmodeindex}{\XTM}(\XRM)& ,
	\end{split}
\end{equation}
where for later convenience we have defined
\begin{align}
    \mathcal{N}^{\mathrm{F}}_{\alpha \beta}(\XRM,\XTM) = \sqrt{N_{\mathrm{F}}(\elecEnergy^v_{\XRM-\dlmassv \XTM,\beta}) - N_{\mathrm{F}}(\elecEnergy^c_{\XRM+\dlmassc \XTM,\alpha})} .
\end{align}
Eq.\ \eqref{eq: temperature-dependent exciton BSE} is, in fact, a temperature-dependent version of the exciton BSE previously given in Eq.\ \eqref{eq: exciton BSE}, and reduces to the latter in the zero-temperature limit provided that the chemical potential lies within the gap, as in this case $\mathcal{N}^{\mathrm{F}} \rightarrow 1$.
Note that to derive Eq.\ \eqref{eq: temperature-dependent exciton BSE} we have formally multiplied by the inverse interaction in momentum space, which is defined via
\begin{equation} \label{eq: inverse interaction identity}
    \frac{1}{\mathcal{V}}\sum_{\boldsymbol{p}}V^{-1}(\XRM-\boldsymbol{p})V(\boldsymbol{p}-\XRM') = \mathcal{V} \delta_{\XRM\XRM'} .
\end{equation}
The fact that the poles of the polarization-field propagator give rise to the exciton BSE should not come as a surprise, since $G^{0,\mathcal{P}}$ essentially describes the propagation of a coupled conduction and valence electron, i.e., an electron--hole pair. 
Its fluctuation spectrum thus naturally contains bound states of these particles.

At first glance it may seem that Eq.\ \eqref{eq: temperature-dependent exciton BSE} is non-Hermitian when $T \neq 0$, because the interaction term is not symmetric under the exchange of $\XRM$ and $\XRM'$. 
However, we may define a new wave function $\tilde{\Phi}$ via
\begin{equation}
    \XWF{\alpha}{\beta}{\Xmodeindex}{\XTM}(\XRM) = \mathcal{N}^{\mathrm{F}}_{\alpha \beta}(\XRM,\XTM) \XWFt{\alpha}{\beta}{\Xmodeindex}{\XTM}(\XRM) ,
\end{equation}
in terms of which the BSE is rendered explicitly Hermitian. 
The effective temperature-dependent interaction then takes the form $\mathcal{N}^{\mathrm{F}}_{\alpha \beta}(\XRM,\XTM) V(\XRM - \XRM') \mathcal{N}^{\mathrm{F}}_{\alpha \beta}(\XRM',\XTM)$.
Thus, for arbitrary temperatures it is $\tilde{\Phi}$, rather than $\Phi$, which satisfies the usual completeness relation of Eq.\ \eqref{eq: wavefunction idendity eigenvalues} and the normalization condition in the form of Eq.\ \eqref{eq: wavefunction idendity momenta and spin}. 
By contrast, the wave function $\Phi$ present in Eq.\ \eqref{eq: temperature-dependent exciton BSE} satisfies slightly modified completeness relations which can easily be derived from its connection to $\tilde{\Phi}$ at nonzero temperatures.
Therefore, in this temperature-dependent setting, we shall use $\tilde{\Phi}$ to describe the exciton states, since it is more computationally convenient given the fulfilment of standard completeness relations.

Since the solutions to Eq.\ \eqref{eq: temperature-dependent exciton BSE} form a complete set, we may expand the polarization field in a new set of fields $X_{\Xmodeindex \XTM}(\mathrm{i}\Omega_{n})$ via
\begin{equation} \label{eq: exciton field}
	\mathcal{P}_{\alpha\beta}(\XRM,\XTM,\mathrm{i}\Omega_{n}) =
    \frac{\mathcal{N}^{\mathrm{F}}_{\alpha \beta}(\XRM,\XTM)}{\sqrt{\mathcal{V}}}\sum_{\Xmodeindex} \XWFt{\alpha}{\beta}{\Xmodeindex}{\XTM}(\XRM) X_{\Xmodeindex \XTM}(\mathrm{i}\Omega_{n}).
\end{equation}
The above is analogous to the definition of the exciton creation operator of Eq.\ \eqref{eq: exciton creation operator} at $T=0$. 
However, to be able to compare the exciton field with the exciton operator at arbitrary temperatures, we introduce temperature dependence in the latter via
\begin{equation}
\label{eq: exciton operator temperature dependent}
    \hat{\TdepX}^{\dagger}_{\Xmodeindex \XTM} = \frac{1}{\sqrt{\mathcal{V}}} \sum_{\alpha\beta\XRM}\frac{\XWFt{\alpha}{\beta}{\Xmodeindex}{\XTM}(\XRM)}{\mathcal{N}^{\mathrm{F}}_{\alpha\beta}(\XRM,\XTM)} \hspace{0.3mm} \hat{c}^{\dagger}_{\alpha, \XRM+\dlmassc \XTM} \hat{v}_{\beta, \XRM-\dlmassv \XTM}.
\end{equation}
With the above operator definition, the $X$ field satisfies the property $\langle X \rangle = \langle \hat{\TdepX} \rangle$. 
Therefore, we may identify this field as the exciton field.
While we could in principle have chosen a different convention for the Fermi factors in these equations, we will see that Eq.\ \eqref{eq: exciton operator temperature dependent} leads to a particularly simple expression for the propagator of the exciton operator.

In what follows it will be convenient to develop a diagrammatic notation.
Firstly, we represent the product of an exciton wave function with a factor $\mathcal{N}^{\mathrm{F}}$ as
\begin{equation}
	\begin{split}
		\diagsymb{\frac{1}{\sqrt{\mathcal{V}}} 
	    \big[\mathcal{N}^{\mathrm{F}}_{\alpha\beta}(\XRM,\XTM)\big]^{\pm 1}
	    \XWFt{\alpha}{\beta}{\Xmodeindex}{\XTM}(\XRM) 
	    \delta_{\XTM\XTM'}\delta_{n,n_c-n_v} =&}
		\\
		\includepdfmt{1}{.4}{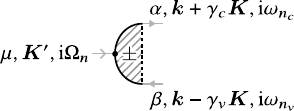} \diagsymb{&.}
	\end{split}
\end{equation}
In this way, the wave function can be interpreted as ``breaking up'' an exciton into a forward-propagating conduction electron and a backward-propagating valence electron, which is equivalent to a forward-propagating hole.
Likewise, its complex conjugate ``forms'' an exciton, which will be represented by an oppositely facing semicircle.
Furthermore, we can rewrite the exciton temperature-dependent BSE as
\begin{equation} \label{eq: temperature-dependent exciton BSE rewrite}
	\begin{split} 	
		\frac{1}{\mathcal{V}}\sum_{\XRM'}
	    V(\XRM - \XRM') \mathcal{N}^{\mathrm{F}}_{\alpha\beta}(\XRM',\XTM)
	    \XWFt{\alpha}{\beta}{\Xmodeindex}{\XTM}(\XRM') &=
	    \\
	    (\Delta^{\alpha\beta}_{\XTM\XRM} - \varepsilon^{\Xmodeindex}_{\XTM})
	    \big[\mathcal{N}^{\mathrm{F}}_{\alpha\beta}(\XRM,\XTM)&\big]^{- 1}
	    \XWFt{\alpha}{\beta}{\Xmodeindex}{\XTM}(\XRM),
	\end{split}
\end{equation}
and diagrammatically represent it as
\begin{equation} \label{diag: BSE}
	\includepdfmt{1}{.43}{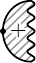} \diagsymb{=} \includepdfmt{1}{0.43}{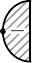} \diagsymb{.}
\end{equation}
The left-hand semicircle with a wiggly line represents the exciton wave function accompanied by an interaction and a $\mathcal{N}^{\mathrm{F}}$ factor. 
The right-hand semicircle with a solid line depicts the exciton wave function bolstered with an inverse factor of $\mathcal{N}^{\mathrm{F}}$ and the energy term ($\Delta-\varepsilon$).
A detailed description of all the rules for the upcoming Feynman diagrams, including the positioning of the factors of $\mathcal{V}$ and $\beta$, is given in Sec.\ S.III of the Supplemental Material \cite{supp}.
In the main text we focus on the intuitive meaning of the processes depicted in the diagrams.
The noninteracting inverse Green's function for the exciton field is obtained from that of the polarization field via the change of basis
\begin{widetext}
	\begin{align} \nonumber 
		{-}&G^{-1}_{0,X;\Xmodeindex  \Xmodeindex'}(\XTM,\mathrm{i}\Omega_n) \nonumber \\
		&= -\frac{1}{\mathcal{V}}\sum_{\alpha \beta\XRM } \sum_{\alpha'\beta'\XRM' }
	    \big[\XWFt{\alpha}{\beta}{\Xmodeindex}{\XTM}(\XRM)\big]^{*} \mathcal{N}^{\mathrm{F}}_{\alpha\beta}(\XRM,\XTM) 
	    G^{-1}_{0,\mathcal{P};\alpha\beta,\alpha'\beta'}(\XRM, \XRM'; \XTM,\mathrm{i}\Omega_{n}) \mathcal{N}^{\mathrm{F}}_{\alpha' \beta'}(\XRM',\XTM) 
	    \XWFt{\alpha'}{\beta'}{\Xmodeindex'}{\XTM}(\XRM') \nonumber
	    \\ \nonumber 
	    &= \frac{1}{\mathcal{V}^{2}} \sum_{\alpha \beta}\sum_{\XRM \XRM'} 
	    \big[\XWFt{\alpha}{\beta}{\Xmodeindex}{\XTM}(\XRM)\big]^{*} \mathcal{N}^{\mathrm{F}}_{\alpha\beta}(\XRM,\XTM)
	    \Bigg\{V(\XRM - \XRM') - \frac{1}{\mathcal{V}} \sum_{\vec{p}} 
	    V(\XRM - \vec{p}) \frac{\big[\mathcal{N}^{\mathrm{F}}_{\alpha\beta}(\vec{p}, \XTM)\big]^2}{\Delta^{\alpha\beta}_{\XTM, \vec{p}} - \mathrm{i} \Omega_{n}} V(\vec{p} - \XRM') \Bigg\} 
	    \mathcal{N}^{\mathrm{F}}_{\alpha\beta}(\XRM',\XTM) \XWFt{\alpha}{\beta}{\Xmodeindex'}{\XTM}(\XRM')
		\\ \label{eq: exciton inverse propagator}
	    \diagsymb{&=} \includepdfmt{1}{.4}{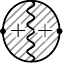} \diagsymb{+} \includepdfmt{1}{.4}{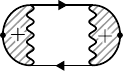} \diagsymb{.}
	\end{align}
	This object is ``noninteracting'' in the sense that it does not contain any interactions between other excitons.
	The above Green's function can be inverted to obtain a Dyson equation for the free exciton-field propagator, namely
	\begin{align} \label{eq: exciton propagator}
		\begin{split}
			G^{0,X}_{\Xmodeindex\Xmodeindex'}(\XTM,\mathrm{i}\Omega_{n}) =
			-&\frac{1}{\mathcal{V}^2} \sum_{\alpha\beta}\sum_{\XRM\XRM'}
			\frac{\big[\XWFt{\alpha}{\beta}{\Xmodeindex}{\XTM}(\XRM)\big]^{*}}{\mathcal{N}^{\mathrm{F}}_{\alpha \beta}(\XTM, \XRM)} V^{-1}(\XRM - \XRM') \frac{\XWFt{\alpha}{\beta}{\Xmodeindex'}{\XTM}(\XRM')}{\mathcal{N}^{\mathrm{F}}_{\alpha \beta}(\XTM, \XRM')}
			\\
		    + \, &\frac{1}{\mathcal{V}} \sum_{\alpha\beta\XRM} \sum_{\bar{\Xmodeindex}} \big[\XWFt{\alpha}{\beta}{\Xmodeindex}{\XTM}(\XRM)\big]^{*}
			\,\frac{\Delta^{\alpha\beta}_{\XTM\XRM}-\varepsilon^{\bar{\Xmodeindex}}_{\XTM}}{\Delta^{\alpha\beta}_{\XTM\XRM} - \mathrm{i}\Omega_{n}}\,
			\XWFt{\alpha}{\beta}{\bar{\Xmodeindex}}{\XTM}(\XRM)
			G^{0,X}_{\bar{\Xmodeindex}\Xmodeindex'}(\XTM,\mathrm{i}\Omega_{n}) ,
		\end{split}
	\end{align}
	where Eq.\ \eqref{eq: temperature-dependent exciton BSE rewrite} was used for the second term.
	The above equation for $G^{0,X}$ is diagrammatically represented as
	\begin{equation} \label{diag: exciton field propagator}
		\begin{split} 
			\includepdfmt{1}{.30}{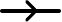} \diagsymb{&=-}
			\includepdfmt{1}{.35}{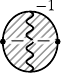} \diagsymb{+}
			\includepdfmt{1}{.43}{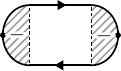} \diagsymb{-}
			\includepdfmt{1}{.43}{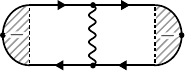} \diagsymb{+}
		    \includepdfmt{1}{.43}{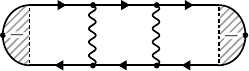} \diagsymb{+\cdots}
		    \\ 
		    \diagsymb{&=-}
		    \includepdfmt{1}{.35}{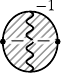} \diagsymb{-}
		    \includepdfmt{1}{.43}{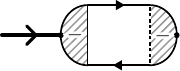} \diagsymb{.}
		\end{split}
	\end{equation}
\end{widetext}
In these Feynman diagrams, the imaginary time is always understood to flow from left to right.
As in Eq.\ \eqref{diag: exciton field propagator}, the arrows on conduction-electron propagators always point in the direction of the flow of time, while those of valence-electron propagators point in the opposite direction. 
The latter are then equivalent to a forward-propagating hole.

Before moving on to the field-theoretical version of the exciton--exciton interaction, we briefly discuss the difference between the propagator of the exciton field,
\begin{subequations}
	\begin{equation}
	    G^{X}_{\mu\mu'}(\XTM, \mathrm{i} \Omega_{n}) = {-}\langle X_{\Xmodeindex \XTM}(\mathrm{i}\Omega_{n}) X^{*}_{\Xmodeindex' \XTM}(\mathrm{i}\Omega_{n})\rangle,
	\end{equation}
and that of the exciton operator, 
	\begin{equation}
	    G^{\hat{X}}_{\mu\mu'}(\XTM, \mathrm{i} \Omega_{n}) = {-}\langle \hat{X}_{\Xmodeindex \XTM}(\mathrm{i}\Omega_{n}) \hat{X}^{\dagger}_{\Xmodeindex' \XTM}(\mathrm{i}\Omega_{n})\rangle.
	\end{equation}
\end{subequations}
As explained in Ref.\ \cite{Stoof_2008}, while the correlator of the exciton operator coincides with that of the exciton field, $\langle X\rangle = \langle \hat{X}\rangle$, it is not true that $\langle X X^{*}\rangle = \langle \hat{X}\hat{X}^{\dagger}\rangle$.
To derive a relation between the latter averages one introduces appropriate functional sources in Eq.\ \eqref{eq: partition function conduction valence}, as worked out in Sec.\ S.V\ A of the Supplemental Material \cite{supp}.
This leads to
\begin{equation} \label{eq: relation operator and field correlator}
	\begin{split} 
	    &G^{\hat{X}}_{\mu\mu'}(\XTM, \mathrm{i} \Omega_{n}) = 
	    G^{X}_{\mu\mu'}(\XTM, \mathrm{i} \Omega_{n}) 
	    \\
	    &+ \frac{1}{\mathcal{V}^2} \sum_{\alpha\beta}\sum_{\XRM\XRM'} 
	    \frac{\big[\XWFt{\alpha}{\beta}{\Xmodeindex}{\XTM}(\XRM) \big]^{*}}{\mathcal{N}^{\mathrm{F}}_{\alpha\beta}(\XRM,\XTM)}
	    V^{-1}(\XRM - \XRM') 
	    \frac{\XWFt{\alpha}{\beta}{\Xmodeindex'}{\XTM}(\XRM')}{\mathcal{N}^{\mathrm{F}}_{\alpha\beta}(\XRM',\XTM)} .
	\end{split}
\end{equation}
If we consider this equation for the free exciton-field propagator $G^{0,X}$, the term on the second line is precisely the first term on the right-hand side of Eqs. \ \eqref{eq: exciton propagator} and \eqref{diag: exciton field propagator}. 
Thus, the correlation function corresponding to the exciton operators is diagrammatically given exactly by the ladder series of Eq.\ \eqref{diag: exciton field propagator}, without the inverse-potential term.
It can be shown that the Dyson equation for the noninteracting operator propagator reads
\begin{widetext}
	\begin{equation}
		\label{eq: Dyson equation G hat X}
	    G^{0,\hat{X}}_{\Xmodeindex\Xmodeindex'}(\XTM,\mathrm{i}\Omega_{n}) =
	    -\frac{1}{\mathcal{V}} \sum_{\alpha \beta \XRM} \sum_{\bar{\Xmodeindex}}
	    \big[\XWFt{\alpha}{\beta}{\Xmodeindex}{\XTM}(\XRM)\big]^{*} 
	    \frac{1}{\Delta^{\alpha\beta}_{\XTM\XRM}  - \mathrm{i} \Omega_{n}}
	    \bigg\{ \delta_{\bar{\Xmodeindex} \Xmodeindex'}
	    - ( \Delta^{\alpha\beta}_{\XTM\XRM} - \varepsilon^{\bar{\Xmodeindex}}_{\XTM} )
	    G^{0,\hat{X}}_{\bar{\Xmodeindex}\Xmodeindex'}(\XTM,\mathrm{i}\Omega_{n}) 
	    \bigg\} \XWFt{\alpha}{\beta}{\bar{\Xmodeindex}}{\XTM}(\XRM) .
	\end{equation}
\end{widetext}
The solution to this series, at all temperatures, is given by 
\begin{equation} \label{eq: exciton operator free Green's function final}
	G^{0,\hat{X}}_{\mu\mu'}(\XTM, \mathrm{i} \Omega_{n}) = \frac{1}{\mathrm{i} \Omega_{n} - \varepsilon^{\mu}_{\XTM}} \delta_{\mu\mu'},
\end{equation}
where we stress that $\varepsilon^{\mu}_{\XTM}$ are the eigenenergies arising from the temperature-dependent BSE.
Thus, the propagator associated with the exciton operator has the standard bosonic form, in particular featuring poles at the exciton bound-state energies.
In contrast, the propagator of the exciton field itself differs from the standard bosonic form by the $V^{-1}$ term of Eq.\ \eqref{eq: exciton propagator}, i.e.
\begin{equation} \label{eq: exciton field free Green's function final}
	\begin{split}
	    &G^{0,X}_{\mu\mu'}(\XTM, \mathrm{i} \Omega_{n}) = \frac{1}{\mathrm{i} \Omega_{n} - \varepsilon^{\mu}_{\XTM}} \delta_{\mu\mu'}
	    \\
	    &- \frac{1}{\mathcal{V}^2} \sum_{\alpha\beta}\sum_{\XRM\XRM'} 
	    \frac{\big[\XWFt{\alpha}{\beta}{\Xmodeindex}{\XTM}(\XRM)\big]^*}{\mathcal{N}^{\mathrm{F}}_{\alpha\beta}(\XRM,\XTM)}
	     V^{-1}(\XRM - \XRM') 
	     \frac{\XWFt{\alpha}{\beta}{\Xmodeindex'}{\XTM}(\XRM')}{\mathcal{N}^{\mathrm{F}}_{\alpha\beta}(\XRM',\XTM)} .
	\end{split}
\end{equation}
At first glance, the inverse interaction in the above propagator appears to be dominated by the pole located at the exciton energy. 
Therefore, one might argue that this former term may be neglected. 
However, as discussed in detail in Sec.\ S.V\ B of the Supplemental Material \cite{supp}, removing this term from the correlation function is equivalent to assuming excitons to be exact bosons, which is the assumption we wanted to avoid.
It turns out that, because of the presence of this additional inverse-potential term, the composite nature of the excitons is correctly taken into account by the field theory.
In the upcoming section, we will show this fact by considering the equal-time, two-exciton propagator.

\subsection{Effective exciton action} \label{sec: Effective exciton action}

Using the above definition for the exciton field, we now reduce the formal polarization action to an effective action for excitons.
As stated before, owing to the nature of the path-integral formalism, the action of Eq.\ \eqref{eq: action exact polarization} gives rise to many-body interaction vertices of arbitrary order in the electrostatic interaction. 
For the effective description, we only take into account the simplest interactions that can occur between two excitons. 
These interactions will predominantly be of first order in the electrostatic potential, i.e., up to order $V$ \footnote{In principle, because of the Hubbard--Stratonovich transformation, every polarization field always comes ``attached'' to a single-particle interaction. Essentially, this is an interaction between the conduction and valence electrons of the same exciton. This kind of interaction can always be rewritten into energy terms via the temperature-dependent BSE. Hence, this single-particle interaction is no longer explicitly present, neither in the upcoming Feynman diagrams nor in the mathematical expressions. Therefore, when there is a mention of an exciton--exciton interaction being of the order $\mathcal{O}(V)$, this will refer to a single-particle interaction that occurs between the electrons of different excitons.}.
Moreover, as stated earlier, we effectively neglect any corrections to the electron propagators.
Under these conditions, there will be 14 resulting interaction processes. 
These include electron--electron, hole--hole, and electron--hole interactions.
Furthermore, part of these processes occur in combination with electron exchange, hole exchange, or both, i.e., exciton exchange.

To obtain the effective theory, the third and fourth terms of Eq.\ \eqref{eq: action exact polarization} are expanded up to fourth order in the polarization field, while the second term is developed to sixth order.
Once the polarization field $\mathcal{P}$ is replaced in favor of the exciton field $X$, the former terms will give contributions arising from the electron--electron and hole--hole interactions, while the quartic part of the latter gives rise to the electron-exchange interaction, i.e., two excitons interchange their conduction electrons without the occurrence of an interaction.
These quartic components do not give rise to interaction processes with electron--hole interactions.
Instead, these terms indirectly stem from the aforementioned sixth-order term.
The reason for this, despite this term being of sixth order, is that the perturbative diagram of this three-body interaction will be partially closed by a (noninteracting) exciton Green's function when computing the two-exciton propagator.
This effectively results in a two-body interaction.
The explicit expression of each interaction component and more details on their derivation are found in Sec.\ S.IV of the Supplemental Material \cite{supp}.

Combining all interactions found via the procedure above, we can write down an effective exciton action as
\begin{equation} \label{eq: action effective exciton}
    S_{\mathrm{eff}}[X^*,X] = \; {-}(X | G^{-1}_{0,X} | X) + \frac{1}{2\beta\mathcal{V}}(XX \lVert \mathcal{W} \rVert XX) ,
\end{equation}
where the functional inner products are in momentum and frequency space. 
The two-body interaction $\mathcal{W}$ is symmetric under the exchange of in- and outgoing excitons, namely 
\begin{subequations} \label{eq: interaction W symmetries}
\begin{align}
    \mathcal{W}(z_1,z_2;z_1',z_2') &= \mathcal{W}(z_1,z_2;z_2',z_1'),
    \\
    &=\mathcal{W}(z_2,z_1;z_1',z_2'),
    \\
    &=\mathcal{W}(z_2,z_1;z_2',z_1'),
\end{align}
\end{subequations}
where $z=(\Xmodeindex,\XTM,\mathrm{i}\Omega_{n})$. 
Each component of the interaction conserves total momentum and frequency. 
Furthermore, the interaction is explicitly dependent on Matsubara frequencies, implying that retardation effects are included. 
The exciton--exciton interaction is diagrammatically defined as
\begin{equation}
    \diagsymb{\frac{1}{\beta\mathcal{V}} \mathcal{W} =} \includepdfmt{0.78}{0.42}{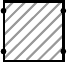} \diagsymb{,}
\end{equation}
which represents the sum of the interaction vertices
\begin{widetext}
\begin{subequations} \label{diag: exciton--exciton interaction field theory}
	\begin{align} \label{diag: W0}
		\diagsymb{2} \includepdfmt{0.78}{0.42}{W} \diagsymb{&=} 
		\includepdfmt{0.78}{.46}{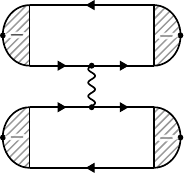} \diagsymb{+} 
		\includepdfmt{0.78}{.46}{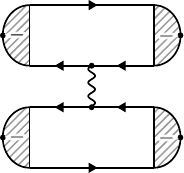} \diagsymb{-} 
		\includepdfmt{0.78}{.46}{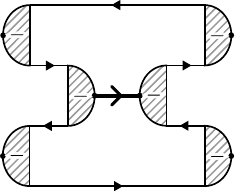} \diagsymb{-} 
		\includepdfmt{0.78}{.46}{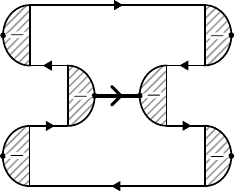}
	    \\ \label{diag: WX}
	    \diagsymb{&+} 
	    \includepdfmt{0.78}{.46}{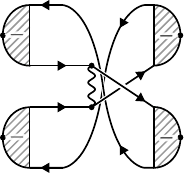} \diagsymb{+} 
	    \includepdfmt{0.78}{.46}{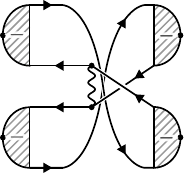} \diagsymb{-} 
	    \includepdfmt{0.78}{.46}{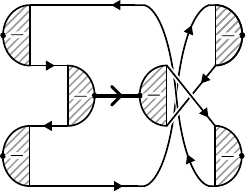} \diagsymb{-} 
	    \includepdfmt{0.78}{.46}{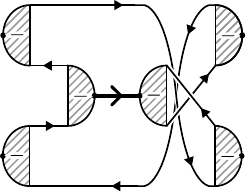}
		\\ \label{diag: Wc}
	    \diagsymb{&+} 
	    \includepdfmt{0.78}{.46}{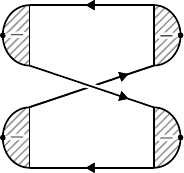} \diagsymb{-} 
	    \includepdfmt{0.78}{.46}{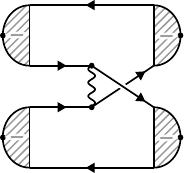} \diagsymb{-} 
	    \includepdfmt{0.78}{.46}{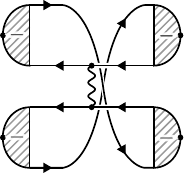}
		\\ \label{diag: Wv}
		\diagsymb{&+} 
		\includepdfmt{0.78}{.46}{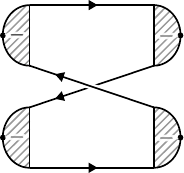} \diagsymb{-} 
		\includepdfmt{0.78}{.46}{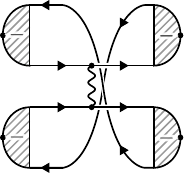} \diagsymb{-} 
		\includepdfmt{0.78}{.46}{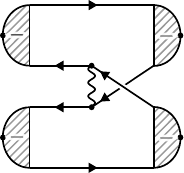}\,.
	\end{align}
\end{subequations}
\end{widetext}
The factor of 2 is incorporated into $\mathcal{W}$ for convenience, such that the quartic term in Eq.\ \eqref{eq: action effective exciton} carries the conventional prefactor of $1/2$.
The third and fourth diagrams in Eqs.\ \eqref{diag: W0} and \eqref{diag: WX} may appear out of place compared to the other vertices in $\mathcal{W}$ because they internally contain $G^{0,X}$.
Nevertheless, when we expand these propagators to their lowest order, i.e, to the first term in Eq.\ \eqref{diag: exciton field propagator}, then these four diagrams reduce to
\begin{subequations}
\begin{align} \label{diag: W0_cv expansion}
	\includepdfmt{0.78}{0.46}{W0_cv_with_propagator} \diagsymb{&\approx -} \includepdfmt{0.78}{0.46}{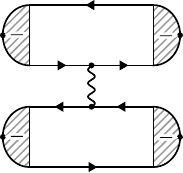} \diagsymb{,}
	\\ \label{diag: W0_vc expansion}
	\includepdfmt{0.78}{0.46}{W0_vc_with_propagator} \diagsymb{&\approx -} \includepdfmt{0.78}{0.46}{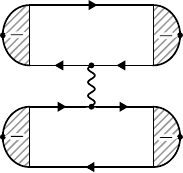} \diagsymb{,}
	\\ \label{diag: WX_cv expansion}
	\includepdfmt{0.78}{0.46}{WX_cv_with_propagator} \diagsymb{&\approx -} \includepdfmt{0.78}{0.46}{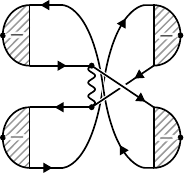} \diagsymb{,}
	\\ \label{diag: WX_vc expansion}
	\includepdfmt{0.78}{0.46}{WX_vc_with_propagator} \diagsymb{&\approx -} \includepdfmt{0.78}{0.46}{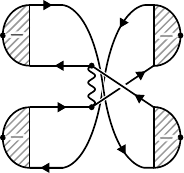} \diagsymb{.}
\end{align}
\end{subequations}
These diagrams are of order $V$ and represent electron--hole interactions, both with and without exciton exchange.
However, we must note that these simplifications cannot be directly incorporated into 
$\mathcal{W}$, as this would cause the two-exciton propagator to exhibit incorrect symmetry properties.
This point will be discussed shortly.

When the field-theoretical interaction $\mathcal{W}$ is evaluated on shell at $T = 0$, and the simplifications of Eqs.\ \eqref{diag: W0_cv expansion}-\eqref{diag: WX_vc expansion} are made, we precisely obtain the exciton--exciton interaction of Eq.\ \eqref{eq: exciton--exciton interaction} which we derived via the variational principle. 
In other words, we evaluate the four Matsubara frequencies at their respective exciton energy, such that $\mathrm{i}\Omega_{n_i}\rightarrow\varepsilon^{\Xmodeindex_i}_{\XTM_i}$.
This results in
\begin{equation}
\label{eq: on shell interaction}
	\mathcal{W}^{\Xmodeindex_1'\Xmodeindex_2'}_{\Xmodeindex_1\Xmodeindex_2}(\{\vec{K}_{i}, \mathrm{i} \Omega_{n_{i}}\}) \approx \mathcal{U}^{\Xmodeindex_1'\Xmodeindex_2'}_{\Xmodeindex_1\Xmodeindex_2}(\BXTM, \BXRM, \BXRM') \delta_{\BXTM \BXTM'} \delta_{n_{1} + n_{2}, n_{1}' + n_{2}'} ,
\end{equation}
where the individual exciton momenta are defined in terms of the total and relative biexciton momenta as in Eq.\ \eqref{eq: CoM momentum coords (biexcitons)}, namely
\begin{equation}
\label{eq: exciton momenta shorthnad}
	\begin{split}
		\XTM_1 &= \frac{1}{2}\BXTM + \BXRM, \qquad\hspace{0.22mm}
        \XTM_1' = \frac{1}{2}\BXTM' + \BXRM',
        \\
        \XTM_2 &= \frac{1}{2}\BXTM - \BXRM, \qquad
        \XTM_2' =  \frac{1}{2}\BXTM' - \BXRM' .
	\end{split}
\end{equation}
The sum of diagrams in Eqs.\ \eqref{diag: W0}--\eqref{diag: Wv} reduce to $\mathcal{U}^{0}$, $\mathcal{U}^{\mathrm{X}}$, $\mathcal{U}^{\mathrm{c}}$, and $\mathcal{U}^{\mathrm{v}}$, respectively. 
The derivation of the above results is discussed in more detail in Sec.\ S.III of the Supplemental Material \cite{supp}.
Eq.\ \eqref{eq: on shell interaction} further confirms that $\mathcal{W}$ indeed describes the interaction that occurs between two excitons. 
Moreover, because these field-theory interactions incorporate retardation and finite-temperature effects, they include corrections with respect to the interaction found via the variational approach.

From the effective action it is straightforward to derive, for instance, the Dyson equation for the four-point correlation function $G^{X}_{2} \equiv \langle X X X^* X^* \rangle$. 
For simplicity, we will neglect any corrections to the single-exciton propagators, i.e., we let $G^{X} \approx G^{0,X}$, and consider only $\mathcal{W}$ as the irreducible part of the Dyson series.
The resulting equation takes the diagrammatic form
\begin{equation} \label{diag: G2_X dyson}
	\includepdfmt{0.78}{0.28}{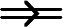} \diagsymb{=} 
    \includepdfmt{0.78}{0.4}{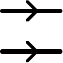} \diagsymb{+} 
    \includepdfmt{0.78}{0.4}{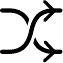} \diagsymb{-} 
    \includepdfmt{0.78}{0.4}{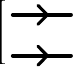} \diagsymb{+} 
    \includepdfmt{0.78}{0.4}{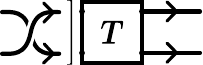} \diagsymb{.} 
\end{equation}
Here we have introduced the two-exciton $T$ matrix, which satisfies the diagrammatic equation
\begin{align} \label{diag: T-matrix equation}
    \includepdfmt{0.78}{0.42}{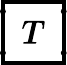} \diagsymb{=} 
    \includepdfmt{0.78}{0.42}{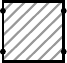} \diagsymb{-}
    \includepdfmt{0.78}{0.42}{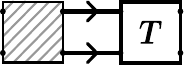} \diagsymb{.}
\end{align}
In the series of Eq.\ \eqref{diag: G2_X dyson}, the behavior of the exciton propagators displays the fundamental difference between products of exciton fields and operators. 
Namely, a product of two exciton operators possesses the symmetry of Eq.\ \eqref{eq: projector symmetry exciton operator product} with the antisymmetrizer $\mathcal{A}$ of Eq.\ \eqref{eq: biexciton projector}, which besides exciton exchange also contains the exchange between the electrons and the holes.
Meanwhile, two bosonic exciton fields can only display exciton exchange because, unlike their operator counterpart, they do not themselves contain any information on the internal structure of the composite state. 
For instance, let us consider the equal-time, two-exciton propagator
\begin{equation} \label{eq: equal-time two-exciton operator propagator}
    \begin{split}
    	&\bigl[G^{\hat{X}}_{2}\bigr]_{\Xmodeindex_1\Xmodeindex_2}^{\Xmodeindex_1'\Xmodeindex_2'}(\BXRM,\BXTM,\tau;\BXRM',\BXTM',\tau')
	    \\
	    &= \Big\langle \hat{\mathcal{T}} \Big[ 
	    \hat{\TdepX}_{\Xmodeindex_1 \XTM_{1}}(\tau)
	    \hat{\TdepX}_{\Xmodeindex_2 \XTM_{2}}(\tau) 
	    \hat{\TdepX}^{\dagger}_{\Xmodeindex_1' \XTM_{1}'}(\tau')
	    \hat{\TdepX}^{\dagger}_{\Xmodeindex_2' \XTM_{2}'}(\tau')
	    \Big]\Big\rangle,
    \end{split}
\end{equation}
where the ingoing exciton operators (in the Heisenberg picture) are taken at the same imaginary time $\tau'$ and the two outgoing ones at a time $\tau$, and we have used the definitions of Eq.\ \eqref{eq: exciton momenta shorthnad} for the momenta.
As a consequence of the exciton operators being defined at pairwise equal times, the time-ordering operator $\hat{\mathcal{T}}$ does not dictate the positions of the operators within each pair.
Therefore, both operator products possess the aforementioned symmetry under $\mathcal{A}$, and thus the propagator must similarly be invariant under $\mathcal{A}$, namely $G^{\hat{X}}_{2} = G^{\hat{X}}_{2} \cdot \mathcal{A} = \mathcal{A} \cdot G^{\hat{X}}_{2}$.
At first glance, because of this invariance, there appears to be a significant difference between $G^{\hat{X}}_{2}$ and $G^{X}_{2}$, since the latter seems not to have these symmetries. 
However, similar to the relation between the exciton-field and exciton-operator propagators of Eq.\ \eqref{eq: relation operator and field correlator}, a corresponding equality can be formulated between the above two-exciton field propagator and its exciton-operator counterpart, as worked out in Sec.\ S.V\ A of the Supplemental Material \cite{supp}.

Starting from Eq.\ \eqref{diag: G2_X dyson} we can derive the expression for the two-exciton operator propagator as
\begin{equation}
\label{diag: G2_X hat dyson}
    \includepdfmt{0.78}{0.21}{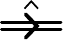} \diagsymb{=8} \includepdfmt{0.78}{0.35}{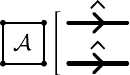} \diagsymb{-} \includepdfmt{0.78}{0.35}{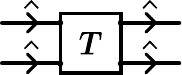} \diagsymb{+\cdots} \includepdfmt{0.78}{0.41}{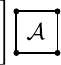} \diagsymb{.}
\end{equation}
Here, the single- and double-lined objects with a hat represent $G^{0,\hat{X}}$ and $G^{\hat{X}}_{2}$, respectively. 
The key takeaway from this result is that all terms on the right-hand side are encased by the antisymmetrizer, correctly showing that $G^{\hat{X}}_{2}$ is invariant when acted upon by $\mathcal{A}$.
This implies that before and after propagation of the exciton pair it does not matter which electrons and holes form the two excitons. 
Furthermore, setting the $T$ matrix to zero in Eq.\ \eqref{diag: G2_X hat dyson} gives the expected result, i.e., a pole at the sum of two exciton energies, as we obtain
\begin{equation}
	\begin{split}
		\big[G^{\hat{X}}_{2}\big]_{\Xmodeindex_1\Xmodeindex_2}^{\Xmodeindex_1'\Xmodeindex_2'}(\BXRM,\BXRM';\BXTM,\mathrm{i}\Omega_{n}) \bigg|_{\vec{T} = 0}
		= 8 \sum_{\bar{\Xmodeindex}\bar{\Xmodeindex}'\vec{p}}
		\mathcal{A}_{\Xmodeindex_1\Xmodeindex_2}^{\bar{\Xmodeindex}\bar{\Xmodeindex}'}(\BXTM,\BXRM,\vec{p})& \\
		\times \,
	    \frac{1 + N_{\mathrm{B}}(\varepsilon^{\bar{\Xmodeindex}}_{\BXTM/2+\vec{p}}) + N_{\mathrm{B}}(\varepsilon^{\bar{\Xmodeindex}'}_{\BXTM/2-\vec{p}})}{\varepsilon^{\bar{\Xmodeindex}}_{\BXTM/2+\vec{p}} + \varepsilon^{\bar{\Xmodeindex}'}_{\BXTM/2-\vec{p}} - \mathrm{i}\Omega_{n}}
	    \mathcal{A}_{\bar{\Xmodeindex}\bar{\Xmodeindex}'}^{\Xmodeindex_1'\Xmodeindex_2'}(\BXTM,\vec{p},\BXRM')& .
	\end{split}
\end{equation}
Here, $N_{\mathrm{B}}(x) = (\mathrm{e}^{\beta x} - 1)^{-1}$ is the Bose--Einstein distribution, which appears after performing the summation over Matsubara frequencies.  

In the derivation of the Dyson series for Eq.\ \eqref{eq: equal-time two-exciton operator propagator}, two-body interaction vertices appear that are not included in $\mathcal{W}$. 
Since the goal of the effective description of Eq.\ \eqref{eq: action effective exciton} was to have excitons interact only via processes in $\mathcal{W}$; the additional vertices are omitted from the series.
With this aspect in mind, the reason for the emergence of the antisymmetrizer in the above equations is twofold. 
Firstly, because of the fact that when relating $G^{\hat{X}}_{2}$ to $G^{X}_{2}$ the allocation of the time variables is the same.
That is to say, within each term of the series in Eq.\ \eqref{diag: exciton field propagator}, the paired propagators always share the same start time $\tau'$ and end time $\tau$.
Secondly, owing to the presence of the inverse-interaction term of Eq.\ \eqref{diag: exciton field propagator}, the electron-exchange and hole-exchange interactions can respectively be written in terms of $\mathcal{K}^{\mathrm{c}}$ and $\mathcal{K}^{\mathrm{v}}$.
In particular, when the first terms of Eqs.\ \eqref{diag: Wc} and \eqref{diag: Wv} are acted upon by two (equal-time) inverse interactions, they can be reduced to the particle exchanges given in Eqs.\ \eqref{eq: conduction electron exchange element} and \eqref{eq: relation conduction and valence exchange}, respectively.
These particle-exchange terms can in turn be rewritten in terms of $\mathcal{A}$.
An additional consequence of this rewriting is that terms that initially belonged to the $T$-matrix part of Eq.\ \eqref{diag: G2_X dyson} ended up contributing to the noninteracting component of $G^{\hat{\TdepX}}_{2}$.
Notably, this applies to the third and fourth diagrams in Eqs.\ \eqref{diag: W0} and \eqref{diag: WX}, and for the first terms of Eqs.\ \eqref{diag: Wc} and \eqref{diag: Wv}.
The details of the derivation, as well as the full expression of Eq.\ \eqref{diag: G2_X hat dyson}, are given in Sec.\ S.VI of the Supplemental Material \cite{supp}.

Some additional remarks are in place.
Firstly, when deriving the effective exciton action, we have limited ourselves to the 14 interaction processes present in Eq.\ \eqref{diag: exciton--exciton interaction field theory}.
Naturally, it is possible to add more processes into $S_{\text{eff}}$, like processes that are of higher order in the electrostatic interaction $V$ or three-body interactions.
Owing to the reduction in the number of ways for excitons to interact, the effective two-body description is mainly valid for a dilute gas of excitons. 
At higher densities, the simultaneous Coulomb and exchange processes between more than two excitons become more likely, and one must ideally keep terms up to higher orders in the exciton field. 
Secondly, if each term in the four-point correlators were to be taken at different times, then the antisymmetrizer would not appear.
From the perspective of Eq.\ \eqref{eq: equal-time two-exciton operator propagator}, if all the operators were defined at different times, then the time-ordering operator would fix their position within the correlator. 
Consequently, no exchange of the exciton operators would be possible, and there would be no invariance with respect to the antisymmetrizer.
Likewise, within the $T$-matrix equation, the antisymmetrizer cannot appear either, because each term in Eq.\ \eqref{diag: T-matrix equation} involves four distinct time coordinates. 
Lastly, if we were to consider a three-exciton propagator with the in- and outgoing excitons again taken at the same respective initial and final times, then the quartic effective action of Eq.\ \eqref{eq: action effective exciton} would not produce the correct symmetries.
This is the case for any $N$-exciton propagator with $N \geq 3$.
The electron-exchange and hole-exchange vertices are crucial for ensuring that the correct dependence on $\mathcal{A}$ appears in $G^{\hat{X}}_{2}$. 
In order to ensure that the three-exciton propagator satisfies the correct symmetries, specific three-body interaction vertices need to be present in the effective action; the two-body vertices are not enough to achieve a result with the correct symmetries. 
This particular aspect is discussed in more detail towards the end of Sec.\ S.VI of the Supplemental Material \cite{supp}.

In summary, we have reduced the formal action of Eq.\ \eqref{eq: action exact polarization} to one that is quartic in the exciton field. 
The two-body interaction present in this action is equal to the interaction calculated from the second-quantization approach when considered up to first order in $V$, set on shell, and at $T = 0$.
Furthermore, using the effective action, we showed that the equal-time, two-exciton propagator incorporates the expected invariance with respect to the antisymmetrizer and the correct pole for the lowest-order term.
Our results highlight how the composite nature of the excitons is reflected in an effective bosonic field theory.

\section{Conclusion and outlook}
\label{sec: Conclusion}


In this work we have studied the interactions between Wannier excitons. 
Specifically, we have derived an effective potential between two ground-state excitons as well as a many-body description for an interacting gas of excitons. 
Via a variational approach we have obtained a generalized eigenvalue problem for the biexciton states which can be rewritten as a Schr\"odinger equation, from where we could identify an effective exciton--exciton potential. 
This potential is nonlocal in position space and depends on the spin states of the (combined) conduction and valence electrons. 
We have computed this potential for the specific case of 2D hydrogen-like excitons in the heavy-hole limit, where it becomes local, and shown that it exactly reproduces the singlet and triplet potentials first obtained by Heitler and London in the treatment of the dihydrogen molecule.
We have also used the same theory to derive the correct van der Waals behavior of two $s$-wave excitons in the limit of large separation.

The potential we have derived takes into account the internal structure of the excitons, and in particular includes the effects of all possible exchange processes that can take place between the constituent electrons and holes.
As a result, it is reliable also at short distances.
This is not the case for existing empirical approaches often based on classical point-charge or dipole--dipole treatments of the exciton--exciton coupling, which usually only correctly model the long-distance physics of the interacting system.
As the physics responsible for the binding of two excitons are sensitive to the short- and intermediate-distance features of the exciton--exciton interaction (cf. Fig.\ \ref{fig: Heitler--London potential}, where the position of the minimum of the singlet potential is of the order of the exciton radius), our theory is expected to deliver much more accurate results for this and similar problems.

This effective exciton--exciton potential could be used to study the role of these interactions in the annihilation of excitons \cite{Dostal_2018}.
Additionally, the effective biexciton eigenvalue equation can be applied to the study of biexciton spectra in experimentally relevant material platforms such as transition-metal dichalcogenides and moiré heterostructures.
The potential we have derived serves as a proof of concept and embodies the physical transparency of the variational approach.
Owing to the generality of the latter, by considering an appropriate variational subspace it is possible to obtain an effective exciton potential matrix for more complex excitonic systems, such as those with multiple (overlapping) bands or nonparabolic band structures.
We emphasize that, even though we have exemplified the use of our method via analytical calculations for hydrogen-like excitons, the theory is not restricted to this case and could be for instance used for more realistic platforms such as modern layered semiconductors with a Rytova--Keldysh interaction.
The only current limitation is that we do not consider the dressing of the electron--hole interaction by form factors arising from the diagonalization of the single-particle Hamiltonian \cite{Wu_2015,MaiselLiceran_2023}.
While this is safe for many conventional semiconducting systems, their inclusion becomes crucial whenever one deals with significant band-geometrical, topological, Berry-curvature, or moiré effects \cite{liceran2025unconventional}.
In a forthcoming paper, we will present an extension of our approach which will naturally incorporate these form factors and thus constitute a generalization of the method to systems with arbitrary single-particle Hamiltonians.

Regarding the many-body treatment of excitons, we have used a finite-temperature path-integral formalism to obtain a formal action for a bosonic exciton field introduced via a Hubbard--Stratonovich transformation. 
This many-body theory provides many-body exciton--exciton interactions up to arbitrary order, which are temperature-dependent and incorporate retardation effects.
From this formal result we have derived an effective quartic excitonic action.
The corresponding two-exciton interaction term, when considered on shell and in the zero-temperature limit, reduces to the same exciton--exciton interaction components obtained from the variational approach.
Furthermore, this effective action produces the correct expression for the equal-time two-exciton propagator.
The effective exciton potential and the derived exciton action are the main results of this work.

Beyond being able to study a gas of interacting excitons, it is also possible to provide a description of a Bose--Einstein condensate of excitons.
As of this writing, the discovery of a condensate of (quasi)bosonic Wannier excitons remains elusive \cite{Perali_2013,Haque_2024}.
By using the field-theoretical result, a Gross--Pitaevskii equation can be derived for the description of the exciton condensate.
While such an equation has been introduced before \cite{Elistratov_2016,Berman_2016}, our result allows for a more detailed description of the interactions via a systematic treatment of the exchange processes that arise from the composite nature of the excitons.

Our methods are more generally applicable to many other systems and problems of interest as long as the excitons under consideration are of Wannier type, i.e., their typical size is significantly larger than the underlying atomic lattice spacing.
It is in this regime where the BSE treatment of excitons (on which our derivations rely) is most useful.
Our theory can for instance be applied to charge-transfer excitons at interfaces if the former are extended enough, but we expect it to be impractical (from the point of view of applicability) for very tightly bound Frenkel excitons residing on the same lattice site or for charge-transfer excitons in organic systems or molecular semiconductors \cite{lin2010charge,bardeen2014structure,nematiaram2021bright}.
We note that, while localized Frenkel or charge-transfer excitons can also in principle be modeled via the BSE, this is typically impractical given their high localization.
In systems such as molecular crystals and organic semiconductors, this leads to alternative treatments such as tight-binding models.

As another example, Ref.\ \cite{gotting2022moire} derives an excitonic moiré Bose--Hubbard model for interlayer excitons in twisted transition-metal dichalcogenide bilayers using a toy interaction between excitons derived from a simple point-charge model.
The calculation of the interaction strength between two sites could be refined by instead using the matrix elements of the exciton--exciton interaction $\mathcal{U}$ of Eq.\ \eqref{eq: exciton--exciton interaction} via
\begin{equation}
	U_{\vec{R} \vec{R}'} = \int \mathrm{d}^{2} r \, \mathrm{d}^{2} r' \, |w_{\vec{R}}(\vec{r})|^{2} \hspace{0.3mm} \mathcal{U}_{\vec{R} - \vec{R}'}(\vec{r}, \vec{r}') |w_{\vec{R}'}(\vec{r}')|^{2} .
\end{equation}
Here, $w_{\vec{R}}(\vec{r})$ are excitonic Wannier functions, and $\mathcal{U}_{\vec{R} - \vec{R}'}(\vec{r}, \vec{r}')$ is the coordinate-space counterpart of $\mathcal{U}(\BXTM, \BXRM, \BXRM')$, where in particular the $\BXTM$ label is transformed via wannierization analogously to the excitonic wave functions \cite{haber2023maximally}.
We note, however, that this requires our extended theory that will be presented in the sequel to this article.
This is so because evaluating $\mathcal{U}$ for the moiré system first requires diagonalizing the single-particle moiré Hamiltonian, which dresses the electron--hole interaction with form factors that we have not considered in this article \cite{chang2023continuum,peng2025many}.
These will introduce into the BSE all the information about the underlying moiré potential and we expect a significant effect on the exciton--exciton interaction matrix elements, especially in the regime of very flat bands.
Note, also, that it would not be correct to compute $U_{\vec{R} \vec{R}'}$ by using the effective potential obtained from rewriting the biexciton eigenvalue problem as a Schr\"odinger equation as we did in Eq.\ \eqref{eq: exciton--exciton potential effective}, because the resulting potential is explicitly non-Hermitian.
This is not an issue for the (effective) biexciton problem, whose eigenvalues remain real.
Meanwhile, $\mathcal{U}$ is Hermitian and is a good first approximation to be used for this purpose \footnote{In fact, $\mathcal{U}$ corresponds to the exciton--exciton interaction that results from setting the overlap integrals $\mathcal{K}^{\mathrm{c}}$ and $\mathcal{K}^{\mathrm{v}}$ to zero in the generalized eigenvalue problem of Eq.\ \eqref{eq: biexciton eigenvalue equation}.
For this reason, it only partially includes the effects of the composite nature of the excitons, but we nevertheless expect it to be a better approximation for this problem than the use of a simple potential between point charges.
Additional exchange effects can be included by adding also the exchange part of the kinetic term of Eq.\ \eqref{eq: Hamiltonian exciton--exciton}, if desired.}.
This illustrates how our variational treatment of two-exciton states yields straightforward and readily applicable approximations to other many-body problems.

In conclusion, the study of exciton--exciton interactions is an involved subject because of the intricate ways in which excitons can interact and rearrange their constituents.
The two-exciton potential and the many-body framework derived here lay down a solid groundwork for studying the interactions between these quasiparticles and pave the way for a more complete understanding of their dynamical behavior.

\section{Acknowledgments}

This work is supported by the Delta-ITP consortium, which is part of the Netherlands Organisation for Scientific Research (NWO).
We acknowledge the research program ``Materials for the Quantum Age'' (QuMat) for financial support.
This program (registration number 024.005.006) is part of the Gravitation program financed by the Dutch Ministry of Education, Culture and Science (OCW).

\begin{appendix}
    \section{Derivation details}
    \label{app: derivation details variational}
    

In this appendix we outline some details, intermediate results, and useful relations concerning the derivation of the biexciton eigenvalue equation and the associated effective exciton--exciton potential.

\subsection{Antisymmetrizer relations}
\label{app: A relations}

When the sums over two-particle modes $\Xmodeindex$ run over the entire space of solutions to the BSE (i.e., both bound and scattering states), the product of two exciton creation operators satisfies the following consistency relation in terms of $\mathcal{A}$:
\begin{equation}
\label{eq: projector symmetry exciton operator product}
	\begin{split}
	    &\hat{X}^{\dagger}_{\Xmodeindex_{1}, \BXTM /2 + \BXRM}\hat{X}^{\dagger}_{\Xmodeindex_2, \BXTM /2 - \BXRM} 
	    \\
	    &= \sum_{\Xmodeindex_1'\Xmodeindex_2'\BXRM'} \! 
	    \mathcal{A}^{\Xmodeindex_1'\Xmodeindex_2'}_{\Xmodeindex_1\Xmodeindex_2}(\BXTM ,\BXRM ,\BXRM ')
	    \hat{X}^{\dagger}_{\Xmodeindex_1', \BXTM /2 + \BXRM '}\hat{X}^{\dagger}_{\Xmodeindex_2', \BXTM /2 - \BXRM '} .
	\end{split}
\end{equation}
This relation can also be found in Ref.\ \cite{Combescot_2007}.
Furthermore, Eq.\ \eqref{eq: projector symmetry exciton operator product} in turn imposes the same constraint on the biexciton wave function, i.e.,
\begin{equation}
\label{eq: biexciton wavefunction projector relation}
	\BXWF{\Xmodeindex_1}{\Xmodeindex_2}{}{\BXTM}(\BXRM ) = \sum_{\Xmodeindex_1'\Xmodeindex_2'\BXRM'} \! \mathcal{A}^{\Xmodeindex_1'\Xmodeindex_2'}_{\Xmodeindex_1\Xmodeindex_2}(\BXTM ,\BXRM ,\BXRM ') \BXWF{\Xmodeindex_{1}'}{\Xmodeindex_{2}'}{}{\BXTM}(\BXRM') .
\end{equation}
In this way, the antisymmetrizer $\mathcal{A}$ implements the fermionic antisymmetry under exchange of identical particles into the biexciton wave function. 
With the help of Eqs.\ \eqref{eq: wavefunction idendity eigenvalues} and \eqref{eq: wavefunction idendity momenta and spin} it can be shown that $\mathcal{K}^{\mathrm{c}}$ and $\mathcal{K}^{\mathrm{v}}$ are involutory, i.e., they both satisfy
\begin{equation}
	\frac{1}{\mathcal{V}} \sum_{\mathclap{\bar{\Xmodeindex}_{1} \bar{\Xmodeindex}_{2} \vec{p}}} \mathcal{K}^{\bar{\Xmodeindex}_{1} \bar{\Xmodeindex}_{2}}_{\Xmodeindex_{1} \Xmodeindex_{2}}(\BXTM, \BXRM, \vec{p}) \mathcal{K}_{\bar{\Xmodeindex}_{1} \bar{\Xmodeindex}_{2}}^{\Xmodeindex_{1}' \Xmodeindex_{2}'}(\BXTM, \vec{p}, \BXRM') = \mathcal{V} \delta_{\BXRM \BXRM'} \delta_{\Xmodeindex_{1} \Xmodeindex_{1}'} \delta_{\Xmodeindex_{2} \Xmodeindex_{2}'} .
\end{equation}
This implies that the antisymmetrizer satisfies $\mathcal{A}^2 = \mathcal{A}$, confirming that it acts as a projector on the biexciton wave function $\Psi$. 
We stress that these statements are only true if the summations run over the entire space of particle-hole states; only in this case will $\mathcal{A}$ act like a projector.

Finally, the electron- and hole-exchange terms of $\mathcal{A}$ satisfy the mutual relations
\begin{subequations}
\label{eq: relation conduction and valence exchange}
	\begin{align}
		[\mathcal{K}^{\mathrm{v}}]^{\Xmodeindex_1'\Xmodeindex_2'}_{\Xmodeindex_1\Xmodeindex_2}(\BXTM ,\BXRM ,\BXRM ') &= 
	    [\mathcal{K}^{\mathrm{c}}]^{\Xmodeindex_2'\Xmodeindex_1'}_{\Xmodeindex_1\Xmodeindex_2}(\BXTM ,\BXRM ,-\BXRM '),
		\\ 
		&= [\mathcal{K}^{\mathrm{c}}]^{\Xmodeindex_1'\Xmodeindex_2'}_{\Xmodeindex_2\Xmodeindex_1}(\BXTM ,-\BXRM ,\BXRM '),
	\end{align}
\end{subequations}

\subsection{Biexciton eigenvalue problem}
\label{app: biexciton eigenvalue problem derivation}

To evaluate the first term on the right-hand side of Eq.\ \eqref{eq: biexciton energy functional}, we require the matrix element
\begin{equation}
\label{eq: Hamiltonian matrix element}
	\begin{split}
		&\frac{1}{2} \langle G| \hat{X}_{\Xmodeindex_1, \BXTM /2+\BXRM}\hat{X}_{\Xmodeindex_2, \BXTM /2-\BXRM}
		\hat{\mathcal{H}}
		\hat{X}^\dagger_{\Xmodeindex_2', \BXTM '/2-\BXRM '}\hat{X}^\dagger_{\Xmodeindex_1', \BXTM '/2+\BXRM '}|G\rangle 
		\\
		&= 
		E_{\Xmodeindex_1 \Xmodeindex_2}^{\Xmodeindex_1' \Xmodeindex_2'} (\BXTM, \BXRM, \BXRM') \mathcal{A}_{\Xmodeindex_1 \Xmodeindex_2}^{\Xmodeindex_1' \Xmodeindex_2'}(\BXTM ,\BXRM ,\BXRM ') \delta_{\BXTM \BXTM '} 
		\\
		&+ \frac{1}{\mathcal{V}} \mathcal{U}_{\Xmodeindex_1 \Xmodeindex_2}^{\Xmodeindex_1' \Xmodeindex_2'}(\BXTM ,\BXRM ,\BXRM)\delta_{\BXTM \BXTM'} . 
	\end{split}
\end{equation}
A useful intermediate result is the following relation between the correlator of four exciton operators and the antisymmetrizer $\mathcal{A}$:
\begin{equation}
\label{eq: projector matrix element connection}
	\begin{split}
		\langle G| \hat{X}_{\Xmodeindex_1, \BXTM /2+\BXRM}\hat{X}_{\Xmodeindex_2, \BXTM /2-\BXRM}
	    \hat{X}^\dagger_{\Xmodeindex_2', \BXTM/2-\BXRM '}\hat{X}^\dagger_{\Xmodeindex_1', \BXTM/2+\BXRM '}|G\rangle 
	    =&
	    \\
	    4 \mathcal{A}_{\Xmodeindex_1 \Xmodeindex_2}^{\Xmodeindex_1' \Xmodeindex_2'}(\BXTM, \BXRM, \BXRM')& .
	\end{split}
\end{equation}
We note that Eq.\ \eqref{eq: Hamiltonian matrix element} is diagonal in the total biexciton momentum, as expected, and that we have neglected vacuum terms that only contribute to the energy of the ground state $| G \rangle$ as they do not play a role in the description of the biexcitons \cite{Singh_1994}.
We also note that in the derivation of Eq.\ \eqref{eq: Hamiltonian matrix element} there are many ways to split the matrix element into a ``kinetic'' and an ``interaction'' part, depending on how one manipulates the matrix element using the single-exciton BSE to obtain the term containing the exciton eigenenergies.
The choice made in the main text is convenient because both parts are individually Hermitian.

It is possible to show that the Hamiltonian of Eq.\ \eqref{eq: Hamiltonian exciton--exciton} commutes with the antisymmetrizer and remains invariant under its action.
This allows us to rewrite the eigenvalue problem of Eq.\ \eqref{eq: biexciton eigenvalue equation} as
\begin{align}
\label{eq: biexciton BSE full variational freedom}
	&(\varepsilon^{\Xmodeindex_1}_{\BXTM /2 + \BXRM } + \varepsilon^{\Xmodeindex_2}_{\BXTM /2 - \BXRM }) \BXWF{\Xmodeindex_1}{\Xmodeindex_2}{}{\BXTM}(\BXRM )\\
	&+ \frac{1}{\mathcal{V}} \! \sum_{\Xmodeindex_1' \Xmodeindex_2' \BXRM'} \! [\mathcal{U}^0]_{\Xmodeindex_1 \Xmodeindex_2}^{\Xmodeindex_1' \Xmodeindex_2'} (\BXTM, \BXRM, \BXRM') \BXWF{\Xmodeindex_1'}{\Xmodeindex_2'}{}{\BXTM}(\BXRM')
	= \mathcal{E}_{\BXTM } \BXWF{\Xmodeindex_1}{\Xmodeindex_2}{}{\BXTM}(\BXRM ) . \nonumber
\end{align}
To clarify, it is not the case that the single-exciton and the interaction term in Eq.\ \eqref{eq: Hamiltonian exciton--exciton} separately commute with $\mathcal{A}$, only the combination of both does.

At first glance, Eq.\ \eqref{eq: biexciton BSE full variational freedom} may seem paradoxical as it does not explicitly contain any exchange processes between the underlying fermions, nor does it distinguish between singlet and triplet spin states.
That the latter should be a feature of the theory at least in some cases is known from the Heitler--London approach to the hydrogen atom, a system equivalent to that of our problem in the case of one particle being much heavier than the other.
This apparent paradox is simply a direct consequence of having written the biexciton as a superposition of all possible particle--hole states (i.e., including both exciton and scattering states) in the ansatz of Eq.\ \eqref{eq: biexciton creation operator}.
Doing so has allowed us to employ the completeness relations of the particle--hole wave functions to reduce the problem to the form given above.
The reduction to this form is actually not surprising: as mentioned in the main text, the eigenvalue problem of Eq.\ \eqref{eq: biexciton BSE full variational freedom} is precisely the two-electron, two-hole Schrödinger equation projected on the coupled particle--hole basis.
Consequently, the solutions to Eq.\ \eqref{eq: biexciton BSE full variational freedom} will generally not directly correspond to the different biexciton states.
Rather, one must project the solution back to the single-particle basis and antisymmetrize the resulting state in agreement with the Pauli exclusion principle for the two electrons and holes.
In particular, this will lead to energy splittings depending on the spin states of the underlying particles, thus no paradox exists.

\subsection{Effective interaction with electron and hole exchange} \label{app: effective int with e and h exchange}

When deriving the effective exciton--exciton potential, we started from the reduced Hamiltonian of Eq.\ \eqref{eq: Hamiltonian exciton--exciton reduced} and subsequently restricted it to the ground state in Eq.\ \eqref{eq: Hamiltonian exciton--exciton ground state}. 
However, it is also possible to make this restriction directly in Eq.\ \eqref{eq: Hamiltonian exciton--exciton}, without first using the exciton-exchange symmetry of the biexciton wave function. 
We can then write the antisymmetrizer in the excitonic ground state as
\begin{equation}
	\mathcal{A}_{S_{\mathrm{c}}S_{\mathrm{v}}}(\BXTM, \BXRM, \BXRM') = \frac{1}{\mathcal{V}}\sum_{\vec{p}}\NormalizationFunction_{S_{\mathrm{c}}}(\BXTM ,\BXRM ,\vec{p}) 
	P_{S_{\mathrm{c}}S_{\mathrm{v}}}(\vec{p}, \BXRM') ,
\end{equation}
where we have introduced the projector
\begin{equation}
	P_{S_{\mathrm{c}}S_{\mathrm{v}}}(\BXRM, \BXRM') = \frac{1}{2} [\delta_{\BXRM \BXRM'} + (-1)^{S_{\mathrm{c}} + S_{\mathrm{v}}} \delta_{\BXRM, -\BXRM'}] .
\end{equation}
This projector maps onto the subspace that is symmetric under exciton exchange. 
The biexciton wave function is evidently invariant under the action of $P$, i.e.,
\begin{equation} 
	\label{eq: biexciton WF exciton exchange projector}
	\sum_{\BXRM'} P_{S_{\mathrm{c}}S_{\mathrm{v}}}(\BXRM, \BXRM') \BXWF{S_{\mathrm{c}}}{S_{\mathrm{v}}}{}{\BXTM}(\BXRM') = \BXWF{S_{\mathrm{c}}}{S_{\mathrm{v}}}{}{\BXTM}(\BXRM) ,
\end{equation}
which is equivalent to Eq.\ \eqref{eq: biexciton effective wavefunction exciton exchange}.
Since $\mathcal{A}$ depends on $P$, a projector, it does not have a true inverse. 
However, a \emph{pseudoinverse} $\mathcal{A}^{-}$ can be defined via
\begin{equation}
	\sum_{\vec{p}}\mathcal{A}^{-}_{S_{\mathrm{c}}S_{\mathrm{v}}}(\BXTM, \BXRM, \vec{p})\mathcal{A}_{S_{\mathrm{c}}S_{\mathrm{v}}}(\BXTM, \vec{p}, \BXRM') = P_{S_{\mathrm{c}}S_{\mathrm{v}}}(\BXRM, \BXRM') ,
\end{equation}
such that
\begin{equation}
	\mathcal{A}^{-}_{S_{\mathrm{c}}S_{\mathrm{v}}}(\BXTM, \BXRM, \BXRM') = \frac{2}{\mathcal{V}}\sum_{\vec{p}}\NormalizationFunction^{-1}_{S_{\mathrm{c}}}(\BXTM ,\BXRM ,\vec{p}) 
	P_{S_{\mathrm{c}}S_{\mathrm{v}}}(\vec{p}, \BXRM') .
\end{equation} 
Here we have used the definitions of $\NormalizationFunction$ and its inverse found in Eqs.\ \eqref{eq: biexciton projector ground state} and \eqref{eq: inverse of P}, respectively.
Then, using the pseudoinverse $\mathcal{A}^{-}$, the biexciton eigenvalue problem can be written as 
\begin{equation}
\label{eq:BiexcEVProblemPseudoinverse}
	\begin{split}
		\sum_{\BXRM'} \bigg[ (\varepsilon_{\BXTM /2+\BXRM }+\varepsilon_{\BXTM /2-\BXRM } - \mathcal{E}_{\BXTM }) P_{S_{\mathrm{c}}S_{\mathrm{v}}}(\BXRM, \BXRM') &
		\\
		+ \, \frac{1}{\mathcal{V}} V^{\mathrm{eff}}_{S_{\mathrm{c}}S_{\mathrm{v}}}(\BXTM ,\BXRM ,\BXRM ') \bigg] \BXWF{S_{\mathrm{c}}}{S_{\mathrm{v}}}{}{\BXTM}(\BXRM') &
		= 0 .
	\end{split}
\end{equation}
which will attain the typical Schr\"odinger form upon the use of Eq.\ \eqref{eq: biexciton WF exciton exchange projector} to ``remove'' the remaining projector term, analogous to what we did to derive Eq.\ \eqref{eq: biexction effective BSE}.
The effective potential arising from this slightly different approach reads
\begin{equation}
	\label{eq: effective potential full exchange}
		\begin{split}
		&V^{\mathrm{eff}}_{S_{\mathrm{c}}S_{\mathrm{v}}}(\BXTM ,\BXRM ,\BXRM ') = 
		\frac{1}{\mathcal{V}}\sum_{\boldsymbol{p}} \NormalizationFunction^{-1}_{S_{\mathrm{c}}}(\BXTM ,\BXRM ,\boldsymbol{p})
		\\
		&\times \, \bigg\{ \frac{1}{4}
		(\varepsilon_{\BXTM /2+\boldsymbol{p}} + \varepsilon_{\BXTM /2-\boldsymbol{p}} - \varepsilon_{\BXTM /2+\BXRM '} - \varepsilon_{\BXTM /2-\BXRM '}) \\
		&\times \, [(-1)^{S_{\mathrm{c}}}\mathcal{K}^{\mathrm{c}} + (-1)^{S_{\mathrm{v}}}\mathcal{K}^{\mathrm{v}}](\BXTM ,\boldsymbol{p},\BXRM ')  + \, \mathcal{U}_{S_{\mathrm{c}}S_{\mathrm{v}}} (\BXTM ,\boldsymbol{p},\BXRM ') \bigg\} .
	\end{split}
\end{equation}
In this form, the potential explicitly contains the exchange terms related to electron and hole exchange, namely both overlap integrals $\mathcal{K}^{\mathrm{c}}$ and $\mathcal{K}^{\mathrm{v}}$ are present, as well as the electrostatic interaction of Eq.\ \eqref{eq: exciton--exciton interaction spin-basis}. 
Furthermore, when the potential is written in this form, it is also invariant when acted upon by $P$.
We emphasize that both Eq.\ \eqref{eq: biexction effective BSE} and Eq.\ \eqref{eq:BiexcEVProblemPseudoinverse} are different ways of rewriting the same generalized eigenvalue problem for the biexcitons and ultimately yield the same spectrum.

\subsection{Corrections from excited states}
\label{app: corrections excited}

Here we compute the correction of Eq.\ \eqref{eq: delta h 0000} to the reduced Hamiltonian of the generalized biexciton eigenvalue problem.
Starting from the general Eq.\ \eqref{eq: biexciton eigenvalue equation reduced}, we separate the contributions of the ground state ``0'' and the excited states $\nu$.
Since we seek a dipole--dipole contribution, we neglect the contribution of the wave-function components $\Psi^{0 \nu}$ and directly look at the effect of $\Psi^{\nu \nu'}$.
It can be shown that including the components $\Psi^{0 \nu}$ does not give significant corrections at long distances, which is physically clear owing to the fact that these correspond to a single exciton in an excited state instead of two instantaneous dipoles.
Eliminating the excited components $\Psi^{\nu \nu'}$ from Eq.\ \eqref{eq: biexciton eigenvalue equation reduced} in favor of $\Psi^{00}$, we find that they contribute to the biexciton problem for $\Psi^{00}$ (as before named simply $\Psi$) via a perturbative term $\delta h$, i.e.
\begin{equation}
\label{eq: h plus dh ev problem}
	\frac{1}{2 \mathcal{V}}\sum_{\BXRM'} [h + \delta h - \mathcal{E}_{\BXTM} \NormalizationFunction]_{00}^{00}(\BXTM, \BXRM, \BXRM') \BXWF{}{}{}{\BXTM}(\BXRM') = 0 .
\end{equation}
Defining $\tilde{h} \equiv h - \mathcal{E}_{\BXTM} R$, this term reads
\begin{equation}
\label{eq: dh0000 first expr}
	\begin{split}
		\delta h_{00}^{00}(\BXTM, \BXRM, \BXRM') = {-} \frac{1}{\mathcal{V}^2} \sum_{\vec{p} \vec{p}'} \sum_{\nu \nu'} \sum_{\bar{\nu} \bar{\nu}'} \tilde{h}_{00}^{\nu \nu'}(\BXTM, \BXRM, \vec{p})& \\
		\times \, [\tilde{h}^{-1}]_{\nu \nu'}^{\bar{\nu} \bar{\nu}'}(\BXTM, \vec{p}, \vec{p}') \tilde{h}_{\bar{\nu} \bar{\nu}'}^{00}(\BXTM, \vec{p}', \BXRM')& ,
	\end{split}
\end{equation} 
where the inverse of $\tilde{h}$ is defined via
\begin{equation}
	\frac{1}{\mathcal{V}} \hspace{0.3mm} \sum_{\mathclap{\bar{\nu}_{1} \bar{\nu}_{2} \vec{p}}} \hspace{0.3mm} \tilde{h}_{\nu_{1} \nu_{2}}^{\bar{\nu}_{1} \bar{\nu}_{2}}(\BXTM, \BXRM, \vec{p}) [\tilde{h}^{-1}]^{\nu_{1}' \nu_{2}'}_{\bar{\nu}_{1} \bar{\nu}_{2}}(\BXTM, \vec{p}, \BXRM') = \mathcal{V} \delta_{\BXRM \BXRM'} \delta_{\nu_{1} \nu_{1}'} \delta_{\nu_{2} \nu_{2}'} .
\end{equation}
We emphasize that $h^{00}_{00}$ and $h_{00}^{\nu \nu'}$ do not show up in these summations, as $\nu$ and $\nu'$ run strictly over excited states.
Eq.\ \eqref{eq: dh0000 first expr} is valid at large distances.
In this case it is safe to neglect all exchange processes, meaning that we set $\mathcal{K}^{\mathrm{c}}$ and $\mathcal{U}^{\mathrm{c}}$ to zero.
Then, $\NormalizationFunction_{00}^{\nu \nu'} \approx 0$, while $\NormalizationFunction_{\nu \nu'}^{\bar{\nu} \bar{\nu}'}(\BXTM, \BXRM, \BXRM') \approx \mathcal{V} \delta_{\BXRM \BXRM'} \delta_{\nu \bar{\nu}} \delta_{\nu' \bar{\nu}'}$.
When plugged into Eq.\ \eqref{eq: h plus dh ev problem}, this results in an eigenvalue problem in the usual form.
Now, in the denominator of Eq.\ \eqref{eq: dh0000 first expr} we neglect the interaction term, as we assume that the biexciton binding energies are much smaller than the energy of the constituent excitons.
This results in
\begin{equation}
	\begin{split}
		&[\tilde{h}^{-1}]_{\nu \nu'}^{\bar{\nu} \bar{\nu}'}(\BXTM, \BXRM, \BXRM') \\
		&\approx \mathcal{V} \delta_{\BXRM \BXRM'} \delta_{\nu \bar{\nu}} \delta_{\nu' \bar{\nu}'} (\varepsilon^{\nu}_{\BXTM/2 + \BXRM} + \varepsilon^{\nu'}_{\BXTM/2 - \BXRM} - \mathcal{E}_{\BXTM})^{-1} .
	\end{split}
\end{equation}
Meanwhile, in the numerators it is precisely the (direct) interaction term that dominates, leading to Eq.\ \eqref{eq: delta h 0000} of the main text.

    \section{Second-quantization interaction components} 
    \label{app: Second Quantization Interaction Components}
    

The explicit expressions for the different components of the exciton--exciton interaction are given below.
For legibility, the exciton CoM momentum labels on the wave function are defined in terms of biexciton momenta via Eq.\ \eqref{eq: CoM momentum coords (biexcitons)}, where an (un)primed exciton momentum will be related to an (un)primed relative biexciton momentum via
\begin{equation}
\begin{split}
	\XTM_1 &= \frac{1}{2}\BXTM + \BXRM, \qquad\hspace{0.22mm}
    \XTM_1' = \frac{1}{2}\BXTM' + \BXRM',
    \\
    \XTM_2 &= \frac{1}{2}\BXTM - \BXRM, \qquad
    \XTM_2' =  \frac{1}{2}\BXTM' - \BXRM' .
\end{split}
\end{equation}
Note that because of momentum conservation $\BXTM=\BXTM'$.
All expressions below are presented in full generality and thus depend on the exciton labels $\Xmodeindex$ and contain spin sums.
However, when considered as part of the potential between ground-state excitons derived in Sec.\ \ref{sec: ground state excitons} for a spin-degenerate system, one must remember to disregard the spin sums (which have been performed explicitly by following the procedure outlined in Appendix\ \ref{app: Spin-Basis Transformation}) and set all exciton modes to the desired ground state.

The direct interaction term takes the form
\begin{widetext}
    \begin{equation}
    \label{eqappx: direct term}
        \begin{split}
            &\hspace{0pt-\widthof{$= V(\BXRM-\BXRM') \,\, \Bigg\{$}}
            [\mathcal{U}^0]^{\Xmodeindex_1'\Xmodeindex_2'}_{\Xmodeindex_1\Xmodeindex_2}(\BXTM,\BXRM,\BXRM')
            \\ 
            = V(\BXRM-\BXRM') \, \Bigg\{&
            \bigg[\frac{1}{\mathcal{V}}\sum_{\mathclap{\alpha\beta\XRM}}
            \big[\XWF{\alpha}{\beta}{\Xmodeindex_1}{\XTM_1}(\XRM)\big]^*
            \XWF{\alpha}{\beta}{\Xmodeindex_1'}{\XTM_1'}(\XRM - \dlmassv (\BXRM-\BXRM'))\bigg]
            \bigg[\frac{1}{\mathcal{V}}\sum_{\mathclap{\alpha'\beta'\XRM'}}
            \big[\XWF{\alpha'}{\beta'}{\Xmodeindex_2}{\XTM_2}(\XRM')\big]^*
            \XWF{\alpha'}{\beta'}{\Xmodeindex_2'}{\XTM_2'}(\XRM' + \dlmassv (\BXRM-\BXRM'))\bigg]
            \\ 
            +\, &
            \bigg[\frac{1}{\mathcal{V}}\sum_{\mathclap{\alpha\beta\XRM}}
            \big[\XWF{\alpha}{\beta}{\Xmodeindex_1}{\XTM_1}(\XRM)\big]^*
            \XWF{\alpha}{\beta}{\Xmodeindex_1'}{\XTM_1'}(\XRM + \dlmassc (\BXRM-\BXRM'))\bigg]
            \bigg[\frac{1}{\mathcal{V}}\sum_{\mathclap{\alpha'\beta'\XRM'}}
            \big[\XWF{\alpha'}{\beta'}{\Xmodeindex_2}{\XTM_2}(\XRM')\big]^*
            \XWF{\alpha'}{\beta'}{\Xmodeindex_2'}{\XTM_2'}(\XRM' - \dlmassc (\BXRM-\BXRM'))\bigg]
            \\ 
            -\, &
            \bigg[\frac{1}{\mathcal{V}}\sum_{\mathclap{\alpha\beta\XRM}}
            \big[\XWF{\alpha}{\beta}{\Xmodeindex_1}{\XTM_1}(\XRM)\big]^*
            \XWF{\alpha}{\beta}{\Xmodeindex_1'}{\XTM_1'}(\XRM - \dlmassv (\BXRM-\BXRM'))\bigg]
            \bigg[\frac{1}{\mathcal{V}}\sum_{\mathclap{\alpha'\beta'\XRM'}}
            \big[\XWF{\alpha'}{\beta'}{\Xmodeindex_2}{\XTM_2}(\XRM')\big]^*
            \XWF{\alpha'}{\beta'}{\Xmodeindex_2'}{\XTM_2'}(\XRM' - \dlmassc (\BXRM-\BXRM'))\bigg]
            \\ 
            -\, &
            \bigg[\frac{1}{\mathcal{V}}\sum_{\mathclap{\alpha\beta\XRM}}
            \big[\XWF{\alpha}{\beta}{\Xmodeindex_1}{\XTM_1}(\XRM)\big]^*
            \XWF{\alpha}{\beta}{\Xmodeindex_1'}{\XTM_1'}(\XRM + \dlmassc (\BXRM-\BXRM'))\bigg]
            \bigg[\frac{1}{\mathcal{V}}\sum_{\mathclap{\alpha'\beta'\XRM'}}
            \big[\XWF{\alpha'}{\beta'}{\Xmodeindex_2}{\XTM_2}(\XRM')\big]^*
            \XWF{\alpha'}{\beta'}{\Xmodeindex_2'}{\XTM_2'}(\XRM' + \dlmassv (\BXRM-\BXRM'))\bigg]
            \Bigg\}.
        \end{split}
    \end{equation}
\end{widetext}
We note that this can be compactly written as
    \begin{equation}
    \label{eqappx: direct term alt}
        \begin{split}
            &[\mathcal{U}^0]^{\Xmodeindex_1'\Xmodeindex_2'}_{\Xmodeindex_1\Xmodeindex_2}(\BXTM,\BXRM,\BXRM') = V(\BXRM-\BXRM') \\
            &\times \big[\Gamma_{1}(\gamma_{c} (\BXRM - \BXRM')) - \Gamma_{1}({-}\gamma_{v} (\BXRM - \BXRM')) \big] \\
            &\times \big[\Gamma_{2}({-}\gamma_{c}(\BXRM - \BXRM')) - \Gamma_{2}(\gamma_{v} (\BXRM - \BXRM')) \big] ,
        \end{split}
    \end{equation}
    where
    \begin{equation}
        \Gamma_{i}(\vec{p}) = \frac{1}{\mathcal{V}}\sum_{\mathclap{\alpha\beta\XRM}}
            \big[\XWF{\alpha}{\beta}{\Xmodeindex_i}{\XTM_i}(\XRM)\big]^*
            \XWF{\alpha}{\beta}{\Xmodeindex_i'}{\XTM_i'}(\XRM + \vec{p})
    \end{equation}
    for $i = 1, 2$.
The exciton exchange term is simply given in terms of $\mathcal{U}^{0}$ by
\begin{subequations}
\label{eqappx: exciton exchange term all}
    \begin{align}
        \label{eqappx: exciton exchange term}
        [\mathcal{U}^{\mathrm{X}}]^{\mu_{1}' \mu_{2}'}_{\mu_{1} \mu_{2}}(\BXTM, \BXRM, \BXRM') &= [\mathcal{U}^{0}]^{\mu_{2}' \mu_{1}'}_{\mu_{1} \mu_{2}}(\BXTM, \BXRM, {-}\BXRM') \\
        &= [\mathcal{U}^{0}]^{\mu_{1}' \mu_{2}'}_{\mu_{2} \mu_{1}}(\BXTM, {-}\BXRM, \BXRM') .
    \end{align}
\end{subequations}
These expressions contain overlap integrals with specific shifts in some of the arguments of the wave functions.
The presence of $\dlmassc $ ($\dlmassv $) in the argument signifies that the electron (hole) does not take part in the Coulomb scattering.
Furthermore, the multiplicative momentum factor of $\BXRM + \BXRM'$ or $\BXRM - \BXRM'$ indicates whether that particle exchanges between excitons or not, respectively.
For example, the first term of Eq.\ \eqref{eqappx: direct term} contains twice the factor $\dlmassv (\BXRM-\BXRM')$.
This implies that the interaction is between the conduction electrons and that no hole exchange takes place.
Moreover, the fact that the Coulomb interaction has $\BXRM - \BXRM'$ as its argument indicates that there is no exchange between excitons, whereas this is the case for Eq.\ \eqref{eqappx: exciton exchange term}.
For both $\mathcal{U}^{0}$ and $\mathcal{U}^{\mathrm{X}}$, the first two terms correspond to electron--electron and hole--hole scatterings, respectively, while the last two correspond to electron--hole scatterings.

On the other hand, the electron-exchange component reads
\begin{widetext}
    \begin{equation}
    \label{eqappx: conduction exchange term}
        \begin{split}
            &\hspace{0pt-\widthof{$\displaystyle = \frac{1}{2\mathcal{V}}\sum_{\alpha\beta\XRM} \, \sum_{\alpha'\beta'} \,\,$}}[\mathcal{U}^{\mathrm{c}}]^{\Xmodeindex_1'\Xmodeindex_2'}_{\Xmodeindex_1\Xmodeindex_2}(\BXTM,\BXRM,\BXRM') \\
            = \frac{1}{2\mathcal{V}}\sum_{\alpha\beta\XRM} \, \sum_{\alpha'\beta'} \,
            &\big[\XWF{\alpha}{\beta}{\Xmodeindex_1}{\XTM_1}(\XRM)\big]^*
            \big[\XWF{\alpha'}{\beta'}{\Xmodeindex_2}{\XTM_2}(\XRM+\dlmassc (\BXRM+\BXRM') - \dlmassv (\BXRM-\BXRM'))\big]^*
            \XWF{\alpha'}{\beta}{\Xmodeindex_1'}{\XTM_1'}(\XRM - \dlmassv (\BXRM-\BXRM'))
            \XWF{\alpha}{\beta'}{\Xmodeindex_2'}{\XTM_2'}(\XRM + \dlmassc (\BXRM+\BXRM'))
            \\[-1mm]
            \times\, &\Big(
            \Delta^{\alpha\beta}_{\XTM_1 \XRM}
            + \Delta^{\alpha'\beta'}_{\XTM_2, \XRM+\dlmassc (\BXRM+\BXRM') - \dlmassv (\BXRM-\BXRM')} 
            + \Delta^{\alpha'\beta}_{\XTM_1', \XRM - \dlmassv (\BXRM-\BXRM')}
            +\Delta^{\alpha\beta'}_{\XTM_2', \XRM+\dlmassc (\BXRM+\BXRM')}
            - \varepsilon^{\Xmodeindex_1}_{\XTM_1}
            - \varepsilon^{\Xmodeindex_2}_{\XTM_2}
            - \varepsilon^{\Xmodeindex_1'}_{\XTM_1'}
            - \varepsilon^{\Xmodeindex_2'}_{\XTM_2'}
            \Big)
            \\ 
            -\frac{1}{\mathcal{V}^2}\sum_{\alpha\beta\XRM}\sum_{\alpha'\beta'\XRM'}
            &\bigl[\XWF{\alpha}{\beta}{\Xmodeindex_1}{\XTM_1}(\XRM)\big]^*
            \XWF{\alpha'}{\beta}{\Xmodeindex_1'}{\XTM_1'}(\XRM - \dlmassv (\BXRM-\BXRM')) 
            V(\XRM-\XRM')
            \\
            \times \, &\big[\XWF{\alpha'}{\beta'}{\Xmodeindex_2}{\XTM_2}(\XRM' + \dlmassc (\BXRM+\BXRM') - \dlmassv (\BXRM-\BXRM'))\big]^*
            \XWF{\alpha}{\beta'}{\Xmodeindex_2'}{\XTM_2'}(\XRM' + \dlmassc (\BXRM+\BXRM'))
            \\ 
            -\frac{1}{\mathcal{V}^2}\sum_{\alpha\beta\XRM}\sum_{\alpha'\beta'\XRM'}
            &\bigl[\XWF{\alpha}{\beta}{\Xmodeindex_1}{\XTM_1}(\XRM )\big]^*
            \XWF{\alpha}{\beta'}{\Xmodeindex_2'}{\XTM_2'}(\XRM + \dlmassc (\BXRM+\BXRM'))
            V(\XRM-\XRM')
            \\
            \times \, &\big[\XWF{\alpha'}{\beta'}{\Xmodeindex_2}{\XTM_2}(\XRM' + \dlmassc (\BXRM+\BXRM') - \dlmassv (\BXRM-\BXRM'))\big]^*
            \XWF{\alpha'}{\beta}{\Xmodeindex_1'}{\XTM_1'}(\XRM' - \dlmassv (\BXRM-\BXRM')).
        \end{split}
    \end{equation}
\end{widetext}
The hole-exchange component is straightforwardly found as
\begin{subequations}
\label{eqappx: valence exchange term all}
	\begin{align}
		\label{eqappx: valence exchange term}
		[\mathcal{U}^{\mathrm{v}}]^{\mu_{1}' \mu_{2}'}_{\mu_{1} \mu_{2}}(\BXTM, \BXRM, \BXRM') &= [\mathcal{U}^{\mathrm{c}}]^{\mu_{2}' \mu_{1}'}_{\mu_{1} \mu_{2}}(\BXTM, \BXRM, {-}\BXRM'), 
		\\ 
		&= [\mathcal{U}^{\mathrm{c}}]^{\mu_{1}' \mu_{2}'}_{\mu_{2} \mu_{1}}(\BXTM, {-}\BXRM, \BXRM').
	\end{align}
\end{subequations}
The first term of Eq.\ \eqref{eqappx: conduction exchange term} relates to the electron--hole scattering, the second to the electron--electron scattering, and the last to the hole--hole scattering.
Moreover, note the similarity of the distribution of relative momenta and exciton eigenvalues over the four wave functions in $\mathcal{U}^{\mathrm{c}}$ and $\mathcal{U}^{\mathrm{v}}$ compared with $\mathcal{K}^{\mathrm{c}}$ and $\mathcal{K}^{\mathrm{v}}$, respectively.
This resemblance is one way to see that these terms indeed correspond to the exchange of one of the particle types.
We also draw attention to the fact that the first term of Eq.\ \eqref{eqappx: conduction exchange term} does not explicitly contain the Coulomb interaction.
This is because in this term it was possible to rewrite the single-particle interaction in terms of energy factors using the exciton BSE,
\begin{equation}
	\frac{1}{\mathcal{V}} \sum_{\XRM'} 
	V(\XRM - \XRM')
	\Phi^{\alpha\beta}_{\Xmodeindex\XTM}(\XRM')
	= (\Delta^{\alpha\beta}_{\XTM\XRM} - \varepsilon^{\Xmodeindex}_{\XTM}) \Phi^{\alpha\beta}_{\Xmodeindex \XTM}(\XRM).
\end{equation}
This transformation is not possible for the other terms.
The sum of the terms of Eqs.\ \eqref{eqappx: direct term}--\eqref{eqappx: valence exchange term} results in the (nonlocal) exciton--exciton interaction of Eq.\ \eqref{eq: exciton--exciton interaction}.

    \section{Spin-basis transformation} \label{app: Spin-Basis Transformation}
    

In this appendix we derive the behavior under exciton exchange of the biexciton wave function and the components of the exciton--exciton interaction.
To start we define the biexciton wave function in the basis of excitons labeled by the individual spins of the conduction and valence electrons as
\begin{equation}
\label{eqappx: biexciton wf individual spin basis}
    \BXWF{\alpha}{\alpha'}{\beta}{\beta'} (\BXRM) = \langle{\alpha \alpha' \beta \beta'} \vert \BXWF{}{}{}{}(\BXRM) \rangle .
\end{equation}
As explained in the main text, this basis can be chosen because of the spin degeneracy in our system.
Here $\alpha, \beta, \dots$ stand for the spin projections, and the total spin is omitted.
The results in this section are independent of the latter, as we clarify below.
In Eq.\ \eqref{eqappx: biexciton wf individual spin basis} and below we omit the dependence on the total momentum for simplicity.
In this basis, the exciton-exchange operation results in the symmetry requirement $\BXWF{\alpha}{\alpha'}{\beta}{\beta'} (\BXRM) = \BXWF{\alpha'}{\alpha}{\beta'}{\beta} ({-}\BXRM)$.
The biexciton wave function in the pairwise-coupled conduction and valence basis is denoted by $\Psi^{\vec{S}_{\mathrm{c}} \vec{S}_{\mathrm{v}}}(\BXRM) = \langle \vec{S}_{\mathrm{c}} \vec{S}_{\mathrm{v}} \vert \Psi(\BXRM) \rangle$, where $\vec{S}_{\mathrm{c}}$ and $\vec{S}_{\mathrm{v}}$ contain both the total-spin and the spin-projection quantum numbers.
We will see that the theory only depends on the former.

We derive the behavior of the biexciton wave function in this new basis under reflection as follows:
\begin{align}
    \Psi^{\vec{S}_{\mathrm{c}} \vec{S}_{\mathrm{v}}}({-}\BXRM) &= \sum_{\alpha \alpha'} \sum_{\beta \beta'} \langle \vec{S}_{\mathrm{c}} \vert \alpha \alpha' \rangle \langle \vec{S}_{\mathrm{v}} \vert \beta \beta' \rangle \Psi^{\alpha \alpha'}_{\beta \beta'}({-} \BXRM) \nonumber \\
    &= (-1)^{S_{\mathrm{c}} + S_{\mathrm{v}}} \hspace{0.3mm} \Psi^{\vec{S}_{\mathrm{c}} \vec{S}_{\mathrm{v}}}(\BXRM) .
\end{align}
Here we have used $\langle \vec{S}_{\mathrm{c}} \vert \alpha \alpha' \rangle = (-1)^{S_{\mathrm{c}} + 1} \langle \vec{S}_{\mathrm{c}} \vert \alpha' \alpha \rangle$ and similarly for $\langle \vec{S}_{\mathrm{v}} \vert \beta \beta' \rangle$, which follows from the general property $\langle J M \vert j_{1} m_{1} j_{2} m_{2} \rangle = (-1)^{J - j_{1} - j_{2}} \langle J M \vert j_{2} m_{2} j_{1} m_{1} \rangle$ of the Clebsch-Gordan coefficients when $j_{1} = j_{2}$ is a half integer.

Now we turn our attention to the interaction terms of the previous appendix.
We will do the calculation for $\mathcal{U}^{\mathrm{c}}$, with the rest of the terms following in a similar fashion.
As in the main text, we assume separability of the spin and orbital parts of the exciton wave function via Eq.\ \eqref{eq: exciton wavefunction spin seperation}.
For the calculation below it is now convenient to assume that the exciton spin is written in the basis of the individual particles, which is possible because of the spin degeneracy of the system under consideration.
Then we simply have $\XWF{\alpha}{\beta}{\alpha_{X} \beta_{X}}{}(\XRM) = \delta_{\alpha \alpha_{X}} \delta_{\beta \beta_{X}} \XWF{}{}{}{}(\XRM)$ and the spin sums in Eq.\ \eqref{eqappx: conduction exchange term} become trivial, giving
\begin{equation}
    [\mathcal{U}^{\mathrm{c}}]^{\alpha_{1}' \beta_{1}' \alpha_{2}' \beta_{2}'}_{\alpha_{1} \beta_{1} \alpha_{2} \beta_{2}}(\BXTM, \BXRM, \BXRM') = \delta^{\alpha_{2}'}_{\alpha_{1}} \delta^{\alpha_{1}'}_{\alpha_{2}} \delta^{\beta_{1}'}_{\beta_{1}}  \delta^{\beta_{2}'}_{\beta_{2}} \hspace{0.3mm} \mathcal{U}^{\mathrm{c}}(\BXTM, \BXRM, \BXRM') .
\end{equation}
Here, $\mathcal{U}^{\mathrm{c}}(\BXTM, \BXRM, \BXRM')$ is the electron-exchange interaction term after separating the spin-dependent part, which is now common for all spin states.
We now write the interaction term in the pairwise-coupled basis (omitting the momentum dependence for compactness) as
\begin{align}
    &[\mathcal{U}^{\mathrm{c}}]^{\vec{S}_{\mathrm{c}}' \vec{S}_{\mathrm{v}}'}_{\vec{S}_{\mathrm{c}} \vec{S}_{\mathrm{v}}} \nonumber \\
    &= \sum_{\mathclap{\text{spins}}} \, \langle \vec{S}_{\mathrm{c}} \vert \alpha_{1} \alpha_{2} \rangle \langle \vec{S}_{\mathrm{v}} \vert \beta_{1}  \beta_{2} \rangle [\mathcal{U}^{\mathrm{c}}]^{\alpha_{1}' \beta_{1}' \alpha_{2}' \beta_{2}'}_{\alpha_{1} \beta_{1} \alpha_{2} \beta_{2}} \langle \alpha_{1}' \alpha_{2}' \vert \vec{S}_{\mathrm{c}}' \rangle \langle \beta_{1}'  \beta_{2}' \vert \vec{S}_{\mathrm{v}}' \rangle \nonumber \\
    &= \mathcal{U}^{\mathrm{c}} \sum_{\mathclap{\alpha_{1} \alpha_{2}}} \langle \vec{S}_{\mathrm{c}} \vert \alpha_{1} \alpha_{2} \rangle \langle \alpha_{2} \alpha_{1} \vert \vec{S}_{\mathrm{c}}' \rangle \sum_{\beta_{1} \beta_{2}} \langle \vec{S}_{\mathrm{v}} \vert \beta_{1}  \beta_{2} \rangle \langle  \beta_{1}  \beta_{2} \vert \vec{S}_{\mathrm{v}}' \rangle \nonumber \\
    &= (-1)^{S_{\mathrm{c}} + 1} \delta_{\vec{S}_{\mathrm{c}} \vec{S}_{\mathrm{c}}'} \hspace{0.3mm} \delta_{\vec{S}_{\mathrm{v}} \vec{S}_{\mathrm{v}}'} \hspace{0.3mm} \mathcal{U}^{\mathrm{c}} ,
\end{align}
where again we have made use of the aforementioned property of the Clebsch-Gordan coefficients.
This shows that the electron-exchange term of the interaction is fully diagonal in this particular coupled basis and how the spin-dependent prefactor found in the main text appears.
A similar calculation for the other terms yields the combination that appears in Eq.\ \eqref{eq: exciton--exciton interaction spin-basis}.
Furthermore, as the spin-degenerate exciton states have the same spatial wave function, we have
\begin{subequations}
    \begin{align}
        \mathcal{U}^{0}(\BXTM, \BXRM, \BXRM') &= \mathcal{U}^{\mathrm{X}}(\BXTM, \BXRM, {-}\BXRM') , \\
        \mathcal{U}^{\mathrm{c}}(\BXTM, \BXRM, \BXRM') &= \mathcal{U}^{\mathrm{v}}(\BXTM, \BXRM, {-}\BXRM') ,
    \end{align}
\end{subequations}
which follows from Eqs.\ \eqref{eqappx: exciton exchange term all} and \eqref{eqappx: valence exchange term all} when the spin sums are ignored.
From here it follows that the matrix elements satisfy
\begin{subequations}
    \begin{align}
        [\mathcal{U}^{0}]^{\vec{S}_{\mathrm{c}}' \vec{S}_{\mathrm{v}}'}_{\vec{S}_{\mathrm{c}} \vec{S}_{\mathrm{v}}}(\BXTM, \BXRM, \BXRM') &= (-1)^{S_{\mathrm{c}} + S_{\mathrm{v}}} [\mathcal{U}^{\mathrm{X}}]^{\vec{S}_{\mathrm{c}}' \vec{S}_{\mathrm{v}}'}_{\vec{S}_{\mathrm{c}} \vec{S}_{\mathrm{v}}}(\BXTM, \BXRM, {-}\BXRM') , \\
        [\mathcal{U}^{\mathrm{c}}]^{\vec{S}_{\mathrm{c}}' \vec{S}_{\mathrm{v}}'}_{\vec{S}_{\mathrm{c}} \vec{S}_{\mathrm{v}}}(\BXTM, \BXRM, \BXRM') &= (-1)^{S_{\mathrm{c}} + S_{\mathrm{v}}} [\mathcal{U}^{\mathrm{v}}]^{\vec{S}_{\mathrm{c}}' \vec{S}_{\mathrm{v}}'}_{\vec{S}_{\mathrm{c}} \vec{S}_{\mathrm{v}}}(\BXTM, \BXRM, {-}\BXRM') .
    \end{align}
\end{subequations}
Finally, since the interaction depends on the total value of the coupled spins only, it is justified to write the biexciton wave function in this basis as $\BXWF{S_{\mathrm{c}}}{S_{\mathrm{v}}}{}{}$ like we have done in the main text, in the understanding that each of these states has the appropriate degeneracy.

    \section{Heavy-hole integrals} \label{app: Heavy-Hole Integrals}
    

Here we show the derivation of the hydrogenic, heavy-hole exciton--exciton potential between two ground-state excitons.
We note that we will work in the thermodynamic limit, such that $(1/\mathcal{V})\sum_{\BXRM} \rightarrow \int \mathrm{d}^{d} k / (2 \pi)^{d}$ and $\mathcal{V} \delta_{\BXRM \BXRM'} \rightarrow (2 \pi)^{d} \delta(\BXRM - \BXRM')$.
The starting point is the momentum-space expression of the potential within this limit, namely
\begin{align}
\label{eqappx: Heitler--London momentum space}
    V^{\mathrm{eff}}_{S_{\mathrm{c}}}(\BXRM,\BXRM') = 
    \int\frac{\mathrm{d}^{2} p}{(2\pi)^2}\, \NormalizationFunctionVeff^{-1}_{S_{\mathrm{c}}}(\BXRM,\boldsymbol{p}) 
    [\mathcal{U}^0 - (-1)^{S_{\mathrm{c}}} \mathcal{U}^{\mathrm{c}}](\boldsymbol{p},\BXRM').
\end{align}
In this appendix, the necessary integrals will be computed to solve for the heavy-hole potential in position space.
Specifically, the normalization $\NormalizationFunctionVeff^{-1}_{S_{\mathrm{c}}}$ and direct interaction $\mathcal{U}^{0}$ can be solved analytically.
The electron-exchange interaction can be solved mostly analytically, except for one of its integrals, which is solved numerically.
We first briefly introduce dimensionless quantities to be used throughout.
In short, the goal of this appendix is, starting from Eq.\ \eqref{eqappx: Heitler--London momentum space}, to obtain the different position-space components for the Heitler--London potential of Eq.\ \eqref{eq: Heitler--London potential real space}, 
\begin{align} \label{eqappx: Heitler--London real space}
    V^{\mathrm{HL}}_{S_{\mathrm{c}}}(r) =
    \frac{\mathcal{U}^0(r) - (-1)^{S_{\mathrm{c}}}\,\mathcal{U}^{\mathrm{c}}(r)}{1 + (-1)^{S_{\mathrm{c}}} \mathcal{K}^{\mathrm{c}}(r)}.
\end{align}

\subsection{Dimensionless quantities} \label{secsupp: Dimensionless Quantities}

Here we discuss the dimensionless quantities used in the calculation of the exciton--exciton potential in the heavy-hole limit.
For clarity, in this appendix we do not set $\hbar$ to unity.
We consider hydrogenic excitons with a potential $V(r) = e^{2} / 4 \pi \epsilon r$, where $\epsilon = \epsilon_{0} \epsilon_{\mathrm{r}}$ is the total effective dielectric constant of the system written in terms of the effective relative dielectric constant $\epsilon_{\mathrm{r}}$.
This is a well-known problem with an analytical solution \cite{chao1991analytical,yang1991analytic,Parfitt_2002,efimkin2021electron}.
The binding energies of such hydrogen-like excitons in 2D are given by
\begin{align}
    \varepsilon^{\mathrm{b}}_n = \frac{\varepsilon^{\mathrm{b}}_{X}}{(2n + 1)^2} ,
\end{align}
where $n = 0,1,2,\dots$ is the principal quantum number and $\varepsilon^{\mathrm{b}}_{0} = \varepsilon^{\mathrm{b}}_{X} = \hbar^{2} / 2 \mu_{X} a_{X}^{2}$ is the binding energy of the ground-state excitons.
Here, $a_{\genericXLabel} = a_{0} / 2$ is the mean radius of a ground-state exciton, i.e., $\langle r \rangle = a_{X}$ in the ground state.
The Bohr radius $a_{0}$ is defined as
\begin{align}
    \aBohr = \frac{4\pi\epsilon\hbar^2}{\mu_{X}  e^2} ,
\end{align}
with $\mu_{\genericXLabel} = m_{c} m_{v} / (m_{c} + m_{v})$ the reduced mass of the system.
Note that the binding energies $\varepsilon^{b}_{n}$ do not depend on the azimuthal quantum number, which is the result of the accidental degeneracy associated with the conservation of the so-called Runge--Lenz vector \cite{Parfitt_2002}.
The total exciton energies thus read $\varepsilon^{n}_{\vec{Q}} = E_{\mathrm{g}} - \varepsilon^{b}_{n} + \vec{Q}^{2} / 2 M_{X}$.

In the discussion below and in Sec.\ \ref{sec: Hydrogenic Example} we take $a_{X}$ as the unit of length and $\varepsilon^{b}_{X}$ as the unit of energy.
In particular we then have $\Delta_{\XTM \XRM} - \varepsilon_{\XTM} = 1 + \XRMM^{2}$.
Furthermore, the Coulomb potential in position and momentum space reads $V(r) = 1 / r$ and $V(\XRMM) = 2 \pi / \XRMM$, respectively. 
Similarly, the ground-state exciton wave function is $\Phi(r) = \sqrt{2 / \pi} \hspace{0.3mm} \mathrm{e}^{{-}r}$ in position space and $\Phi(\XRMM) = 2 \sqrt{2 \pi} / (1 + \XRMM^{2})^{\nicefrac{3}{2}}$ in momentum space.

\subsection{Preliminary definitions}

To begin we introduce the function $f(\XRM) \equiv [\Phi(\XRM)]^2$, as it occurs often in the upcoming expressions.
We also define $\int_{\vec{p}} \equiv \int \mathrm{d}^{2} p / (2 \pi)^{2}$ for the sake of brevity.
In position space, $f$ takes the form
\begin{align}
	f(\boldsymbol{r}) &= \int_{\XRM} f(\vec{k}) \hspace{0.3mm}\mathrm{e}^{\mathrm{i}\XRM\boldsymbol{\cdot}\boldsymbol{r}} \nonumber
	\\ \nonumber
	&= 4 \int^{\infty}_{0} \mathrm{d} \XRMM\, \frac{\XRMM J_{0}(\XRMM r)}{(1 + \XRMM^2)^{3}}
	\\
	&= \frac{1}{2} r^2 K_2(r),
\end{align}
which similarly only depends on the magnitude of $\vec{r}$.
Here, $J_n(x)$ and $K_n(x)$ are the $n$th order Bessel function of the first kind and modified Bessel function of the second kind, respectively. Furthermore, the Fourier transform of a two-variable function $g$ is defined as
\begin{align}
    g(\boldsymbol{r},\boldsymbol{r}') = \int_{\vec{p} \vec{p}'}
    \mathrm{e}^{\mathrm{i}\boldsymbol{p}\boldsymbol{\cdot}\boldsymbol{r}} g(\boldsymbol{p},\boldsymbol{p}') \hspace{0.3mm} \mathrm{e}^{-\mathrm{i}\boldsymbol{p}'\boldsymbol{\cdot}\boldsymbol{r}'}.
\end{align}
If $g(\boldsymbol{p},\boldsymbol{p}') = g(\boldsymbol{p}-\boldsymbol{p}')$, such that $g$ is a local function, the transformation into position space can be written as
\begin{align}
    g(\boldsymbol{r},\boldsymbol{r}') = \delta(\boldsymbol{r} - \boldsymbol{r}') \underbrace{
    \int_{\vec{p}} g(\boldsymbol{p}) \hspace{0.3mm} \mathrm{e}^{\mathrm{i}\boldsymbol{p}\boldsymbol{\cdot}\boldsymbol{r}}
    }_{\equiv \,g(\boldsymbol{r})}.
\end{align}

\subsection{Normalization factor}
The electron-exchange overlap integral for ground-state hydrogen wave functions is
\begin{equation}
    \begin{split}
        &\hspace{0pt-\widthof{$\displaystyle = \int_{\XRM} \Phi(\XRM) \Phi(\XRM + \dlmassc(\BXRM + \BXRM') - \dlmassv(\BXRM - \BXRM')) \,$}}
        \mathcal{K}^{\mathrm{c}}(\BXRM,\BXRM') \\
        = \int_{\XRM}
        \Phi(\XRM) \Phi(\XRM + \dlmassc(\BXRM + \BXRM') - \dlmassv(\BXRM - \BXRM'))& \\
        \times \, \Phi(\XRM - \dlmassv(\BXRM - \BXRM')) \Phi(\XRM + \dlmassc(\BXRM + \BXRM'))& ,
    \end{split}
\end{equation}
where $\Phi(\XRM)$ is real in this case.
In the heavy-hole limit we have $\dlmassc=0$ and $\dlmassv=1$, such that $\mathcal{K}^{\mathrm{c}}$ only depends on $\BXRM-\BXRM'$, making it a local function.
The position space expression of $\mathcal{K}^{\mathrm{c}}$ can be computed exactly by the convolution theorem as
\begin{align} \nonumber
    \mathcal{K}^{\mathrm{c}}(\boldsymbol{r}) &= \int_{\BXRM \vec{k}} f(\XRM)f(\XRM-\BXRM) 
    \mathrm{e}^{\mathrm{i}\BXRM \boldsymbol{\cdot} \boldsymbol{r}} 
    \\ \nonumber
    &= [f(\boldsymbol{r})]^2
    \\ \label{eqappx: Kc overlap integral}
    &= \frac{1}{4} r^4 [K_2(r)]^2 .
\end{align}
As it also depends only on the magnitude of $\vec{r}$, in what follows we write it as $\mathcal{K}^{\mathrm{c}}(r)$.
Also, $\mathcal{K}^{\mathrm{c}}(\boldsymbol{r},\boldsymbol{r}') = \mathcal{K}^{\mathrm{c}}(r)\delta(\boldsymbol{r}-\boldsymbol{r}')$.
With the above expression, the function $R_{S_{\mathrm{c}}}$ and its inverse read
\begin{subequations}
    \begin{align}
        \NormalizationFunctionVeff_{S_{\mathrm{c}}}(\boldsymbol{r},\boldsymbol{r}') &= [1 + (-1)^{S_{\mathrm{c}}} \mathcal{K}^{\mathrm{c}}(r)]\delta(\boldsymbol{r}-\boldsymbol{r}') , \\
        \NormalizationFunctionVeff^{-1}_{S_{\mathrm{c}}}(\boldsymbol{r},\boldsymbol{r}') &= [1 + (-1)^{S_{\mathrm{c}}} \mathcal{K}^{\mathrm{c}}(r)]^{-1} \delta(\boldsymbol{r}-\boldsymbol{r}').
    \end{align}
\end{subequations}
The latter effectively acts as a normalization factor in the effective potential.

\subsection{Direct exciton--exciton interaction}
Next, we consider the local direct interaction, which for general electron and hole masses takes the form
\begin{equation} \label{eq: hydrogen direct term}
    \mathcal{U}^0(\BXRMM) = V(\BXRMM)
    \bigg(
    \frac{1}{(1 + \dlmassc^{2} \BXRMM^{2} / 4)^{\nicefrac{3}{2}}}
    - \frac{1}{(1 + \dlmassv^{2} \BXRMM^{2} / 4)^{\nicefrac{3}{2}}}
    \bigg)^2 .
\end{equation} 
We have taken into account that it only depends on the magnitude of $\BXRM$.
Similarly, its position-space expression only depends on the magnitude of $\vec{r}$, and we write it as $\mathcal{U}^{0}(r)$.
In the heavy-hole limit this becomes
\begin{align}
    \mathcal{U}^{0}(r) &= \int^{\infty}_{0} \mathrm{d}\BXRMM \, \bigg(1 - \frac{1}{(1 + \BXRMM^{2} / 4)^{\nicefrac{3}{2}}} \bigg)^2 J_{0}(\BXRMM r) \nonumber
    \\
    &= \frac{1}{r} - 2\mathcal{I}_{3/2}(r) + \mathcal{I}_{2}(r),
\end{align}
where
\begin{equation}
    \mathcal{I}_{\lambda}(x) = \int_{0}^{\infty} \mathrm{d} u \, \frac{J_{0}(ux)}{(1 + u^{2} / 4)^{\lambda}} .
\end{equation}
In our case, the integrals of interest are
\begin{subequations}
\begin{align}
    \label{eqappx: meijerG1}
    \mathcal{I}_{3/2}(r) &= 
    \frac{2}{\sqrt{\pi}} G^{2,1}_{1,3}\bigg(
    \begin{matrix}
    \nicefrac{1}{2} 
    \\
    0, 1, 0
    \end{matrix}\,
    \bigg\vert\, 
    r^2
    \bigg),
    \\ \nonumber
    \mathcal{I}_{3}(r) &=\frac{4r}{5} - \frac{2r^3}{5} + \bigg( \frac{3\pi}{8} + \frac{\pi r^{2}}{2}\bigg) I_{0}(2r) - \pi r I_{1}(2r) \nonumber \\
    &+ \bigg(\frac{5\pi}{8} - \frac{\pi r^{2}}{2} - \frac{3\pi}{4 r^2}\bigg) \textbf{L}_{2}(2r) + \bigg(\frac{\pi r}{2} - \frac{3 \pi}{8 r}\bigg) \textbf{L}_{3}(2r) . \label{eqappx: struveL}
\end{align}
\end{subequations}
Here, $I_{n}$ is the $n$th order modified Bessel function of the first kind, $\textbf{L}_n(x)$ the $n$th order modified Struve function, and $G$ the Meijer G-function. 

\subsection{Electron-exchange interaction}
For general effective electron and hole masses, the electron-exchange interaction is nonlocal.
However, when $\dlmassc = 0$ and $\dlmassv =1$, the interaction becomes local and reads
\begin{align}
    \mathcal{U}^{\mathrm{c}}(\BXRM) = \mathcal{U}^{\mathrm{c}}_{\mathrm{cv}}(\BXRM) - \mathcal{U}^{\mathrm{c}}_{\mathrm{vv}}(\BXRM) - \mathcal{U}^{\mathrm{c}}_{\mathrm{cc}}(\BXRM) ,
\end{align}
with the three components defined as
\begin{subequations}
\begin{align} \label{eqappx: Uccv integral}
    \mathcal{U}^{\mathrm{c}}_{\mathrm{cv}}(\BXRM) &= 2 \int_{\XRM} (1 + \XRMM^2) f(\XRM) f(\XRM-\BXRM),
    \\ \label{eqappx: Ucvv integral}
    \mathcal{U}^{\mathrm{c}}_{\mathrm{vv}}(\BXRM) &= \int_{\XRM \XRM'}  f(\XRM) V(\XRM - \XRM') f(\XRM' - \BXRM),
    \\ \label{eqappx: Uccc integral}
    \mathcal{U}^{\mathrm{c}}_{\mathrm{cc}}(\BXRM) &= \int_{\XRM \XRM'} \Phi(\XRM) \Phi(\XRM-\BXRM) V(\XRM-\XRM') \Phi(\XRM')\Phi(\XRM'-\BXRM).
\end{align}
\end{subequations}
It is easy to see that the three terms depend only on the magnitude of $\vec{q}$ by observing that all involved functions depend only on the magnitude of their argument, invoking the law of cosines, and subsequently shifting both polar angles $\phi_{\vec{k}}$ and $\phi_{\vec{k}'}$ by $\phi_{\vec{q}}$, whence the dependence on the latter completely drops out.
In position space, this implies that the entire effective potential depends only on the magnitude of $\vec{r}$. 
The position-space expressions of the electron--hole and the hole--hole scattering terms have the analytical form
\begin{align} \nonumber
    \mathcal{U}^{\mathrm{c}}_{\mathrm{cv}}(r) &= 2 \int_{\BXRM \XRM} (1 + \XRMM^2) f(\XRM) f(\XRM-\BXRM) \mathrm{e}^{\mathrm{i}\BXRM \boldsymbol{\cdot} \boldsymbol{r}}
    \\ \nonumber
    &= 2 [f(\boldsymbol{r})]^2 - 2 f(\boldsymbol{r}) \nabla^2f(\boldsymbol{r})
    \\
    &= 2 r^3 K_1(r) K_2(r), \\ \nonumber
    \mathcal{U}^{\mathrm{c}}_{\mathrm{vv}}(r) &= \int_{\BXRM \XRM \XRM'} f(\XRM)V(\XRM - \XRM') 
    f(\XRM' - \BXRM) \mathrm{e}^{\mathrm{i}\BXRM \boldsymbol{\cdot}\boldsymbol{r}}
    \\ \nonumber
    &= V(r)[f(\boldsymbol{r})]^2
    \\ 
    &= \frac{1}{4} r^{3} [K_2(r)]^2.
\end{align}
The electron--electron component of the interaction is the only one which cannot be computed analytically.
Its position-space expression is
\begin{equation}
\label{eqappx: integral}
    \mathcal{U}^{\mathrm{c}}_{\mathrm{cc}}(\boldsymbol{r}) = \int_{\vec{x} \vec{x}'} \Phi(\boldsymbol{x})\Phi(\boldsymbol{x}-\boldsymbol{r}) V(\boldsymbol{x}-\boldsymbol{x}') \Phi(\boldsymbol{x}')\Phi(\boldsymbol{x}'-\boldsymbol{r}) ,
\end{equation}
where $\int_{\vec{x}} \equiv \int \mathrm{d}^{2} x$.
This is a four-dimensional integral that has to be solved numerically.
Sec.\ I of the Supplemental Material \cite{supp} details our numerical procedure (which uses the function of Ref.\ \cite{Young_2024}), and the resulting potential is plotted in Fig.\ \ref{fig: Heitler--London potential}.

    \section{Field-theoretical notation}
    \label{app: ft notation}
    

In this appendix we summarize the notation used throughout the field-theoretical derivations of Sec.\ \ref{sec: Field Theoretic Approach}.

\subsection{Functional inner-product notation}
\label{app: inner product notation}

Following Ref.\ \cite{Stoof_2008}, we define the following shorthand notations for the functional inner products of vectors and matrices:
\begin{subequations} \label{eq: inner product notation}
\begin{align} \label{eq: inner product}
    (A|B) &= \sum_{i} A^*(i)  B(i) ,
    \\ \label{eq: inner product single line}
    (A|M|B) &= \sum_{ii'} A^*(i) M(i,i') B(i') ,
    \\ \label{eq: inner product double line}
    (A \lVert M \rVert B) &= 
    \\ \nonumber
    \sum_{\substack{i,\dots,i' \\ j,\dots, j'}} &A^*(i,\dots,i') M(i,\dots,i';j,\dots,j') B(j,\dots,j') ,
\end{align}
where $A$ and $B$ represent fields and $M$ is some matrix with the appropriate number of variables.
The dummy variables $i, j, \dots$ represent the combined spin, position, and imaginary-time variables; in the latter two cases the sum is understood as an integral.
If $A$ and $B$ are products of fields, then the single vertical line notation of Eq.\ \eqref{eq: inner product single line} implies that all fields are evaluated for the same variables, as in the second term on the second line of Eq.\ \eqref{eq: action electrons suggestive}.
Meanwhile, inner products with the double vertical line of Eq.\ \eqref{eq: inner product double line} imply that all fields are evaluated at different variables, as in the final term in Eq.\ \eqref{eq: action electrons suggestive}.
Furthermore, we also define the functional trace and multiplication as
\begin{align}
	\label{eq: functional trace}
	\Tr M &= \sum_{j} M(j,j),
	\\
	\label{eq: functional product}
	[M\cdot M'](i,i') &= \sum_{j} M(i,j)M'(j,i'),
	\\ \label{eq: functional field product}
	[A\cdot M] (i) &= \sum_{j} A(j)M(j,i).
\end{align}
\end{subequations}
In the shorthand notation, we leave the band indices of the fields visible explicitly because it will be necessary to distinguish between both types of electrons.

\subsection{Elements of the polarization-field action}
\label{app: polarization field action notation}

Here we give the explicit expressions of various elements present in the polarization-field action of Eq.\ \eqref{eq: action exact polarization}.
Firstly, $\boldsymbol{\eta}$ is a real vector quantity in the band and spin spaces, defined as
\begin{subequations}
\begin{align} \label{eq: Hartree self-energy generalization simplification}
	\boldsymbol{\eta} &= \sum^\infty_{n=1} \boldsymbol{\eta}^{(n)}, 
	\\ \label{eq: Hartree self-energy generalized}
	\eta^{(n)}_{a;  \sigma}(x) &= 
	[(\mathbf{G}^{0} \cdot \boldsymbol{\Sigma}^{\mathcal{P}})^{2n} \cdot \mathbf{G}^{0}]_{a a; \sigma \sigma}(x, x^{+}) .
\end{align}
\end{subequations}
In this equation and below, the functional multiplication of boldface symbols is understood to also include a matrix product over the band space introduced in Eqs.\ \eqref{eq: Green's function non-interacting matrix}--\eqref{eq: selfenergy polarization matrix}.
The coordinate $x^{+}$ indicates that the corresponding time argument of the Green's function is evaluated at $\tau^{+} \equiv \tau + 0^{+}$, which is needed to ensure the correct time ordering.
Furthermore, the matrix quantity $\boldsymbol{\pi}$ is defined as
\begin{subequations}
	\begin{align} \label{eq: bubble diagram generalization simplification}
		\boldsymbol{\pi} &= \sum^{\infty}_{n=0}\sum^{2n}_{i=0} \boldsymbol{\pi}^{(n,i)} , \\
		\begin{split}
			\pi^{(n,i)}_{a b; \sigma \sigma'}(x, x') &=
			[(\mathbf{G}^{0} \cdot \boldsymbol{\Sigma}^{\mathcal{P}})^{i} \cdot \mathbf{G}^{0}]_{a b; \sigma \sigma'}(x, x')
			\\ \label{eq: bubble diagram generalized}
			&\times
			[(\mathbf{G}^{0} \cdot \boldsymbol{\Sigma}^{\mathcal{P}})^{2n-i} \cdot \mathbf{G}^{0}]_{ba; \sigma' \sigma}(x' ,x) ,
		\end{split}
	\end{align}
\end{subequations}
which possesses the following symmetry:
\begin{equation}\label{eq: Pi symmetry}
	\pi^{(n,i)}_{a b; \sigma \sigma'}(x, x') = \pi^{(n, 2n-i)}_{b a; \sigma' \sigma}(x', x) .
\end{equation}
The object $\eta^{(n)}_{a}$ may be interpreted as a correction to the Hartree self-energy arising from the polarization field.
Furthermore, the lowest-order matrix component $\boldsymbol{\pi}^{(0,0)}$ corresponds to the so-called bubble diagram \cite{Stoof_2008} and is present for both species of electron.
The $\mathcal{P}$ field is present in the higher orders of $\boldsymbol{\pi}^{(n,i)}$, thus one can interpret $\boldsymbol{\pi}$ as containing corrections to the bubble diagram by the polarization field.
The $\mathcal{P}$-dependent, inverse free propagator for the density fluctuations reads 
\begin{equation}
	\begin{split}
		\mathbf{G}^{-1}_{0,\rho'; \sigma \sigma'}(x, x') &= V(x - x') \, \mathbf{I} \, \delta_{\sigma \sigma'}
		\\ 
		- \int_{zz'}& \, V(x - z)
		\vec{\pi}_{\sigma \sigma'}(z, z') V(z' - x') ,
	\end{split}
\end{equation}
with $\mathbf{I}$ the $2 \times 2$ identity matrix in the band space.
Since both $V$ and $\boldsymbol{\pi}$ are symmetric, $\mathbf{G}^{0, \rho'}$ is symmetric also.
\end{appendix}

\twocolumngrid
\bibliography{lib.bib}

\begin{thebibliography}{125}%
\makeatletter
\providecommand \@ifxundefined [1]{%
 \@ifx{#1\undefined}
}%
\providecommand \@ifnum [1]{%
 \ifnum #1\expandafter \@firstoftwo
 \else \expandafter \@secondoftwo
 \fi
}%
\providecommand \@ifx [1]{%
 \ifx #1\expandafter \@firstoftwo
 \else \expandafter \@secondoftwo
 \fi
}%
\providecommand \natexlab [1]{#1}%
\providecommand \enquote  [1]{``#1''}%
\providecommand \bibnamefont  [1]{#1}%
\providecommand \bibfnamefont [1]{#1}%
\providecommand \citenamefont [1]{#1}%
\providecommand \href@noop [0]{\@secondoftwo}%
\providecommand \href [0]{\begingroup \@sanitize@url \@href}%
\providecommand \@href[1]{\@@startlink{#1}\@@href}%
\providecommand \@@href[1]{\endgroup#1\@@endlink}%
\providecommand \@sanitize@url [0]{\catcode `\\12\catcode `\$12\catcode
  `\&12\catcode `\#12\catcode `\^12\catcode `\_12\catcode `\%12\relax}%
\providecommand \@@startlink[1]{}%
\providecommand \@@endlink[0]{}%
\providecommand \url  [0]{\begingroup\@sanitize@url \@url }%
\providecommand \@url [1]{\endgroup\@href {#1}{\urlprefix }}%
\providecommand \urlprefix  [0]{URL }%
\providecommand \Eprint [0]{\href }%
\providecommand \doibase [0]{https://doi.org/}%
\providecommand \selectlanguage [0]{\@gobble}%
\providecommand \bibinfo  [0]{\@secondoftwo}%
\providecommand \bibfield  [0]{\@secondoftwo}%
\providecommand \translation [1]{[#1]}%
\providecommand \BibitemOpen [0]{}%
\providecommand \bibitemStop [0]{}%
\providecommand \bibitemNoStop [0]{.\EOS\space}%
\providecommand \EOS [0]{\spacefactor3000\relax}%
\providecommand \BibitemShut  [1]{\csname bibitem#1\endcsname}%
\let\auto@bib@innerbib\@empty
\bibitem [{\citenamefont {Frenkel}(1931)}]{Frenkel_1931}%
  \BibitemOpen
  \bibfield  {author} {\bibinfo {author} {\bibfnamefont {J.}~\bibnamefont
  {Frenkel}},\ }\bibfield  {title} {\bibinfo {title} {On the transformation of
  light into heat in solids. {I}},\ }\href
  {https://doi.org/10.1103/PhysRev.37.17} {\bibfield  {journal} {\bibinfo
  {journal} {Physical Review}\ }\textbf {\bibinfo {volume} {37}},\ \bibinfo
  {pages} {17} (\bibinfo {year} {1931})}\BibitemShut {NoStop}%
\bibitem [{\citenamefont {Wannier}(1937)}]{Wannier_1937}%
  \BibitemOpen
  \bibfield  {author} {\bibinfo {author} {\bibfnamefont {G.~H.}\ \bibnamefont
  {Wannier}},\ }\bibfield  {title} {\bibinfo {title} {The structure of
  electronic excitation levels in insulating crystals},\ }\href
  {https://doi.org/10.1103/PhysRev.52.191} {\bibfield  {journal} {\bibinfo
  {journal} {Physical Review}\ }\textbf {\bibinfo {volume} {52}},\ \bibinfo
  {pages} {191} (\bibinfo {year} {1937})}\BibitemShut {NoStop}%
\bibitem [{\citenamefont {Mak}\ and\ \citenamefont
  {Shan}(2016)}]{mak2016photonics}%
  \BibitemOpen
  \bibfield  {author} {\bibinfo {author} {\bibfnamefont {K.~F.}\ \bibnamefont
  {Mak}}\ and\ \bibinfo {author} {\bibfnamefont {J.}~\bibnamefont {Shan}},\
  }\bibfield  {title} {\bibinfo {title} {Photonics and optoelectronics of 2{D}
  semiconductor transition metal dichalcogenides},\ }\href
  {https://doi.org/10.1038/nphoton.2015.282} {\bibfield  {journal} {\bibinfo
  {journal} {Nature Photonics}\ }\textbf {\bibinfo {volume} {10}},\ \bibinfo
  {pages} {216} (\bibinfo {year} {2016})}\BibitemShut {NoStop}%
\bibitem [{\citenamefont {Wang}\ \emph {et~al.}(2018)\citenamefont {Wang},
  \citenamefont {Chernikov}, \citenamefont {Glazov}, \citenamefont {Heinz},
  \citenamefont {Marie}, \citenamefont {Amand},\ and\ \citenamefont
  {Urbaszek}}]{wang2018colloquium}%
  \BibitemOpen
  \bibfield  {author} {\bibinfo {author} {\bibfnamefont {G.}~\bibnamefont
  {Wang}}, \bibinfo {author} {\bibfnamefont {A.}~\bibnamefont {Chernikov}},
  \bibinfo {author} {\bibfnamefont {M.~M.}\ \bibnamefont {Glazov}}, \bibinfo
  {author} {\bibfnamefont {T.~F.}\ \bibnamefont {Heinz}}, \bibinfo {author}
  {\bibfnamefont {X.}~\bibnamefont {Marie}}, \bibinfo {author} {\bibfnamefont
  {T.}~\bibnamefont {Amand}},\ and\ \bibinfo {author} {\bibfnamefont
  {B.}~\bibnamefont {Urbaszek}},\ }\bibfield  {title} {\bibinfo {title}
  {Colloquium: Excitons in atomically thin transition metal dichalcogenides},\
  }\href {https://doi.org/10.1103/RevModPhys.90.021001} {\bibfield  {journal}
  {\bibinfo  {journal} {Reviews of Modern Physics}\ }\textbf {\bibinfo {volume}
  {90}},\ \bibinfo {pages} {021001} (\bibinfo {year} {2018})}\BibitemShut
  {NoStop}%
\bibitem [{\citenamefont {Zheng}\ \emph {et~al.}(2018)\citenamefont {Zheng},
  \citenamefont {Jiang}, \citenamefont {Hu}, \citenamefont {Li}, \citenamefont
  {Zeng}, \citenamefont {Wang},\ and\ \citenamefont {Pan}}]{zheng2018light}%
  \BibitemOpen
  \bibfield  {author} {\bibinfo {author} {\bibfnamefont {W.}~\bibnamefont
  {Zheng}}, \bibinfo {author} {\bibfnamefont {Y.}~\bibnamefont {Jiang}},
  \bibinfo {author} {\bibfnamefont {X.}~\bibnamefont {Hu}}, \bibinfo {author}
  {\bibfnamefont {H.}~\bibnamefont {Li}}, \bibinfo {author} {\bibfnamefont
  {Z.}~\bibnamefont {Zeng}}, \bibinfo {author} {\bibfnamefont {X.}~\bibnamefont
  {Wang}},\ and\ \bibinfo {author} {\bibfnamefont {A.}~\bibnamefont {Pan}},\
  }\bibfield  {title} {\bibinfo {title} {Light emission properties of 2{D}
  transition metal dichalcogenides: fundamentals and applications},\ }\href
  {https://doi.org/10.1002/adom.201800420} {\bibfield  {journal} {\bibinfo
  {journal} {Advanced Optical Materials}\ }\textbf {\bibinfo {volume} {6}},\
  \bibinfo {pages} {1800420} (\bibinfo {year} {2018})}\BibitemShut {NoStop}%
\bibitem [{\citenamefont {Ugeda}\ \emph {et~al.}(2014)\citenamefont {Ugeda},
  \citenamefont {Bradley}, \citenamefont {Shi}, \citenamefont {Da~Jornada},
  \citenamefont {Zhang}, \citenamefont {Qiu}, \citenamefont {Ruan},
  \citenamefont {Mo}, \citenamefont {Hussain}, \citenamefont {Shen} \emph
  {et~al.}}]{ugeda2014giant}%
  \BibitemOpen
  \bibfield  {author} {\bibinfo {author} {\bibfnamefont {M.~M.}\ \bibnamefont
  {Ugeda}}, \bibinfo {author} {\bibfnamefont {A.~J.}\ \bibnamefont {Bradley}},
  \bibinfo {author} {\bibfnamefont {S.-F.}\ \bibnamefont {Shi}}, \bibinfo
  {author} {\bibfnamefont {F.~H.}\ \bibnamefont {Da~Jornada}}, \bibinfo
  {author} {\bibfnamefont {Y.}~\bibnamefont {Zhang}}, \bibinfo {author}
  {\bibfnamefont {D.~Y.}\ \bibnamefont {Qiu}}, \bibinfo {author} {\bibfnamefont
  {W.}~\bibnamefont {Ruan}}, \bibinfo {author} {\bibfnamefont {S.-K.}\
  \bibnamefont {Mo}}, \bibinfo {author} {\bibfnamefont {Z.}~\bibnamefont
  {Hussain}}, \bibinfo {author} {\bibfnamefont {Z.-X.}\ \bibnamefont {Shen}},
  \emph {et~al.},\ }\bibfield  {title} {\bibinfo {title} {Giant bandgap
  renormalization and excitonic effects in a monolayer transition metal
  dichalcogenide semiconductor},\ }\href {https://doi.org/10.1038/nmat4061}
  {\bibfield  {journal} {\bibinfo  {journal} {Nature Materials}\ }\textbf
  {\bibinfo {volume} {13}},\ \bibinfo {pages} {1091} (\bibinfo {year}
  {2014})}\BibitemShut {NoStop}%
\bibitem [{\citenamefont {Hanbicki}\ \emph {et~al.}(2015)\citenamefont
  {Hanbicki}, \citenamefont {Currie}, \citenamefont {Kioseoglou}, \citenamefont
  {Friedman},\ and\ \citenamefont {Jonker}}]{hanbicki2015measurement}%
  \BibitemOpen
  \bibfield  {author} {\bibinfo {author} {\bibfnamefont {A.}~\bibnamefont
  {Hanbicki}}, \bibinfo {author} {\bibfnamefont {M.}~\bibnamefont {Currie}},
  \bibinfo {author} {\bibfnamefont {G.}~\bibnamefont {Kioseoglou}}, \bibinfo
  {author} {\bibfnamefont {A.}~\bibnamefont {Friedman}},\ and\ \bibinfo
  {author} {\bibfnamefont {B.}~\bibnamefont {Jonker}},\ }\bibfield  {title}
  {\bibinfo {title} {Measurement of high exciton binding energy in the
  monolayer transition-metal dichalcogenides \ce{WS2} and \ce{WSe2}},\ }\href
  {https://doi.org/10.1016/j.ssc.2014.11.005} {\bibfield  {journal} {\bibinfo
  {journal} {Solid State Communications}\ }\textbf {\bibinfo {volume} {203}},\
  \bibinfo {pages} {16} (\bibinfo {year} {2015})}\BibitemShut {NoStop}%
\bibitem [{\citenamefont {Zhao}\ \emph {et~al.}(2015)\citenamefont {Zhao},
  \citenamefont {Ribeiro},\ and\ \citenamefont {Eda}}]{zhao2015electronic}%
  \BibitemOpen
  \bibfield  {author} {\bibinfo {author} {\bibfnamefont {W.}~\bibnamefont
  {Zhao}}, \bibinfo {author} {\bibfnamefont {R.~M.}\ \bibnamefont {Ribeiro}},\
  and\ \bibinfo {author} {\bibfnamefont {G.}~\bibnamefont {Eda}},\ }\bibfield
  {title} {\bibinfo {title} {Electronic structure and optical signatures of
  semiconducting transition metal dichalcogenide nanosheets},\ }\href
  {https://doi.org/10.1021/ar500303m} {\bibfield  {journal} {\bibinfo
  {journal} {Accounts of Chemical Research}\ }\textbf {\bibinfo {volume}
  {48}},\ \bibinfo {pages} {91} (\bibinfo {year} {2015})}\BibitemShut {NoStop}%
\bibitem [{\citenamefont {Shang}\ \emph {et~al.}(2015)\citenamefont {Shang},
  \citenamefont {Shen}, \citenamefont {Cong}, \citenamefont {Peimyoo},
  \citenamefont {Cao}, \citenamefont {Eginligil},\ and\ \citenamefont
  {Yu}}]{shang2015observation}%
  \BibitemOpen
  \bibfield  {author} {\bibinfo {author} {\bibfnamefont {J.}~\bibnamefont
  {Shang}}, \bibinfo {author} {\bibfnamefont {X.}~\bibnamefont {Shen}},
  \bibinfo {author} {\bibfnamefont {C.}~\bibnamefont {Cong}}, \bibinfo {author}
  {\bibfnamefont {N.}~\bibnamefont {Peimyoo}}, \bibinfo {author} {\bibfnamefont
  {B.}~\bibnamefont {Cao}}, \bibinfo {author} {\bibfnamefont {M.}~\bibnamefont
  {Eginligil}},\ and\ \bibinfo {author} {\bibfnamefont {T.}~\bibnamefont
  {Yu}},\ }\bibfield  {title} {\bibinfo {title} {Observation of excitonic fine
  structure in a 2{D} transition-metal dichalcogenide semiconductor},\ }\href
  {https://doi.org/10.1021/nn5059908} {\bibfield  {journal} {\bibinfo
  {journal} {ACS Nano}\ }\textbf {\bibinfo {volume} {9}},\ \bibinfo {pages}
  {647} (\bibinfo {year} {2015})}\BibitemShut {NoStop}%
\bibitem [{\citenamefont {Palummo}\ \emph {et~al.}(2015)\citenamefont
  {Palummo}, \citenamefont {Bernardi},\ and\ \citenamefont
  {Grossman}}]{palummo2015exciton}%
  \BibitemOpen
  \bibfield  {author} {\bibinfo {author} {\bibfnamefont {M.}~\bibnamefont
  {Palummo}}, \bibinfo {author} {\bibfnamefont {M.}~\bibnamefont {Bernardi}},\
  and\ \bibinfo {author} {\bibfnamefont {J.~C.}\ \bibnamefont {Grossman}},\
  }\bibfield  {title} {\bibinfo {title} {Exciton radiative lifetimes in
  two-dimensional transition metal dichalcogenides},\ }\href
  {https://doi.org/10.1021/nl503799t} {\bibfield  {journal} {\bibinfo
  {journal} {Nano Letters}\ }\textbf {\bibinfo {volume} {15}},\ \bibinfo
  {pages} {2794} (\bibinfo {year} {2015})}\BibitemShut {NoStop}%
\bibitem [{\citenamefont {Moody}\ \emph {et~al.}(2016)\citenamefont {Moody},
  \citenamefont {Schaibley},\ and\ \citenamefont {Xu}}]{moody2016exciton}%
  \BibitemOpen
  \bibfield  {author} {\bibinfo {author} {\bibfnamefont {G.}~\bibnamefont
  {Moody}}, \bibinfo {author} {\bibfnamefont {J.}~\bibnamefont {Schaibley}},\
  and\ \bibinfo {author} {\bibfnamefont {X.}~\bibnamefont {Xu}},\ }\bibfield
  {title} {\bibinfo {title} {Exciton dynamics in monolayer transition metal
  dichalcogenides},\ }\href {https://doi.org/10.1364/JOSAB.33.000C39}
  {\bibfield  {journal} {\bibinfo  {journal} {Journal of the Optical Society of
  America B}\ }\textbf {\bibinfo {volume} {33}},\ \bibinfo {pages} {C39}
  (\bibinfo {year} {2016})}\BibitemShut {NoStop}%
\bibitem [{\citenamefont {Ceballos}\ \emph {et~al.}(2016)\citenamefont
  {Ceballos}, \citenamefont {Cui}, \citenamefont {Bellus},\ and\ \citenamefont
  {Zhao}}]{ceballos2016exciton}%
  \BibitemOpen
  \bibfield  {author} {\bibinfo {author} {\bibfnamefont {F.}~\bibnamefont
  {Ceballos}}, \bibinfo {author} {\bibfnamefont {Q.}~\bibnamefont {Cui}},
  \bibinfo {author} {\bibfnamefont {M.~Z.}\ \bibnamefont {Bellus}},\ and\
  \bibinfo {author} {\bibfnamefont {H.}~\bibnamefont {Zhao}},\ }\bibfield
  {title} {\bibinfo {title} {Exciton formation in monolayer transition metal
  dichalcogenides},\ }\href {https://doi.org/10.1039/C6NR02516A} {\bibfield
  {journal} {\bibinfo  {journal} {Nanoscale}\ }\textbf {\bibinfo {volume}
  {8}},\ \bibinfo {pages} {11681} (\bibinfo {year} {2016})}\BibitemShut
  {NoStop}%
\bibitem [{\citenamefont {Robert}\ \emph {et~al.}(2016)\citenamefont {Robert},
  \citenamefont {Lagarde}, \citenamefont {Cadiz}, \citenamefont {Wang},
  \citenamefont {Lassagne}, \citenamefont {Amand}, \citenamefont {Balocchi},
  \citenamefont {Renucci}, \citenamefont {Tongay}, \citenamefont {Urbaszek}
  \emph {et~al.}}]{robert2016exciton}%
  \BibitemOpen
  \bibfield  {author} {\bibinfo {author} {\bibfnamefont {C.}~\bibnamefont
  {Robert}}, \bibinfo {author} {\bibfnamefont {D.}~\bibnamefont {Lagarde}},
  \bibinfo {author} {\bibfnamefont {F.}~\bibnamefont {Cadiz}}, \bibinfo
  {author} {\bibfnamefont {G.}~\bibnamefont {Wang}}, \bibinfo {author}
  {\bibfnamefont {B.}~\bibnamefont {Lassagne}}, \bibinfo {author}
  {\bibfnamefont {T.}~\bibnamefont {Amand}}, \bibinfo {author} {\bibfnamefont
  {A.}~\bibnamefont {Balocchi}}, \bibinfo {author} {\bibfnamefont
  {P.}~\bibnamefont {Renucci}}, \bibinfo {author} {\bibfnamefont
  {S.}~\bibnamefont {Tongay}}, \bibinfo {author} {\bibfnamefont
  {B.}~\bibnamefont {Urbaszek}}, \emph {et~al.},\ }\bibfield  {title} {\bibinfo
  {title} {Exciton radiative lifetime in transition metal dichalcogenide
  monolayers},\ }\href {https://doi.org/10.1103/PhysRevB.93.205423} {\bibfield
  {journal} {\bibinfo  {journal} {Physical Review B}\ }\textbf {\bibinfo
  {volume} {93}},\ \bibinfo {pages} {205423} (\bibinfo {year}
  {2016})}\BibitemShut {NoStop}%
\bibitem [{\citenamefont {Robert}\ \emph {et~al.}(2017)\citenamefont {Robert},
  \citenamefont {Amand}, \citenamefont {Cadiz}, \citenamefont {Lagarde},
  \citenamefont {Courtade}, \citenamefont {Manca}, \citenamefont {Taniguchi},
  \citenamefont {Watanabe}, \citenamefont {Urbaszek},\ and\ \citenamefont
  {Marie}}]{robert2017fine}%
  \BibitemOpen
  \bibfield  {author} {\bibinfo {author} {\bibfnamefont {C.}~\bibnamefont
  {Robert}}, \bibinfo {author} {\bibfnamefont {T.}~\bibnamefont {Amand}},
  \bibinfo {author} {\bibfnamefont {F.}~\bibnamefont {Cadiz}}, \bibinfo
  {author} {\bibfnamefont {D.}~\bibnamefont {Lagarde}}, \bibinfo {author}
  {\bibfnamefont {E.}~\bibnamefont {Courtade}}, \bibinfo {author}
  {\bibfnamefont {M.}~\bibnamefont {Manca}}, \bibinfo {author} {\bibfnamefont
  {T.}~\bibnamefont {Taniguchi}}, \bibinfo {author} {\bibfnamefont
  {K.}~\bibnamefont {Watanabe}}, \bibinfo {author} {\bibfnamefont
  {B.}~\bibnamefont {Urbaszek}},\ and\ \bibinfo {author} {\bibfnamefont
  {X.}~\bibnamefont {Marie}},\ }\bibfield  {title} {\bibinfo {title} {Fine
  structure and lifetime of dark excitons in transition metal dichalcogenide
  monolayers},\ }\href {https://doi.org/10.1103/PhysRevB.96.155423} {\bibfield
  {journal} {\bibinfo  {journal} {Physical Review B}\ }\textbf {\bibinfo
  {volume} {96}},\ \bibinfo {pages} {155423} (\bibinfo {year}
  {2017})}\BibitemShut {NoStop}%
\bibitem [{\citenamefont {Krasnok}\ \emph {et~al.}(2018)\citenamefont
  {Krasnok}, \citenamefont {Lepeshov},\ and\ \citenamefont
  {Al{\'u}}}]{krasnok2018nanophotonics}%
  \BibitemOpen
  \bibfield  {author} {\bibinfo {author} {\bibfnamefont {A.}~\bibnamefont
  {Krasnok}}, \bibinfo {author} {\bibfnamefont {S.}~\bibnamefont {Lepeshov}},\
  and\ \bibinfo {author} {\bibfnamefont {A.}~\bibnamefont {Al{\'u}}},\
  }\bibfield  {title} {\bibinfo {title} {Nanophotonics with 2{D} transition
  metal dichalcogenides},\ }\href {https://doi.org/10.1364/OE.26.015972}
  {\bibfield  {journal} {\bibinfo  {journal} {Optics Express}\ }\textbf
  {\bibinfo {volume} {26}},\ \bibinfo {pages} {15972} (\bibinfo {year}
  {2018})}\BibitemShut {NoStop}%
\bibitem [{\citenamefont {Guo}\ \emph {et~al.}(2019)\citenamefont {Guo},
  \citenamefont {Wu}, \citenamefont {Cao}, \citenamefont {Monahan},
  \citenamefont {Lee}, \citenamefont {Louie},\ and\ \citenamefont
  {Fleming}}]{guo2019exchange}%
  \BibitemOpen
  \bibfield  {author} {\bibinfo {author} {\bibfnamefont {L.}~\bibnamefont
  {Guo}}, \bibinfo {author} {\bibfnamefont {M.}~\bibnamefont {Wu}}, \bibinfo
  {author} {\bibfnamefont {T.}~\bibnamefont {Cao}}, \bibinfo {author}
  {\bibfnamefont {D.~M.}\ \bibnamefont {Monahan}}, \bibinfo {author}
  {\bibfnamefont {Y.-H.}\ \bibnamefont {Lee}}, \bibinfo {author} {\bibfnamefont
  {S.~G.}\ \bibnamefont {Louie}},\ and\ \bibinfo {author} {\bibfnamefont
  {G.~R.}\ \bibnamefont {Fleming}},\ }\bibfield  {title} {\bibinfo {title}
  {Exchange-driven intravalley mixing of excitons in monolayer transition metal
  dichalcogenides},\ }\href {https://doi.org/10.1038/s41567-018-0362-y}
  {\bibfield  {journal} {\bibinfo  {journal} {Nature Physics}\ }\textbf
  {\bibinfo {volume} {15}},\ \bibinfo {pages} {228} (\bibinfo {year}
  {2019})}\BibitemShut {NoStop}%
\bibitem [{\citenamefont {Jiang}\ \emph {et~al.}(2021)\citenamefont {Jiang},
  \citenamefont {Chen}, \citenamefont {Zheng}, \citenamefont {Zheng},\ and\
  \citenamefont {Pan}}]{Jiang_2021}%
  \BibitemOpen
  \bibfield  {author} {\bibinfo {author} {\bibfnamefont {Y.}~\bibnamefont
  {Jiang}}, \bibinfo {author} {\bibfnamefont {S.}~\bibnamefont {Chen}},
  \bibinfo {author} {\bibfnamefont {W.}~\bibnamefont {Zheng}}, \bibinfo
  {author} {\bibfnamefont {B.}~\bibnamefont {Zheng}},\ and\ \bibinfo {author}
  {\bibfnamefont {A.}~\bibnamefont {Pan}},\ }\bibfield  {title} {\bibinfo
  {title} {Interlayer exciton formation, relaxation, and transport in {TMD} van
  der {W}aals heterostructures},\ }\href
  {https://doi.org/10.1038/s41377-021-00500-1} {\bibfield  {journal} {\bibinfo
  {journal} {Light: Science {\&} Applications}\ }\textbf {\bibinfo {volume}
  {10}},\ \bibinfo {pages} {72} (\bibinfo {year} {2021})}\BibitemShut {NoStop}%
\bibitem [{\citenamefont {Rivera}\ \emph {et~al.}(2018)\citenamefont {Rivera},
  \citenamefont {Yu}, \citenamefont {Seyler}, \citenamefont {Wilson},
  \citenamefont {Yao},\ and\ \citenamefont {Xu}}]{rivera2018interlayer}%
  \BibitemOpen
  \bibfield  {author} {\bibinfo {author} {\bibfnamefont {P.}~\bibnamefont
  {Rivera}}, \bibinfo {author} {\bibfnamefont {H.}~\bibnamefont {Yu}}, \bibinfo
  {author} {\bibfnamefont {K.~L.}\ \bibnamefont {Seyler}}, \bibinfo {author}
  {\bibfnamefont {N.~P.}\ \bibnamefont {Wilson}}, \bibinfo {author}
  {\bibfnamefont {W.}~\bibnamefont {Yao}},\ and\ \bibinfo {author}
  {\bibfnamefont {X.}~\bibnamefont {Xu}},\ }\bibfield  {title} {\bibinfo
  {title} {Interlayer valley excitons in heterobilayers of transition metal
  dichalcogenides},\ }\href {https://doi.org/10.1038/s41565-018-0193-0}
  {\bibfield  {journal} {\bibinfo  {journal} {Nature Nanotechnology}\ }\textbf
  {\bibinfo {volume} {13}},\ \bibinfo {pages} {1004} (\bibinfo {year}
  {2018})}\BibitemShut {NoStop}%
\bibitem [{\citenamefont {Miller}\ \emph {et~al.}(2017)\citenamefont {Miller},
  \citenamefont {Steinhoff}, \citenamefont {Pano}, \citenamefont {Klein},
  \citenamefont {Jahnke}, \citenamefont {Holleitner},\ and\ \citenamefont
  {Wurstbauer}}]{miller2017long}%
  \BibitemOpen
  \bibfield  {author} {\bibinfo {author} {\bibfnamefont {B.}~\bibnamefont
  {Miller}}, \bibinfo {author} {\bibfnamefont {A.}~\bibnamefont {Steinhoff}},
  \bibinfo {author} {\bibfnamefont {B.}~\bibnamefont {Pano}}, \bibinfo {author}
  {\bibfnamefont {J.}~\bibnamefont {Klein}}, \bibinfo {author} {\bibfnamefont
  {F.}~\bibnamefont {Jahnke}}, \bibinfo {author} {\bibfnamefont
  {A.}~\bibnamefont {Holleitner}},\ and\ \bibinfo {author} {\bibfnamefont
  {U.}~\bibnamefont {Wurstbauer}},\ }\bibfield  {title} {\bibinfo {title}
  {Long-lived direct and indirect interlayer excitons in van der waals
  heterostructures},\ }\href {https://doi.org/10.1021/acs.nanolett.7b01304}
  {\bibfield  {journal} {\bibinfo  {journal} {Nano Letters}\ }\textbf {\bibinfo
  {volume} {17}},\ \bibinfo {pages} {5229} (\bibinfo {year}
  {2017})}\BibitemShut {NoStop}%
\bibitem [{\citenamefont {Jin}\ \emph {et~al.}(2018)\citenamefont {Jin},
  \citenamefont {Ma}, \citenamefont {Karni}, \citenamefont {Regan},
  \citenamefont {Wang},\ and\ \citenamefont {Heinz}}]{jin2018ultrafast}%
  \BibitemOpen
  \bibfield  {author} {\bibinfo {author} {\bibfnamefont {C.}~\bibnamefont
  {Jin}}, \bibinfo {author} {\bibfnamefont {E.~Y.}\ \bibnamefont {Ma}},
  \bibinfo {author} {\bibfnamefont {O.}~\bibnamefont {Karni}}, \bibinfo
  {author} {\bibfnamefont {E.~C.}\ \bibnamefont {Regan}}, \bibinfo {author}
  {\bibfnamefont {F.}~\bibnamefont {Wang}},\ and\ \bibinfo {author}
  {\bibfnamefont {T.~F.}\ \bibnamefont {Heinz}},\ }\bibfield  {title} {\bibinfo
  {title} {Ultrafast dynamics in van der {W}aals heterostructures},\ }\href
  {https://doi.org/10.1038/s41565-018-0298-5} {\bibfield  {journal} {\bibinfo
  {journal} {Nature Nanotechnology}\ }\textbf {\bibinfo {volume} {13}},\
  \bibinfo {pages} {994} (\bibinfo {year} {2018})}\BibitemShut {NoStop}%
\bibitem [{\citenamefont {Calman}\ \emph {et~al.}(2018)\citenamefont {Calman},
  \citenamefont {Fogler}, \citenamefont {Butov}, \citenamefont {Hu},
  \citenamefont {Mishchenko},\ and\ \citenamefont {Geim}}]{calman2018indirect}%
  \BibitemOpen
  \bibfield  {author} {\bibinfo {author} {\bibfnamefont {E.}~\bibnamefont
  {Calman}}, \bibinfo {author} {\bibfnamefont {M.}~\bibnamefont {Fogler}},
  \bibinfo {author} {\bibfnamefont {L.}~\bibnamefont {Butov}}, \bibinfo
  {author} {\bibfnamefont {S.}~\bibnamefont {Hu}}, \bibinfo {author}
  {\bibfnamefont {A.}~\bibnamefont {Mishchenko}},\ and\ \bibinfo {author}
  {\bibfnamefont {A.}~\bibnamefont {Geim}},\ }\bibfield  {title} {\bibinfo
  {title} {Indirect excitons in van der {W}aals heterostructures at room
  temperature},\ }\href {https://doi.org/10.1038/s41467-018-04293-7} {\bibfield
   {journal} {\bibinfo  {journal} {Nature Communications}\ }\textbf {\bibinfo
  {volume} {9}},\ \bibinfo {pages} {1895} (\bibinfo {year} {2018})}\BibitemShut
  {NoStop}%
\bibitem [{\citenamefont {Kunstmann}\ \emph {et~al.}(2018)\citenamefont
  {Kunstmann}, \citenamefont {Mooshammer}, \citenamefont {Nagler},
  \citenamefont {Chaves}, \citenamefont {Stein}, \citenamefont {Paradiso},
  \citenamefont {Plechinger}, \citenamefont {Strunk}, \citenamefont
  {Sch{\"u}ller}, \citenamefont {Seifert} \emph
  {et~al.}}]{kunstmann2018momentum}%
  \BibitemOpen
  \bibfield  {author} {\bibinfo {author} {\bibfnamefont {J.}~\bibnamefont
  {Kunstmann}}, \bibinfo {author} {\bibfnamefont {F.}~\bibnamefont
  {Mooshammer}}, \bibinfo {author} {\bibfnamefont {P.}~\bibnamefont {Nagler}},
  \bibinfo {author} {\bibfnamefont {A.}~\bibnamefont {Chaves}}, \bibinfo
  {author} {\bibfnamefont {F.}~\bibnamefont {Stein}}, \bibinfo {author}
  {\bibfnamefont {N.}~\bibnamefont {Paradiso}}, \bibinfo {author}
  {\bibfnamefont {G.}~\bibnamefont {Plechinger}}, \bibinfo {author}
  {\bibfnamefont {C.}~\bibnamefont {Strunk}}, \bibinfo {author} {\bibfnamefont
  {C.}~\bibnamefont {Sch{\"u}ller}}, \bibinfo {author} {\bibfnamefont
  {G.}~\bibnamefont {Seifert}}, \emph {et~al.},\ }\bibfield  {title} {\bibinfo
  {title} {Momentum-space indirect interlayer excitons in transition-metal
  dichalcogenide van der {W}aals heterostructures},\ }\href
  {https://doi.org/10.1038/s41567-018-0123-y} {\bibfield  {journal} {\bibinfo
  {journal} {Nature Physics}\ }\textbf {\bibinfo {volume} {14}},\ \bibinfo
  {pages} {801} (\bibinfo {year} {2018})}\BibitemShut {NoStop}%
\bibitem [{\citenamefont {Alexeev}\ \emph {et~al.}(2019)\citenamefont
  {Alexeev}, \citenamefont {Ruiz-Tijerina}, \citenamefont {Danovich},
  \citenamefont {Hamer}, \citenamefont {Terry}, \citenamefont {Nayak},
  \citenamefont {Ahn}, \citenamefont {Pak}, \citenamefont {Lee}, \citenamefont
  {Sohn} \emph {et~al.}}]{alexeev2019resonantly}%
  \BibitemOpen
  \bibfield  {author} {\bibinfo {author} {\bibfnamefont {E.~M.}\ \bibnamefont
  {Alexeev}}, \bibinfo {author} {\bibfnamefont {D.~A.}\ \bibnamefont
  {Ruiz-Tijerina}}, \bibinfo {author} {\bibfnamefont {M.}~\bibnamefont
  {Danovich}}, \bibinfo {author} {\bibfnamefont {M.~J.}\ \bibnamefont {Hamer}},
  \bibinfo {author} {\bibfnamefont {D.~J.}\ \bibnamefont {Terry}}, \bibinfo
  {author} {\bibfnamefont {P.~K.}\ \bibnamefont {Nayak}}, \bibinfo {author}
  {\bibfnamefont {S.}~\bibnamefont {Ahn}}, \bibinfo {author} {\bibfnamefont
  {S.}~\bibnamefont {Pak}}, \bibinfo {author} {\bibfnamefont {J.}~\bibnamefont
  {Lee}}, \bibinfo {author} {\bibfnamefont {J.~I.}\ \bibnamefont {Sohn}}, \emph
  {et~al.},\ }\bibfield  {title} {\bibinfo {title} {Resonantly hybridized
  excitons in moir{\'e} superlattices in van der {W}aals heterostructures},\
  }\href {https://doi.org/10.1038/s41586-019-0986-9} {\bibfield  {journal}
  {\bibinfo  {journal} {Nature}\ }\textbf {\bibinfo {volume} {567}},\ \bibinfo
  {pages} {81} (\bibinfo {year} {2019})}\BibitemShut {NoStop}%
\bibitem [{\citenamefont {Tran}\ \emph {et~al.}(2019)\citenamefont {Tran},
  \citenamefont {Moody}, \citenamefont {Wu}, \citenamefont {Lu}, \citenamefont
  {Choi}, \citenamefont {Kim}, \citenamefont {Rai}, \citenamefont {Sanchez},
  \citenamefont {Quan}, \citenamefont {Singh} \emph
  {et~al.}}]{tran2019evidence}%
  \BibitemOpen
  \bibfield  {author} {\bibinfo {author} {\bibfnamefont {K.}~\bibnamefont
  {Tran}}, \bibinfo {author} {\bibfnamefont {G.}~\bibnamefont {Moody}},
  \bibinfo {author} {\bibfnamefont {F.}~\bibnamefont {Wu}}, \bibinfo {author}
  {\bibfnamefont {X.}~\bibnamefont {Lu}}, \bibinfo {author} {\bibfnamefont
  {J.}~\bibnamefont {Choi}}, \bibinfo {author} {\bibfnamefont {K.}~\bibnamefont
  {Kim}}, \bibinfo {author} {\bibfnamefont {A.}~\bibnamefont {Rai}}, \bibinfo
  {author} {\bibfnamefont {D.~A.}\ \bibnamefont {Sanchez}}, \bibinfo {author}
  {\bibfnamefont {J.}~\bibnamefont {Quan}}, \bibinfo {author} {\bibfnamefont
  {A.}~\bibnamefont {Singh}}, \emph {et~al.},\ }\bibfield  {title} {\bibinfo
  {title} {Evidence for moir{\'e} excitons in van der {W}aals
  heterostructures},\ }\href {https://doi.org/10.1038/s41586-019-0975-z}
  {\bibfield  {journal} {\bibinfo  {journal} {Nature}\ }\textbf {\bibinfo
  {volume} {567}},\ \bibinfo {pages} {71} (\bibinfo {year} {2019})}\BibitemShut
  {NoStop}%
\bibitem [{\citenamefont {Ciarrocchi}\ \emph {et~al.}(2022)\citenamefont
  {Ciarrocchi}, \citenamefont {Tagarelli}, \citenamefont {Avsar},\ and\
  \citenamefont {Kis}}]{ciarrocchi2022excitonic}%
  \BibitemOpen
  \bibfield  {author} {\bibinfo {author} {\bibfnamefont {A.}~\bibnamefont
  {Ciarrocchi}}, \bibinfo {author} {\bibfnamefont {F.}~\bibnamefont
  {Tagarelli}}, \bibinfo {author} {\bibfnamefont {A.}~\bibnamefont {Avsar}},\
  and\ \bibinfo {author} {\bibfnamefont {A.}~\bibnamefont {Kis}},\ }\bibfield
  {title} {\bibinfo {title} {Excitonic devices with van der {W}aals
  heterostructures: valleytronics meets twistronics},\ }\href
  {https://doi.org/10.1038/s41578-021-00408-7} {\bibfield  {journal} {\bibinfo
  {journal} {Nature Reviews Materials}\ }\textbf {\bibinfo {volume} {7}},\
  \bibinfo {pages} {449} (\bibinfo {year} {2022})}\BibitemShut {NoStop}%
\bibitem [{\citenamefont {Regan}\ \emph {et~al.}(2022)\citenamefont {Regan},
  \citenamefont {Wang}, \citenamefont {Paik}, \citenamefont {Zeng},
  \citenamefont {Zhang}, \citenamefont {Zhu}, \citenamefont {MacDonald},
  \citenamefont {Deng},\ and\ \citenamefont {Wang}}]{Regan_2022}%
  \BibitemOpen
  \bibfield  {author} {\bibinfo {author} {\bibfnamefont {E.~C.}\ \bibnamefont
  {Regan}}, \bibinfo {author} {\bibfnamefont {D.}~\bibnamefont {Wang}},
  \bibinfo {author} {\bibfnamefont {E.~Y.}\ \bibnamefont {Paik}}, \bibinfo
  {author} {\bibfnamefont {Y.}~\bibnamefont {Zeng}}, \bibinfo {author}
  {\bibfnamefont {L.}~\bibnamefont {Zhang}}, \bibinfo {author} {\bibfnamefont
  {J.}~\bibnamefont {Zhu}}, \bibinfo {author} {\bibfnamefont {A.~H.}\
  \bibnamefont {MacDonald}}, \bibinfo {author} {\bibfnamefont {H.}~\bibnamefont
  {Deng}},\ and\ \bibinfo {author} {\bibfnamefont {F.}~\bibnamefont {Wang}},\
  }\bibfield  {title} {\bibinfo {title} {Emerging exciton physics in transition
  metal dichalcogenide heterobilayers},\ }\href
  {https://doi.org/10.1038/s41578-022-00440-1} {\bibfield  {journal} {\bibinfo
  {journal} {Nature Reviews Materials}\ }\textbf {\bibinfo {volume} {7}},\
  \bibinfo {pages} {778} (\bibinfo {year} {2022})}\BibitemShut {NoStop}%
\bibitem [{\citenamefont {Carvalho}\ \emph {et~al.}(2016)\citenamefont
  {Carvalho}, \citenamefont {Wang}, \citenamefont {Zhu}, \citenamefont {Rodin},
  \citenamefont {Su},\ and\ \citenamefont {Castro~Neto}}]{Carvalho_2016}%
  \BibitemOpen
  \bibfield  {author} {\bibinfo {author} {\bibfnamefont {A.}~\bibnamefont
  {Carvalho}}, \bibinfo {author} {\bibfnamefont {M.}~\bibnamefont {Wang}},
  \bibinfo {author} {\bibfnamefont {X.}~\bibnamefont {Zhu}}, \bibinfo {author}
  {\bibfnamefont {A.~S.}\ \bibnamefont {Rodin}}, \bibinfo {author}
  {\bibfnamefont {H.}~\bibnamefont {Su}},\ and\ \bibinfo {author}
  {\bibfnamefont {A.~H.}\ \bibnamefont {Castro~Neto}},\ }\bibfield  {title}
  {\bibinfo {title} {Phosphorene: from theory to applications},\ }\href
  {https://doi.org/10.1038/natrevmats.2016.61} {\bibfield  {journal} {\bibinfo
  {journal} {Nature Reviews Materials}\ }\textbf {\bibinfo {volume} {1}},\
  \bibinfo {pages} {16061} (\bibinfo {year} {2016})}\BibitemShut {NoStop}%
\bibitem [{\citenamefont {Yang}\ \emph {et~al.}(2015)\citenamefont {Yang},
  \citenamefont {Xu}, \citenamefont {Pei}, \citenamefont {Myint}, \citenamefont
  {Wang}, \citenamefont {Wang}, \citenamefont {Zhang}, \citenamefont {Yu},\
  and\ \citenamefont {Lu}}]{yang2015optical}%
  \BibitemOpen
  \bibfield  {author} {\bibinfo {author} {\bibfnamefont {J.}~\bibnamefont
  {Yang}}, \bibinfo {author} {\bibfnamefont {R.}~\bibnamefont {Xu}}, \bibinfo
  {author} {\bibfnamefont {J.}~\bibnamefont {Pei}}, \bibinfo {author}
  {\bibfnamefont {Y.~W.}\ \bibnamefont {Myint}}, \bibinfo {author}
  {\bibfnamefont {F.}~\bibnamefont {Wang}}, \bibinfo {author} {\bibfnamefont
  {Z.}~\bibnamefont {Wang}}, \bibinfo {author} {\bibfnamefont {S.}~\bibnamefont
  {Zhang}}, \bibinfo {author} {\bibfnamefont {Z.}~\bibnamefont {Yu}},\ and\
  \bibinfo {author} {\bibfnamefont {Y.}~\bibnamefont {Lu}},\ }\bibfield
  {title} {\bibinfo {title} {Optical tuning of exciton and trion emissions in
  monolayer phosphorene},\ }\href {https://doi.org/10.1038/lsa.2015.85}
  {\bibfield  {journal} {\bibinfo  {journal} {Light: Science \& Applications}\
  }\textbf {\bibinfo {volume} {4}},\ \bibinfo {pages} {e312} (\bibinfo {year}
  {2015})}\BibitemShut {NoStop}%
\bibitem [{\citenamefont {Lu}\ \emph {et~al.}(2016)\citenamefont {Lu},
  \citenamefont {Yang}, \citenamefont {Carvalho}, \citenamefont {Liu},
  \citenamefont {Lu},\ and\ \citenamefont {Sow}}]{lu2016light}%
  \BibitemOpen
  \bibfield  {author} {\bibinfo {author} {\bibfnamefont {J.}~\bibnamefont
  {Lu}}, \bibinfo {author} {\bibfnamefont {J.}~\bibnamefont {Yang}}, \bibinfo
  {author} {\bibfnamefont {A.}~\bibnamefont {Carvalho}}, \bibinfo {author}
  {\bibfnamefont {H.}~\bibnamefont {Liu}}, \bibinfo {author} {\bibfnamefont
  {Y.}~\bibnamefont {Lu}},\ and\ \bibinfo {author} {\bibfnamefont {C.~H.}\
  \bibnamefont {Sow}},\ }\bibfield  {title} {\bibinfo {title} {Light-matter
  interactions in phosphorene},\ }\href
  {https://doi.org/10.1021/acs.accounts.6b00266} {\bibfield  {journal}
  {\bibinfo  {journal} {Accounts of Chemical Research}\ }\textbf {\bibinfo
  {volume} {49}},\ \bibinfo {pages} {1806} (\bibinfo {year}
  {2016})}\BibitemShut {NoStop}%
\bibitem [{\citenamefont {Xu}\ \emph {et~al.}(2016)\citenamefont {Xu},
  \citenamefont {Zhang}, \citenamefont {Wang}, \citenamefont {Yang},
  \citenamefont {Wang}, \citenamefont {Pei}, \citenamefont {Myint},
  \citenamefont {Xing}, \citenamefont {Yu}, \citenamefont {Fu} \emph
  {et~al.}}]{xu2016extraordinarily}%
  \BibitemOpen
  \bibfield  {author} {\bibinfo {author} {\bibfnamefont {R.}~\bibnamefont
  {Xu}}, \bibinfo {author} {\bibfnamefont {S.}~\bibnamefont {Zhang}}, \bibinfo
  {author} {\bibfnamefont {F.}~\bibnamefont {Wang}}, \bibinfo {author}
  {\bibfnamefont {J.}~\bibnamefont {Yang}}, \bibinfo {author} {\bibfnamefont
  {Z.}~\bibnamefont {Wang}}, \bibinfo {author} {\bibfnamefont {J.}~\bibnamefont
  {Pei}}, \bibinfo {author} {\bibfnamefont {Y.~W.}\ \bibnamefont {Myint}},
  \bibinfo {author} {\bibfnamefont {B.}~\bibnamefont {Xing}}, \bibinfo {author}
  {\bibfnamefont {Z.}~\bibnamefont {Yu}}, \bibinfo {author} {\bibfnamefont
  {L.}~\bibnamefont {Fu}}, \emph {et~al.},\ }\bibfield  {title} {\bibinfo
  {title} {Extraordinarily bound quasi-one-dimensional trions in
  two-dimensional phosphorene atomic semiconductors},\ }\href
  {https://doi.org/10.1021/acsnano.5b06193} {\bibfield  {journal} {\bibinfo
  {journal} {Acs Nano}\ }\textbf {\bibinfo {volume} {10}},\ \bibinfo {pages}
  {2046} (\bibinfo {year} {2016})}\BibitemShut {NoStop}%
\bibitem [{\citenamefont {Akhtar}\ \emph {et~al.}(2017)\citenamefont {Akhtar},
  \citenamefont {Anderson}, \citenamefont {Zhao}, \citenamefont {Alruqi},
  \citenamefont {Mroczkowska}, \citenamefont {Sumanasekera},\ and\
  \citenamefont {Jasinski}}]{akhtar2017recent}%
  \BibitemOpen
  \bibfield  {author} {\bibinfo {author} {\bibfnamefont {M.}~\bibnamefont
  {Akhtar}}, \bibinfo {author} {\bibfnamefont {G.}~\bibnamefont {Anderson}},
  \bibinfo {author} {\bibfnamefont {R.}~\bibnamefont {Zhao}}, \bibinfo {author}
  {\bibfnamefont {A.}~\bibnamefont {Alruqi}}, \bibinfo {author} {\bibfnamefont
  {J.~E.}\ \bibnamefont {Mroczkowska}}, \bibinfo {author} {\bibfnamefont
  {G.}~\bibnamefont {Sumanasekera}},\ and\ \bibinfo {author} {\bibfnamefont
  {J.~B.}\ \bibnamefont {Jasinski}},\ }\bibfield  {title} {\bibinfo {title}
  {Recent advances in synthesis, properties, and applications of phosphorene},\
  }\href {https://doi.org/10.1038/s41699-017-0007-5} {\bibfield  {journal}
  {\bibinfo  {journal} {npj 2D Materials and Applications}\ }\textbf {\bibinfo
  {volume} {1}},\ \bibinfo {pages} {5} (\bibinfo {year} {2017})}\BibitemShut
  {NoStop}%
\bibitem [{\citenamefont {Kung}\ \emph {et~al.}(2019)\citenamefont {Kung},
  \citenamefont {Goyal}, \citenamefont {Maslov}, \citenamefont {Wang},
  \citenamefont {Lee}, \citenamefont {Kemper}, \citenamefont {Cheong},\ and\
  \citenamefont {Blumberg}}]{kung2019observation}%
  \BibitemOpen
  \bibfield  {author} {\bibinfo {author} {\bibfnamefont {H.-H.}\ \bibnamefont
  {Kung}}, \bibinfo {author} {\bibfnamefont {A.}~\bibnamefont {Goyal}},
  \bibinfo {author} {\bibfnamefont {D.}~\bibnamefont {Maslov}}, \bibinfo
  {author} {\bibfnamefont {X.}~\bibnamefont {Wang}}, \bibinfo {author}
  {\bibfnamefont {A.}~\bibnamefont {Lee}}, \bibinfo {author} {\bibfnamefont
  {A.}~\bibnamefont {Kemper}}, \bibinfo {author} {\bibfnamefont {S.-W.}\
  \bibnamefont {Cheong}},\ and\ \bibinfo {author} {\bibfnamefont
  {G.}~\bibnamefont {Blumberg}},\ }\bibfield  {title} {\bibinfo {title}
  {Observation of chiral surface excitons in a topological insulator
  \ce{Bi2Se3}},\ }\href {https://doi.org/10.1073/pnas.1813514116} {\bibfield
  {journal} {\bibinfo  {journal} {Proceedings of the National Academy of
  Sciences}\ }\textbf {\bibinfo {volume} {116}},\ \bibinfo {pages} {4006}
  (\bibinfo {year} {2019})}\BibitemShut {NoStop}%
\bibitem [{\citenamefont {Syperek}\ \emph {et~al.}(2022)\citenamefont
  {Syperek}, \citenamefont {St{\"u}hler}, \citenamefont {Consiglio},
  \citenamefont {Holewa}, \citenamefont {Wyborski}, \citenamefont {Dusanowski},
  \citenamefont {Reis}, \citenamefont {H{\"o}fling}, \citenamefont {Thomale},
  \citenamefont {Hanke} \emph {et~al.}}]{syperek2022observation}%
  \BibitemOpen
  \bibfield  {author} {\bibinfo {author} {\bibfnamefont {M.}~\bibnamefont
  {Syperek}}, \bibinfo {author} {\bibfnamefont {R.}~\bibnamefont
  {St{\"u}hler}}, \bibinfo {author} {\bibfnamefont {A.}~\bibnamefont
  {Consiglio}}, \bibinfo {author} {\bibfnamefont {P.}~\bibnamefont {Holewa}},
  \bibinfo {author} {\bibfnamefont {P.}~\bibnamefont {Wyborski}}, \bibinfo
  {author} {\bibfnamefont {{\L}.}~\bibnamefont {Dusanowski}}, \bibinfo {author}
  {\bibfnamefont {F.}~\bibnamefont {Reis}}, \bibinfo {author} {\bibfnamefont
  {S.}~\bibnamefont {H{\"o}fling}}, \bibinfo {author} {\bibfnamefont
  {R.}~\bibnamefont {Thomale}}, \bibinfo {author} {\bibfnamefont
  {W.}~\bibnamefont {Hanke}}, \emph {et~al.},\ }\bibfield  {title} {\bibinfo
  {title} {Observation of room temperature excitons in an atomically thin
  topological insulator},\ }\href {https://doi.org/10.1038/s41467-022-33822-8}
  {\bibfield  {journal} {\bibinfo  {journal} {Nature Communications}\ }\textbf
  {\bibinfo {volume} {13}},\ \bibinfo {pages} {6313} (\bibinfo {year}
  {2022})}\BibitemShut {NoStop}%
\bibitem [{\citenamefont {Mori}\ \emph {et~al.}(2023)\citenamefont {Mori},
  \citenamefont {Ciocys}, \citenamefont {Takasan}, \citenamefont {Ai},
  \citenamefont {Currier}, \citenamefont {Morimoto}, \citenamefont {Moore},\
  and\ \citenamefont {Lanzara}}]{mori2023spin}%
  \BibitemOpen
  \bibfield  {author} {\bibinfo {author} {\bibfnamefont {R.}~\bibnamefont
  {Mori}}, \bibinfo {author} {\bibfnamefont {S.}~\bibnamefont {Ciocys}},
  \bibinfo {author} {\bibfnamefont {K.}~\bibnamefont {Takasan}}, \bibinfo
  {author} {\bibfnamefont {P.}~\bibnamefont {Ai}}, \bibinfo {author}
  {\bibfnamefont {K.}~\bibnamefont {Currier}}, \bibinfo {author} {\bibfnamefont
  {T.}~\bibnamefont {Morimoto}}, \bibinfo {author} {\bibfnamefont {J.~E.}\
  \bibnamefont {Moore}},\ and\ \bibinfo {author} {\bibfnamefont
  {A.}~\bibnamefont {Lanzara}},\ }\bibfield  {title} {\bibinfo {title}
  {Spin-polarized spatially indirect excitons in a topological insulator},\
  }\href {https://doi.org/10.1038/s41586-022-05567-3} {\bibfield  {journal}
  {\bibinfo  {journal} {Nature}\ }\textbf {\bibinfo {volume} {614}},\ \bibinfo
  {pages} {249} (\bibinfo {year} {2023})}\BibitemShut {NoStop}%
\bibitem [{\citenamefont {Chernikov}\ \emph {et~al.}(2014)\citenamefont
  {Chernikov}, \citenamefont {Berkelbach}, \citenamefont {Hill}, \citenamefont
  {Rigosi}, \citenamefont {Li}, \citenamefont {Aslan}, \citenamefont
  {Reichman}, \citenamefont {Hybertsen},\ and\ \citenamefont
  {Heinz}}]{Chernikov_2014}%
  \BibitemOpen
  \bibfield  {author} {\bibinfo {author} {\bibfnamefont {A.}~\bibnamefont
  {Chernikov}}, \bibinfo {author} {\bibfnamefont {T.~C.}\ \bibnamefont
  {Berkelbach}}, \bibinfo {author} {\bibfnamefont {H.~M.}\ \bibnamefont
  {Hill}}, \bibinfo {author} {\bibfnamefont {A.}~\bibnamefont {Rigosi}},
  \bibinfo {author} {\bibfnamefont {Y.}~\bibnamefont {Li}}, \bibinfo {author}
  {\bibfnamefont {B.}~\bibnamefont {Aslan}}, \bibinfo {author} {\bibfnamefont
  {D.~R.}\ \bibnamefont {Reichman}}, \bibinfo {author} {\bibfnamefont {M.~S.}\
  \bibnamefont {Hybertsen}},\ and\ \bibinfo {author} {\bibfnamefont {T.~F.}\
  \bibnamefont {Heinz}},\ }\bibfield  {title} {\bibinfo {title} {Exciton
  binding energy and nonhydrogenic {R}ydberg series in monolayer \ce{WS2}},\
  }\href {https://doi.org/10.1103/PhysRevLett.113.076802} {\bibfield  {journal}
  {\bibinfo  {journal} {Physical Review Letters}\ }\textbf {\bibinfo {volume}
  {113}},\ \bibinfo {pages} {076802} (\bibinfo {year} {2014})}\BibitemShut
  {NoStop}%
\bibitem [{\citenamefont {Selig}\ \emph {et~al.}(2016)\citenamefont {Selig},
  \citenamefont {Berghäuser}, \citenamefont {Raja}, \citenamefont {Nagler},
  \citenamefont {Schüller}, \citenamefont {Heinz}, \citenamefont {Korn},
  \citenamefont {Chernikov}, \citenamefont {Malic},\ and\ \citenamefont
  {Knorr}}]{Selig_2016}%
  \BibitemOpen
  \bibfield  {author} {\bibinfo {author} {\bibfnamefont {M.}~\bibnamefont
  {Selig}}, \bibinfo {author} {\bibfnamefont {G.}~\bibnamefont {Berghäuser}},
  \bibinfo {author} {\bibfnamefont {A.}~\bibnamefont {Raja}}, \bibinfo {author}
  {\bibfnamefont {P.}~\bibnamefont {Nagler}}, \bibinfo {author} {\bibfnamefont
  {C.}~\bibnamefont {Schüller}}, \bibinfo {author} {\bibfnamefont {T.~F.}\
  \bibnamefont {Heinz}}, \bibinfo {author} {\bibfnamefont {T.}~\bibnamefont
  {Korn}}, \bibinfo {author} {\bibfnamefont {A.}~\bibnamefont {Chernikov}},
  \bibinfo {author} {\bibfnamefont {E.}~\bibnamefont {Malic}},\ and\ \bibinfo
  {author} {\bibfnamefont {A.}~\bibnamefont {Knorr}},\ }\bibfield  {title}
  {\bibinfo {title} {Excitonic linewidth and coherence lifetime in monolayer
  transition metal dichalcogenides},\ }\href
  {https://doi.org/10.1038/ncomms13279} {\bibfield  {journal} {\bibinfo
  {journal} {Nature Communications}\ }\textbf {\bibinfo {volume} {7}},\
  \bibinfo {pages} {13279} (\bibinfo {year} {2016})}\BibitemShut {NoStop}%
\bibitem [{\citenamefont {Wang}\ \emph {et~al.}(2016)\citenamefont {Wang},
  \citenamefont {Zhang}, \citenamefont {Chan}, \citenamefont {Manolatou},
  \citenamefont {Tiwari},\ and\ \citenamefont {Rana}}]{Haining_2016}%
  \BibitemOpen
  \bibfield  {author} {\bibinfo {author} {\bibfnamefont {H.}~\bibnamefont
  {Wang}}, \bibinfo {author} {\bibfnamefont {C.}~\bibnamefont {Zhang}},
  \bibinfo {author} {\bibfnamefont {W.}~\bibnamefont {Chan}}, \bibinfo {author}
  {\bibfnamefont {C.}~\bibnamefont {Manolatou}}, \bibinfo {author}
  {\bibfnamefont {S.}~\bibnamefont {Tiwari}},\ and\ \bibinfo {author}
  {\bibfnamefont {F.}~\bibnamefont {Rana}},\ }\bibfield  {title} {\bibinfo
  {title} {Radiative lifetimes of excitons and trions in monolayers of the
  metal dichalcogenide \ce{MoS2}},\ }\href
  {https://doi.org/10.1103/PhysRevB.93.045407} {\bibfield  {journal} {\bibinfo
  {journal} {Physical Review B}\ }\textbf {\bibinfo {volume} {93}},\ \bibinfo
  {pages} {045407} (\bibinfo {year} {2016})}\BibitemShut {NoStop}%
\bibitem [{\citenamefont {Quan}\ \emph {et~al.}(2016)\citenamefont {Quan},
  \citenamefont {Yuan}, \citenamefont {Comin}, \citenamefont {Voznyy},
  \citenamefont {Beauregard}, \citenamefont {Hoogland}, \citenamefont {Buin},
  \citenamefont {Kirmani}, \citenamefont {Zhao}, \citenamefont {Amassian},
  \citenamefont {Kim},\ and\ \citenamefont {Sargent}}]{Quan_2016}%
  \BibitemOpen
  \bibfield  {author} {\bibinfo {author} {\bibfnamefont {L.~N.}\ \bibnamefont
  {Quan}}, \bibinfo {author} {\bibfnamefont {M.}~\bibnamefont {Yuan}}, \bibinfo
  {author} {\bibfnamefont {R.}~\bibnamefont {Comin}}, \bibinfo {author}
  {\bibfnamefont {O.}~\bibnamefont {Voznyy}}, \bibinfo {author} {\bibfnamefont
  {E.~M.}\ \bibnamefont {Beauregard}}, \bibinfo {author} {\bibfnamefont
  {S.}~\bibnamefont {Hoogland}}, \bibinfo {author} {\bibfnamefont
  {A.}~\bibnamefont {Buin}}, \bibinfo {author} {\bibfnamefont {A.~R.}\
  \bibnamefont {Kirmani}}, \bibinfo {author} {\bibfnamefont {K.}~\bibnamefont
  {Zhao}}, \bibinfo {author} {\bibfnamefont {A.}~\bibnamefont {Amassian}},
  \bibinfo {author} {\bibfnamefont {D.~H.}\ \bibnamefont {Kim}},\ and\ \bibinfo
  {author} {\bibfnamefont {E.~H.}\ \bibnamefont {Sargent}},\ }\bibfield
  {title} {\bibinfo {title} {Ligand-stabilized reduced-dimensionality
  perovskites},\ }\href {https://doi.org/10.1021/jacs.5b11740} {\bibfield
  {journal} {\bibinfo  {journal} {Journal of the American Chemical Society}\
  }\textbf {\bibinfo {volume} {138}},\ \bibinfo {pages} {2649} (\bibinfo {year}
  {2016})}\BibitemShut {NoStop}%
\bibitem [{\citenamefont {Zhang}\ \emph {et~al.}(2017)\citenamefont {Zhang},
  \citenamefont {Cao}, \citenamefont {Lu}, \citenamefont {Lin}, \citenamefont
  {Zhang}, \citenamefont {Wang}, \citenamefont {Li}, \citenamefont {Hone},
  \citenamefont {Robinson}, \citenamefont {Smirnov}, \citenamefont {Louie},\
  and\ \citenamefont {Heinz}}]{Zhang_2017}%
  \BibitemOpen
  \bibfield  {author} {\bibinfo {author} {\bibfnamefont {X.-X.}\ \bibnamefont
  {Zhang}}, \bibinfo {author} {\bibfnamefont {T.}~\bibnamefont {Cao}}, \bibinfo
  {author} {\bibfnamefont {Z.}~\bibnamefont {Lu}}, \bibinfo {author}
  {\bibfnamefont {Y.-C.}\ \bibnamefont {Lin}}, \bibinfo {author} {\bibfnamefont
  {F.}~\bibnamefont {Zhang}}, \bibinfo {author} {\bibfnamefont
  {Y.}~\bibnamefont {Wang}}, \bibinfo {author} {\bibfnamefont {Z.}~\bibnamefont
  {Li}}, \bibinfo {author} {\bibfnamefont {J.~C.}\ \bibnamefont {Hone}},
  \bibinfo {author} {\bibfnamefont {J.~A.}\ \bibnamefont {Robinson}}, \bibinfo
  {author} {\bibfnamefont {D.}~\bibnamefont {Smirnov}}, \bibinfo {author}
  {\bibfnamefont {S.~G.}\ \bibnamefont {Louie}},\ and\ \bibinfo {author}
  {\bibfnamefont {T.~F.}\ \bibnamefont {Heinz}},\ }\bibfield  {title} {\bibinfo
  {title} {Magnetic brightening and control of dark excitons in monolayer
  \ce{WSe2}},\ }\href {https://doi.org/10.1038/nnano.2017.105} {\bibfield
  {journal} {\bibinfo  {journal} {Nature Nanotechnology}\ }\textbf {\bibinfo
  {volume} {12}},\ \bibinfo {pages} {883} (\bibinfo {year} {2017})}\BibitemShut
  {NoStop}%
\bibitem [{\citenamefont {Xiao}\ \emph {et~al.}(2017)\citenamefont {Xiao},
  \citenamefont {Zhao}, \citenamefont {Wang},\ and\ \citenamefont
  {Zhang}}]{Xiao_2017}%
  \BibitemOpen
  \bibfield  {author} {\bibinfo {author} {\bibfnamefont {J.}~\bibnamefont
  {Xiao}}, \bibinfo {author} {\bibfnamefont {M.}~\bibnamefont {Zhao}}, \bibinfo
  {author} {\bibfnamefont {Y.}~\bibnamefont {Wang}},\ and\ \bibinfo {author}
  {\bibfnamefont {X.}~\bibnamefont {Zhang}},\ }\bibfield  {title} {\bibinfo
  {title} {Excitons in atomically thin 2{D} semiconductors and their
  applications},\ }\href {https://doi.org/10.1515/nanoph-2016-0160} {\bibfield
  {journal} {\bibinfo  {journal} {Nanophotonics}\ }\textbf {\bibinfo {volume}
  {6}},\ \bibinfo {pages} {1309} (\bibinfo {year} {2017})}\BibitemShut
  {NoStop}%
\bibitem [{\citenamefont {Mueller}\ and\ \citenamefont
  {Malic}(2018)}]{Mueller_2018}%
  \BibitemOpen
  \bibfield  {author} {\bibinfo {author} {\bibfnamefont {T.}~\bibnamefont
  {Mueller}}\ and\ \bibinfo {author} {\bibfnamefont {E.}~\bibnamefont
  {Malic}},\ }\bibfield  {title} {\bibinfo {title} {Exciton physics and device
  application of two-dimensional transition metal dichalcogenide
  semiconductors},\ }\href {https://doi.org/10.1038/s41699-018-0074-2}
  {\bibfield  {journal} {\bibinfo  {journal} {npj 2D Materials and
  Applications}\ }\textbf {\bibinfo {volume} {2}},\ \bibinfo {pages} {29}
  (\bibinfo {year} {2018})}\BibitemShut {NoStop}%
\bibitem [{\citenamefont {Rodin}\ \emph {et~al.}(2020)\citenamefont {Rodin},
  \citenamefont {Trushin}, \citenamefont {Carvalho},\ and\ \citenamefont
  {Castro~Neto}}]{Rodin_2020}%
  \BibitemOpen
  \bibfield  {author} {\bibinfo {author} {\bibfnamefont {A.}~\bibnamefont
  {Rodin}}, \bibinfo {author} {\bibfnamefont {M.}~\bibnamefont {Trushin}},
  \bibinfo {author} {\bibfnamefont {A.}~\bibnamefont {Carvalho}},\ and\
  \bibinfo {author} {\bibfnamefont {A.~H.}\ \bibnamefont {Castro~Neto}},\
  }\bibfield  {title} {\bibinfo {title} {Collective excitations in {2D}
  materials},\ }\href {https://doi.org/10.1038/s42254-020-0214-4} {\bibfield
  {journal} {\bibinfo  {journal} {Nature Reviews Physics}\ }\textbf {\bibinfo
  {volume} {2}},\ \bibinfo {pages} {524} (\bibinfo {year} {2020})}\BibitemShut
  {NoStop}%
\bibitem [{\citenamefont {Lee}\ \emph {et~al.}(2023)\citenamefont {Lee},
  \citenamefont {Kim}, \citenamefont {Ryu}, \citenamefont {Kim}, \citenamefont
  {Bae}, \citenamefont {Koo}, \citenamefont {Jang},\ and\ \citenamefont
  {Park}}]{Lee_2023}%
  \BibitemOpen
  \bibfield  {author} {\bibinfo {author} {\bibfnamefont {H.}~\bibnamefont
  {Lee}}, \bibinfo {author} {\bibfnamefont {Y.~B.}\ \bibnamefont {Kim}},
  \bibinfo {author} {\bibfnamefont {J.~W.}\ \bibnamefont {Ryu}}, \bibinfo
  {author} {\bibfnamefont {S.}~\bibnamefont {Kim}}, \bibinfo {author}
  {\bibfnamefont {J.}~\bibnamefont {Bae}}, \bibinfo {author} {\bibfnamefont
  {Y.}~\bibnamefont {Koo}}, \bibinfo {author} {\bibfnamefont {D.}~\bibnamefont
  {Jang}},\ and\ \bibinfo {author} {\bibfnamefont {K.-D.}\ \bibnamefont
  {Park}},\ }\bibfield  {title} {\bibinfo {title} {Recent progress of exciton
  transport in two-dimensional semiconductors},\ }\href
  {https://doi.org/10.1186/s40580-023-00404-3} {\bibfield  {journal} {\bibinfo
  {journal} {Nano Convergence}\ }\textbf {\bibinfo {volume} {10}},\ \bibinfo
  {pages} {57} (\bibinfo {year} {2023})}\BibitemShut {NoStop}%
\bibitem [{\citenamefont {Maisel~Licerán}\ \emph {et~al.}(2023)\citenamefont
  {Maisel~Licerán}, \citenamefont {García~Flórez}, \citenamefont
  {Siebbeles},\ and\ \citenamefont {Stoof}}]{MaiselLiceran_2023}%
  \BibitemOpen
  \bibfield  {author} {\bibinfo {author} {\bibfnamefont {L.}~\bibnamefont
  {Maisel~Licerán}}, \bibinfo {author} {\bibfnamefont {F.}~\bibnamefont
  {García~Flórez}}, \bibinfo {author} {\bibfnamefont {L.~D.~A.}\ \bibnamefont
  {Siebbeles}},\ and\ \bibinfo {author} {\bibfnamefont {H.~T.~C.}\ \bibnamefont
  {Stoof}},\ }\bibfield  {title} {\bibinfo {title} {Single-particle properties
  of topological {W}annier excitons in bismuth chalcogenide nanosheets},\
  }\href {https://doi.org/10.1038/s41598-023-32740-z} {\bibfield  {journal}
  {\bibinfo  {journal} {Scientific Reports}\ }\textbf {\bibinfo {volume}
  {13}},\ \bibinfo {pages} {6337} (\bibinfo {year} {2023})}\BibitemShut
  {NoStop}%
\bibitem [{\citenamefont {Glazov}\ \emph {et~al.}(2024)\citenamefont {Glazov},
  \citenamefont {Arora}, \citenamefont {Chaves},\ and\ \citenamefont
  {Gobato}}]{Glazov_2024}%
  \BibitemOpen
  \bibfield  {author} {\bibinfo {author} {\bibfnamefont {M.}~\bibnamefont
  {Glazov}}, \bibinfo {author} {\bibfnamefont {A.}~\bibnamefont {Arora}},
  \bibinfo {author} {\bibfnamefont {A.}~\bibnamefont {Chaves}},\ and\ \bibinfo
  {author} {\bibfnamefont {Y.~G.}\ \bibnamefont {Gobato}},\ }\bibfield  {title}
  {\bibinfo {title} {Excitons in two-dimensional materials and
  heterostructures: {O}ptical and magneto-optical properties},\ }\href
  {https://doi.org/10.1557/s43577-024-00754-1} {\bibfield  {journal} {\bibinfo
  {journal} {MRS Bulletin}\ }\textbf {\bibinfo {volume} {49}},\ \bibinfo
  {pages} {899} (\bibinfo {year} {2024})}\BibitemShut {NoStop}%
\bibitem [{\citenamefont {Peyghambarian}\ \emph {et~al.}(1984)\citenamefont
  {Peyghambarian}, \citenamefont {Gibbs}, \citenamefont {Jewell}, \citenamefont
  {Antonetti}, \citenamefont {Migus}, \citenamefont {Hulin},\ and\
  \citenamefont {Mysyrowicz}}]{peyghambarian1984blue}%
  \BibitemOpen
  \bibfield  {author} {\bibinfo {author} {\bibfnamefont {N.}~\bibnamefont
  {Peyghambarian}}, \bibinfo {author} {\bibfnamefont {H.}~\bibnamefont
  {Gibbs}}, \bibinfo {author} {\bibfnamefont {J.}~\bibnamefont {Jewell}},
  \bibinfo {author} {\bibfnamefont {A.}~\bibnamefont {Antonetti}}, \bibinfo
  {author} {\bibfnamefont {A.}~\bibnamefont {Migus}}, \bibinfo {author}
  {\bibfnamefont {D.}~\bibnamefont {Hulin}},\ and\ \bibinfo {author}
  {\bibfnamefont {A.}~\bibnamefont {Mysyrowicz}},\ }\bibfield  {title}
  {\bibinfo {title} {Blue shift of the exciton resonance due to exciton-exciton
  interactions in a multiple-quantum-well structure},\ }\href
  {https://doi.org/10.1103/PhysRevLett.53.2433} {\bibfield  {journal} {\bibinfo
   {journal} {Physical Review Letters}\ }\textbf {\bibinfo {volume} {53}},\
  \bibinfo {pages} {2433} (\bibinfo {year} {1984})}\BibitemShut {NoStop}%
\bibitem [{\citenamefont {Kossacki}\ \emph {et~al.}(2005)\citenamefont
  {Kossacki}, \citenamefont {P{\l}ochocka}, \citenamefont {Piechal},
  \citenamefont {Ma{\'s}lana}, \citenamefont {Golnik}, \citenamefont {Cibert},
  \citenamefont {Tatarenko},\ and\ \citenamefont {Gaj}}]{kossacki2005exciton}%
  \BibitemOpen
  \bibfield  {author} {\bibinfo {author} {\bibfnamefont {P.}~\bibnamefont
  {Kossacki}}, \bibinfo {author} {\bibfnamefont {P.}~\bibnamefont
  {P{\l}ochocka}}, \bibinfo {author} {\bibfnamefont {B.}~\bibnamefont
  {Piechal}}, \bibinfo {author} {\bibfnamefont {W.}~\bibnamefont
  {Ma{\'s}lana}}, \bibinfo {author} {\bibfnamefont {A.}~\bibnamefont {Golnik}},
  \bibinfo {author} {\bibfnamefont {J.}~\bibnamefont {Cibert}}, \bibinfo
  {author} {\bibfnamefont {S.}~\bibnamefont {Tatarenko}},\ and\ \bibinfo
  {author} {\bibfnamefont {J.~A.}\ \bibnamefont {Gaj}},\ }\bibfield  {title}
  {\bibinfo {title} {Exciton-exciton interaction and biexcitons in the presence
  of spin-polarized carriers},\ }\href
  {https://doi.org/10.1103/PhysRevB.72.035340} {\bibfield  {journal} {\bibinfo
  {journal} {Physical Review B}\ }\textbf {\bibinfo {volume} {72}},\ \bibinfo
  {pages} {035340} (\bibinfo {year} {2005})}\BibitemShut {NoStop}%
\bibitem [{\citenamefont {Stone}\ \emph {et~al.}(2009)\citenamefont {Stone},
  \citenamefont {Turner}, \citenamefont {Gundogdu}, \citenamefont {Cundiff},\
  and\ \citenamefont {Nelson}}]{stone2009exciton}%
  \BibitemOpen
  \bibfield  {author} {\bibinfo {author} {\bibfnamefont {K.~W.}\ \bibnamefont
  {Stone}}, \bibinfo {author} {\bibfnamefont {D.~B.}\ \bibnamefont {Turner}},
  \bibinfo {author} {\bibfnamefont {K.}~\bibnamefont {Gundogdu}}, \bibinfo
  {author} {\bibfnamefont {S.~T.}\ \bibnamefont {Cundiff}},\ and\ \bibinfo
  {author} {\bibfnamefont {K.~A.}\ \bibnamefont {Nelson}},\ }\bibfield  {title}
  {\bibinfo {title} {Exciton-exciton correlations revealed by two-quantum,
  two-dimensional {F}ourier transform optical spectroscopy},\ }\href
  {https://doi.org/10.1021/ar900122k} {\bibfield  {journal} {\bibinfo
  {journal} {Accounts of Chemical Research}\ }\textbf {\bibinfo {volume}
  {42}},\ \bibinfo {pages} {1452} (\bibinfo {year} {2009})}\BibitemShut
  {NoStop}%
\bibitem [{\citenamefont {Sim}\ \emph {et~al.}(2013)\citenamefont {Sim},
  \citenamefont {Park}, \citenamefont {Song}, \citenamefont {In}, \citenamefont
  {Lee}, \citenamefont {Kim},\ and\ \citenamefont {Choi}}]{sim2013exciton}%
  \BibitemOpen
  \bibfield  {author} {\bibinfo {author} {\bibfnamefont {S.}~\bibnamefont
  {Sim}}, \bibinfo {author} {\bibfnamefont {J.}~\bibnamefont {Park}}, \bibinfo
  {author} {\bibfnamefont {J.-G.}\ \bibnamefont {Song}}, \bibinfo {author}
  {\bibfnamefont {C.}~\bibnamefont {In}}, \bibinfo {author} {\bibfnamefont
  {Y.-S.}\ \bibnamefont {Lee}}, \bibinfo {author} {\bibfnamefont
  {H.}~\bibnamefont {Kim}},\ and\ \bibinfo {author} {\bibfnamefont
  {H.}~\bibnamefont {Choi}},\ }\bibfield  {title} {\bibinfo {title} {Exciton
  dynamics in atomically thin \ce{MoS2}: interexcitonic interaction and
  broadening kinetics},\ }\href {https://doi.org/10.1103/PhysRevB.88.075434}
  {\bibfield  {journal} {\bibinfo  {journal} {Physical Review B}\ }\textbf
  {\bibinfo {volume} {88}},\ \bibinfo {pages} {075434} (\bibinfo {year}
  {2013})}\BibitemShut {NoStop}%
\bibitem [{\citenamefont {Mouri}\ \emph {et~al.}(2014)\citenamefont {Mouri},
  \citenamefont {Miyauchi}, \citenamefont {Toh}, \citenamefont {Zhao},
  \citenamefont {Eda},\ and\ \citenamefont {Matsuda}}]{mouri2014nonlinear}%
  \BibitemOpen
  \bibfield  {author} {\bibinfo {author} {\bibfnamefont {S.}~\bibnamefont
  {Mouri}}, \bibinfo {author} {\bibfnamefont {Y.}~\bibnamefont {Miyauchi}},
  \bibinfo {author} {\bibfnamefont {M.}~\bibnamefont {Toh}}, \bibinfo {author}
  {\bibfnamefont {W.}~\bibnamefont {Zhao}}, \bibinfo {author} {\bibfnamefont
  {G.}~\bibnamefont {Eda}},\ and\ \bibinfo {author} {\bibfnamefont
  {K.}~\bibnamefont {Matsuda}},\ }\bibfield  {title} {\bibinfo {title}
  {Nonlinear photoluminescence in atomically thin layered \ce{WSe2} arising
  from diffusion-assisted exciton-exciton annihilation},\ }\href
  {https://doi.org/10.1103/PhysRevB.90.155449} {\bibfield  {journal} {\bibinfo
  {journal} {Physical Review B}\ }\textbf {\bibinfo {volume} {90}},\ \bibinfo
  {pages} {155449} (\bibinfo {year} {2014})}\BibitemShut {NoStop}%
\bibitem [{\citenamefont {Sun}\ \emph {et~al.}(2014)\citenamefont {Sun},
  \citenamefont {Rao}, \citenamefont {Reider}, \citenamefont {Chen},
  \citenamefont {You}, \citenamefont {Br{\'e}zin}, \citenamefont
  {Harutyunyan},\ and\ \citenamefont {Heinz}}]{sun2014observation}%
  \BibitemOpen
  \bibfield  {author} {\bibinfo {author} {\bibfnamefont {D.}~\bibnamefont
  {Sun}}, \bibinfo {author} {\bibfnamefont {Y.}~\bibnamefont {Rao}}, \bibinfo
  {author} {\bibfnamefont {G.~A.}\ \bibnamefont {Reider}}, \bibinfo {author}
  {\bibfnamefont {G.}~\bibnamefont {Chen}}, \bibinfo {author} {\bibfnamefont
  {Y.}~\bibnamefont {You}}, \bibinfo {author} {\bibfnamefont {L.}~\bibnamefont
  {Br{\'e}zin}}, \bibinfo {author} {\bibfnamefont {A.~R.}\ \bibnamefont
  {Harutyunyan}},\ and\ \bibinfo {author} {\bibfnamefont {T.~F.}\ \bibnamefont
  {Heinz}},\ }\bibfield  {title} {\bibinfo {title} {Observation of rapid
  exciton-exciton annihilation in monolayer molybdenum disulfide},\ }\href
  {https://doi.org/10.1021/nl5021975} {\bibfield  {journal} {\bibinfo
  {journal} {Nano Letters}\ }\textbf {\bibinfo {volume} {14}},\ \bibinfo
  {pages} {5625} (\bibinfo {year} {2014})}\BibitemShut {NoStop}%
\bibitem [{\citenamefont {Kumar}\ \emph {et~al.}(2014)\citenamefont {Kumar},
  \citenamefont {Cui}, \citenamefont {Ceballos}, \citenamefont {He},
  \citenamefont {Wang},\ and\ \citenamefont {Zhao}}]{kumar2014exciton}%
  \BibitemOpen
  \bibfield  {author} {\bibinfo {author} {\bibfnamefont {N.}~\bibnamefont
  {Kumar}}, \bibinfo {author} {\bibfnamefont {Q.}~\bibnamefont {Cui}}, \bibinfo
  {author} {\bibfnamefont {F.}~\bibnamefont {Ceballos}}, \bibinfo {author}
  {\bibfnamefont {D.}~\bibnamefont {He}}, \bibinfo {author} {\bibfnamefont
  {Y.}~\bibnamefont {Wang}},\ and\ \bibinfo {author} {\bibfnamefont
  {H.}~\bibnamefont {Zhao}},\ }\bibfield  {title} {\bibinfo {title}
  {Exciton-exciton annihilation in \ce{MoSe2} monolayers},\ }\href
  {https://doi.org/10.1103/PhysRevB.89.125427} {\bibfield  {journal} {\bibinfo
  {journal} {Physical Review B}\ }\textbf {\bibinfo {volume} {89}},\ \bibinfo
  {pages} {125427} (\bibinfo {year} {2014})}\BibitemShut {NoStop}%
\bibitem [{\citenamefont {Soavi}\ \emph {et~al.}(2016)\citenamefont {Soavi},
  \citenamefont {Dal~Conte}, \citenamefont {Manzoni}, \citenamefont {Viola},
  \citenamefont {Narita}, \citenamefont {Hu}, \citenamefont {Feng},
  \citenamefont {Hohenester}, \citenamefont {Molinari}, \citenamefont {Prezzi}
  \emph {et~al.}}]{soavi2016exciton}%
  \BibitemOpen
  \bibfield  {author} {\bibinfo {author} {\bibfnamefont {G.}~\bibnamefont
  {Soavi}}, \bibinfo {author} {\bibfnamefont {S.}~\bibnamefont {Dal~Conte}},
  \bibinfo {author} {\bibfnamefont {C.}~\bibnamefont {Manzoni}}, \bibinfo
  {author} {\bibfnamefont {D.}~\bibnamefont {Viola}}, \bibinfo {author}
  {\bibfnamefont {A.}~\bibnamefont {Narita}}, \bibinfo {author} {\bibfnamefont
  {Y.}~\bibnamefont {Hu}}, \bibinfo {author} {\bibfnamefont {X.}~\bibnamefont
  {Feng}}, \bibinfo {author} {\bibfnamefont {U.}~\bibnamefont {Hohenester}},
  \bibinfo {author} {\bibfnamefont {E.}~\bibnamefont {Molinari}}, \bibinfo
  {author} {\bibfnamefont {D.}~\bibnamefont {Prezzi}}, \emph {et~al.},\
  }\bibfield  {title} {\bibinfo {title} {Exciton-exciton annihilation and
  biexciton stimulated emission in graphene nanoribbons},\ }\href
  {https://doi.org/10.1038/ncomms11010} {\bibfield  {journal} {\bibinfo
  {journal} {Nature Communications}\ }\textbf {\bibinfo {volume} {7}},\
  \bibinfo {pages} {11010} (\bibinfo {year} {2016})}\BibitemShut {NoStop}%
\bibitem [{\citenamefont {Dostál}\ \emph {et~al.}(2018)\citenamefont
  {Dostál}, \citenamefont {Fennel}, \citenamefont {Koch}, \citenamefont
  {Herbst}, \citenamefont {Würthner},\ and\ \citenamefont
  {Brixner}}]{Dostal_2018}%
  \BibitemOpen
  \bibfield  {author} {\bibinfo {author} {\bibfnamefont {J.}~\bibnamefont
  {Dostál}}, \bibinfo {author} {\bibfnamefont {F.}~\bibnamefont {Fennel}},
  \bibinfo {author} {\bibfnamefont {F.}~\bibnamefont {Koch}}, \bibinfo {author}
  {\bibfnamefont {S.}~\bibnamefont {Herbst}}, \bibinfo {author} {\bibfnamefont
  {F.}~\bibnamefont {Würthner}},\ and\ \bibinfo {author} {\bibfnamefont
  {T.}~\bibnamefont {Brixner}},\ }\bibfield  {title} {\bibinfo {title} {Direct
  observation of exciton--exciton interactions},\ }\href
  {https://doi.org/10.1038/s41467-018-04884-4} {\bibfield  {journal} {\bibinfo
  {journal} {Nature Communications}\ }\textbf {\bibinfo {volume} {9}},\
  \bibinfo {pages} {2466} (\bibinfo {year} {2018})}\BibitemShut {NoStop}%
\bibitem [{\citenamefont {Mahmood}\ \emph {et~al.}(2018)\citenamefont
  {Mahmood}, \citenamefont {Alpichshev}, \citenamefont {Lee}, \citenamefont
  {Kong},\ and\ \citenamefont {Gedik}}]{mahmood2018observation}%
  \BibitemOpen
  \bibfield  {author} {\bibinfo {author} {\bibfnamefont {F.}~\bibnamefont
  {Mahmood}}, \bibinfo {author} {\bibfnamefont {Z.}~\bibnamefont {Alpichshev}},
  \bibinfo {author} {\bibfnamefont {Y.-H.}\ \bibnamefont {Lee}}, \bibinfo
  {author} {\bibfnamefont {J.}~\bibnamefont {Kong}},\ and\ \bibinfo {author}
  {\bibfnamefont {N.}~\bibnamefont {Gedik}},\ }\bibfield  {title} {\bibinfo
  {title} {Observation of exciton-exciton interaction mediated valley
  depolarization in monolayer \ce{MoSe2}},\ }\href
  {https://doi.org/10.1021/acs.nanolett.7b03953} {\bibfield  {journal}
  {\bibinfo  {journal} {Nano Letters}\ }\textbf {\bibinfo {volume} {18}},\
  \bibinfo {pages} {223} (\bibinfo {year} {2018})}\BibitemShut {NoStop}%
\bibitem [{\citenamefont {Purz}\ \emph {et~al.}(2021)\citenamefont {Purz},
  \citenamefont {Martin}, \citenamefont {Rivera}, \citenamefont {Holtzmann},
  \citenamefont {Xu},\ and\ \citenamefont {Cundiff}}]{purz2021coherent}%
  \BibitemOpen
  \bibfield  {author} {\bibinfo {author} {\bibfnamefont {T.~L.}\ \bibnamefont
  {Purz}}, \bibinfo {author} {\bibfnamefont {E.~W.}\ \bibnamefont {Martin}},
  \bibinfo {author} {\bibfnamefont {P.}~\bibnamefont {Rivera}}, \bibinfo
  {author} {\bibfnamefont {W.~G.}\ \bibnamefont {Holtzmann}}, \bibinfo {author}
  {\bibfnamefont {X.}~\bibnamefont {Xu}},\ and\ \bibinfo {author}
  {\bibfnamefont {S.~T.}\ \bibnamefont {Cundiff}},\ }\bibfield  {title}
  {\bibinfo {title} {Coherent exciton-exciton interactions and exciton dynamics
  in a \ce{MoSe2}/\ce{WSe2} heterostructure},\ }\href
  {https://doi.org/10.1103/PhysRevB.104.L241302} {\bibfield  {journal}
  {\bibinfo  {journal} {Physical Review B}\ }\textbf {\bibinfo {volume}
  {104}},\ \bibinfo {pages} {L241302} (\bibinfo {year} {2021})}\BibitemShut
  {NoStop}%
\bibitem [{\citenamefont {Birkmeier}\ \emph {et~al.}(2022)\citenamefont
  {Birkmeier}, \citenamefont {Hertel},\ and\ \citenamefont
  {Hartschuh}}]{Birkmeier_2022}%
  \BibitemOpen
  \bibfield  {author} {\bibinfo {author} {\bibfnamefont {K.}~\bibnamefont
  {Birkmeier}}, \bibinfo {author} {\bibfnamefont {T.}~\bibnamefont {Hertel}},\
  and\ \bibinfo {author} {\bibfnamefont {A.}~\bibnamefont {Hartschuh}},\
  }\bibfield  {title} {\bibinfo {title} {Probing the ultrafast dynamics of
  excitons in single semiconducting carbon nanotubes},\ }\href
  {https://doi.org/10.1038/s41467-022-33941-2} {\bibfield  {journal} {\bibinfo
  {journal} {Nature Communications}\ }\textbf {\bibinfo {volume} {13}},\
  \bibinfo {pages} {6290} (\bibinfo {year} {2022})}\BibitemShut {NoStop}%
\bibitem [{\citenamefont {Steinhoff}\ \emph {et~al.}(2024)\citenamefont
  {Steinhoff}, \citenamefont {Wietek}, \citenamefont {Florian}, \citenamefont
  {Schulz}, \citenamefont {Taniguchi}, \citenamefont {Watanabe}, \citenamefont
  {Zhao}, \citenamefont {H{\"o}gele}, \citenamefont {Jahnke},\ and\
  \citenamefont {Chernikov}}]{steinhoff2024exciton}%
  \BibitemOpen
  \bibfield  {author} {\bibinfo {author} {\bibfnamefont {A.}~\bibnamefont
  {Steinhoff}}, \bibinfo {author} {\bibfnamefont {E.}~\bibnamefont {Wietek}},
  \bibinfo {author} {\bibfnamefont {M.}~\bibnamefont {Florian}}, \bibinfo
  {author} {\bibfnamefont {T.}~\bibnamefont {Schulz}}, \bibinfo {author}
  {\bibfnamefont {T.}~\bibnamefont {Taniguchi}}, \bibinfo {author}
  {\bibfnamefont {K.}~\bibnamefont {Watanabe}}, \bibinfo {author}
  {\bibfnamefont {S.}~\bibnamefont {Zhao}}, \bibinfo {author} {\bibfnamefont
  {A.}~\bibnamefont {H{\"o}gele}}, \bibinfo {author} {\bibfnamefont
  {F.}~\bibnamefont {Jahnke}},\ and\ \bibinfo {author} {\bibfnamefont
  {A.}~\bibnamefont {Chernikov}},\ }\bibfield  {title} {\bibinfo {title}
  {Exciton-exciton interactions in van der {W}aals heterobilayers},\ }\href
  {https://doi.org/10.1103/PhysRevX.14.031025} {\bibfield  {journal} {\bibinfo
  {journal} {Physical Review X}\ }\textbf {\bibinfo {volume} {14}},\ \bibinfo
  {pages} {031025} (\bibinfo {year} {2024})}\BibitemShut {NoStop}%
\bibitem [{\citenamefont {B{\'a}nyai}\ \emph {et~al.}(1987)\citenamefont
  {B{\'a}nyai}, \citenamefont {Galbraith}, \citenamefont {Ell},\ and\
  \citenamefont {Haug}}]{banyai1987excitons}%
  \BibitemOpen
  \bibfield  {author} {\bibinfo {author} {\bibfnamefont {L.}~\bibnamefont
  {B{\'a}nyai}}, \bibinfo {author} {\bibfnamefont {I.}~\bibnamefont
  {Galbraith}}, \bibinfo {author} {\bibfnamefont {C.}~\bibnamefont {Ell}},\
  and\ \bibinfo {author} {\bibfnamefont {H.}~\bibnamefont {Haug}},\ }\bibfield
  {title} {\bibinfo {title} {Excitons and biexcitons in semiconductor quantum
  wires},\ }\href {https://doi.org/10.1103/PhysRevB.36.6099} {\bibfield
  {journal} {\bibinfo  {journal} {Physical Review B}\ }\textbf {\bibinfo
  {volume} {36}},\ \bibinfo {pages} {6099} (\bibinfo {year}
  {1987})}\BibitemShut {NoStop}%
\bibitem [{\citenamefont {Bacher}\ \emph {et~al.}(1999)\citenamefont {Bacher},
  \citenamefont {Weigand}, \citenamefont {Seufert}, \citenamefont
  {Kulakovskii}, \citenamefont {Gippius}, \citenamefont {Forchel},
  \citenamefont {Leonardi},\ and\ \citenamefont
  {Hommel}}]{bacher1999biexciton}%
  \BibitemOpen
  \bibfield  {author} {\bibinfo {author} {\bibfnamefont {G.}~\bibnamefont
  {Bacher}}, \bibinfo {author} {\bibfnamefont {R.}~\bibnamefont {Weigand}},
  \bibinfo {author} {\bibfnamefont {J.}~\bibnamefont {Seufert}}, \bibinfo
  {author} {\bibfnamefont {V.}~\bibnamefont {Kulakovskii}}, \bibinfo {author}
  {\bibfnamefont {N.}~\bibnamefont {Gippius}}, \bibinfo {author} {\bibfnamefont
  {A.}~\bibnamefont {Forchel}}, \bibinfo {author} {\bibfnamefont
  {K.}~\bibnamefont {Leonardi}},\ and\ \bibinfo {author} {\bibfnamefont
  {D.}~\bibnamefont {Hommel}},\ }\bibfield  {title} {\bibinfo {title}
  {Biexciton versus exciton lifetime in a single semiconductor quantum dot},\
  }\href {https://doi.org/10.1103/PhysRevLett.83.4417} {\bibfield  {journal}
  {\bibinfo  {journal} {Physical Review Letters}\ }\textbf {\bibinfo {volume}
  {83}},\ \bibinfo {pages} {4417} (\bibinfo {year} {1999})}\BibitemShut
  {NoStop}%
\bibitem [{\citenamefont {Vonk}\ \emph {et~al.}(2021)\citenamefont {Vonk},
  \citenamefont {Heemskerk}, \citenamefont {Keitel}, \citenamefont
  {Hinterding}, \citenamefont {Geuchies}, \citenamefont {Houtepen},\ and\
  \citenamefont {Rabouw}}]{vonk2021biexciton}%
  \BibitemOpen
  \bibfield  {author} {\bibinfo {author} {\bibfnamefont {S.~J.}\ \bibnamefont
  {Vonk}}, \bibinfo {author} {\bibfnamefont {B.~A.}\ \bibnamefont {Heemskerk}},
  \bibinfo {author} {\bibfnamefont {R.~C.}\ \bibnamefont {Keitel}}, \bibinfo
  {author} {\bibfnamefont {S.~O.}\ \bibnamefont {Hinterding}}, \bibinfo
  {author} {\bibfnamefont {J.~J.}\ \bibnamefont {Geuchies}}, \bibinfo {author}
  {\bibfnamefont {A.~J.}\ \bibnamefont {Houtepen}},\ and\ \bibinfo {author}
  {\bibfnamefont {F.~T.}\ \bibnamefont {Rabouw}},\ }\bibfield  {title}
  {\bibinfo {title} {Biexciton binding energy and line width of single quantum
  dots at room temperature},\ }\href
  {https://doi.org/10.1021/acs.nanolett.1c01556} {\bibfield  {journal}
  {\bibinfo  {journal} {Nano Letters}\ }\textbf {\bibinfo {volume} {21}},\
  \bibinfo {pages} {5760} (\bibinfo {year} {2021})}\BibitemShut {NoStop}%
\bibitem [{\citenamefont {Sun}\ \emph {et~al.}(2022)\citenamefont {Sun},
  \citenamefont {Cavanaugh}, \citenamefont {Jen-La~Plante}, \citenamefont
  {Ippen}, \citenamefont {Bautista}, \citenamefont {Ma},\ and\ \citenamefont
  {Kelley}}]{sun2022biexciton}%
  \BibitemOpen
  \bibfield  {author} {\bibinfo {author} {\bibfnamefont {H.}~\bibnamefont
  {Sun}}, \bibinfo {author} {\bibfnamefont {P.}~\bibnamefont {Cavanaugh}},
  \bibinfo {author} {\bibfnamefont {I.}~\bibnamefont {Jen-La~Plante}}, \bibinfo
  {author} {\bibfnamefont {C.}~\bibnamefont {Ippen}}, \bibinfo {author}
  {\bibfnamefont {M.}~\bibnamefont {Bautista}}, \bibinfo {author}
  {\bibfnamefont {R.}~\bibnamefont {Ma}},\ and\ \bibinfo {author}
  {\bibfnamefont {D.~F.}\ \bibnamefont {Kelley}},\ }\bibfield  {title}
  {\bibinfo {title} {Biexciton and trion dynamics in
  \ce{InP}/\ce{ZnSe}/\ce{ZnS} quantum dots},\ }\href
  {https://doi.org/10.1063/5.0082223} {\bibfield  {journal} {\bibinfo
  {journal} {The Journal of Chemical Physics}\ }\textbf {\bibinfo {volume}
  {156}},\ \bibinfo {pages} {054703} (\bibinfo {year} {2022})}\BibitemShut
  {NoStop}%
\bibitem [{\citenamefont {Diroll}\ \emph {et~al.}(2023)\citenamefont {Diroll},
  \citenamefont {Hua}, \citenamefont {Guzelturk}, \citenamefont {P{\'a}lmai},\
  and\ \citenamefont {Tomczak}}]{diroll2023long}%
  \BibitemOpen
  \bibfield  {author} {\bibinfo {author} {\bibfnamefont {B.~T.}\ \bibnamefont
  {Diroll}}, \bibinfo {author} {\bibfnamefont {M.}~\bibnamefont {Hua}},
  \bibinfo {author} {\bibfnamefont {B.}~\bibnamefont {Guzelturk}}, \bibinfo
  {author} {\bibfnamefont {M.}~\bibnamefont {P{\'a}lmai}},\ and\ \bibinfo
  {author} {\bibfnamefont {K.}~\bibnamefont {Tomczak}},\ }\bibfield  {title}
  {\bibinfo {title} {Long-lived and bright biexcitons in quantum dots with
  parabolic band potentials},\ }\href
  {https://doi.org/10.1021/acs.nanolett.3c04361} {\bibfield  {journal}
  {\bibinfo  {journal} {Nano Letters}\ }\textbf {\bibinfo {volume} {23}},\
  \bibinfo {pages} {11975} (\bibinfo {year} {2023})}\BibitemShut {NoStop}%
\bibitem [{\citenamefont {Huang}\ \emph {et~al.}(2023)\citenamefont {Huang},
  \citenamefont {Sun}, \citenamefont {Lei}, \citenamefont {Zhang},
  \citenamefont {Qin},\ and\ \citenamefont {Zhong}}]{huang2023nonlocal}%
  \BibitemOpen
  \bibfield  {author} {\bibinfo {author} {\bibfnamefont {P.}~\bibnamefont
  {Huang}}, \bibinfo {author} {\bibfnamefont {S.}~\bibnamefont {Sun}}, \bibinfo
  {author} {\bibfnamefont {H.}~\bibnamefont {Lei}}, \bibinfo {author}
  {\bibfnamefont {Y.}~\bibnamefont {Zhang}}, \bibinfo {author} {\bibfnamefont
  {H.}~\bibnamefont {Qin}},\ and\ \bibinfo {author} {\bibfnamefont
  {H.}~\bibnamefont {Zhong}},\ }\bibfield  {title} {\bibinfo {title} {Nonlocal
  interaction enhanced biexciton emission in large \ce{CsPbBr3} nanocrystals},\
  }\href {https://doi.org/10.1186/s43593-023-00045-3} {\bibfield  {journal}
  {\bibinfo  {journal} {elight}\ }\textbf {\bibinfo {volume} {3}},\ \bibinfo
  {pages} {10} (\bibinfo {year} {2023})}\BibitemShut {NoStop}%
\bibitem [{\citenamefont {You}\ \emph {et~al.}(2015)\citenamefont {You},
  \citenamefont {Zhang}, \citenamefont {Berkelbach}, \citenamefont {Hybertsen},
  \citenamefont {Reichman},\ and\ \citenamefont {Heinz}}]{you2015observation}%
  \BibitemOpen
  \bibfield  {author} {\bibinfo {author} {\bibfnamefont {Y.}~\bibnamefont
  {You}}, \bibinfo {author} {\bibfnamefont {X.-X.}\ \bibnamefont {Zhang}},
  \bibinfo {author} {\bibfnamefont {T.~C.}\ \bibnamefont {Berkelbach}},
  \bibinfo {author} {\bibfnamefont {M.~S.}\ \bibnamefont {Hybertsen}}, \bibinfo
  {author} {\bibfnamefont {D.~R.}\ \bibnamefont {Reichman}},\ and\ \bibinfo
  {author} {\bibfnamefont {T.~F.}\ \bibnamefont {Heinz}},\ }\bibfield  {title}
  {\bibinfo {title} {Observation of biexcitons in monolayer \ce{WSe2}},\ }\href
  {https://doi.org/https://www.nature.com/articles/nphys3324} {\bibfield
  {journal} {\bibinfo  {journal} {Nature Physics}\ }\textbf {\bibinfo {volume}
  {11}},\ \bibinfo {pages} {477} (\bibinfo {year} {2015})}\BibitemShut
  {NoStop}%
\bibitem [{\citenamefont {Wang}\ \emph {et~al.}(2019)\citenamefont {Wang},
  \citenamefont {Sui}, \citenamefont {Ni}, \citenamefont {Chi}, \citenamefont
  {Pan}, \citenamefont {Zhang}, \citenamefont {Kang}, \citenamefont {Zhou},\
  and\ \citenamefont {Wang}}]{wang2019studying}%
  \BibitemOpen
  \bibfield  {author} {\bibinfo {author} {\bibfnamefont {W.}~\bibnamefont
  {Wang}}, \bibinfo {author} {\bibfnamefont {N.}~\bibnamefont {Sui}}, \bibinfo
  {author} {\bibfnamefont {M.}~\bibnamefont {Ni}}, \bibinfo {author}
  {\bibfnamefont {X.}~\bibnamefont {Chi}}, \bibinfo {author} {\bibfnamefont
  {L.}~\bibnamefont {Pan}}, \bibinfo {author} {\bibfnamefont {H.}~\bibnamefont
  {Zhang}}, \bibinfo {author} {\bibfnamefont {Z.}~\bibnamefont {Kang}},
  \bibinfo {author} {\bibfnamefont {Q.}~\bibnamefont {Zhou}},\ and\ \bibinfo
  {author} {\bibfnamefont {Y.}~\bibnamefont {Wang}},\ }\bibfield  {title}
  {\bibinfo {title} {Studying of the biexciton characteristics in monolayer
  \ce{MoS2}},\ }\href {https://doi.org/10.1021/acs.jpcc.9b11582} {\bibfield
  {journal} {\bibinfo  {journal} {The Journal of Physical Chemistry C}\
  }\textbf {\bibinfo {volume} {124}},\ \bibinfo {pages} {1749} (\bibinfo {year}
  {2019})}\BibitemShut {NoStop}%
\bibitem [{\citenamefont {Pei}\ \emph {et~al.}(2017)\citenamefont {Pei},
  \citenamefont {Yang}, \citenamefont {Wang}, \citenamefont {Wang},
  \citenamefont {Mokkapati}, \citenamefont {Lu}, \citenamefont {Zheng},
  \citenamefont {Qin}, \citenamefont {Neshev}, \citenamefont {Tan} \emph
  {et~al.}}]{pei2017excited}%
  \BibitemOpen
  \bibfield  {author} {\bibinfo {author} {\bibfnamefont {J.}~\bibnamefont
  {Pei}}, \bibinfo {author} {\bibfnamefont {J.}~\bibnamefont {Yang}}, \bibinfo
  {author} {\bibfnamefont {X.}~\bibnamefont {Wang}}, \bibinfo {author}
  {\bibfnamefont {F.}~\bibnamefont {Wang}}, \bibinfo {author} {\bibfnamefont
  {S.}~\bibnamefont {Mokkapati}}, \bibinfo {author} {\bibfnamefont
  {T.}~\bibnamefont {Lu}}, \bibinfo {author} {\bibfnamefont {J.-C.}\
  \bibnamefont {Zheng}}, \bibinfo {author} {\bibfnamefont {Q.}~\bibnamefont
  {Qin}}, \bibinfo {author} {\bibfnamefont {D.}~\bibnamefont {Neshev}},
  \bibinfo {author} {\bibfnamefont {H.~H.}\ \bibnamefont {Tan}}, \emph
  {et~al.},\ }\bibfield  {title} {\bibinfo {title} {Excited state biexcitons in
  atomically thin \ce{MoSe2}},\ }\href
  {https://doi.org/10.1021/acsnano.7b03909} {\bibfield  {journal} {\bibinfo
  {journal} {ACS Nano}\ }\textbf {\bibinfo {volume} {11}},\ \bibinfo {pages}
  {7468} (\bibinfo {year} {2017})}\BibitemShut {NoStop}%
\bibitem [{\citenamefont {Steinhoff}\ \emph {et~al.}(2018)\citenamefont
  {Steinhoff}, \citenamefont {Florian}, \citenamefont {Singh}, \citenamefont
  {Tran}, \citenamefont {Kolarczik}, \citenamefont {Helmrich}, \citenamefont
  {Achtstein}, \citenamefont {Woggon}, \citenamefont {Owschimikow},
  \citenamefont {Jahnke} \emph {et~al.}}]{steinhoff2018biexciton}%
  \BibitemOpen
  \bibfield  {author} {\bibinfo {author} {\bibfnamefont {A.}~\bibnamefont
  {Steinhoff}}, \bibinfo {author} {\bibfnamefont {M.}~\bibnamefont {Florian}},
  \bibinfo {author} {\bibfnamefont {A.}~\bibnamefont {Singh}}, \bibinfo
  {author} {\bibfnamefont {K.}~\bibnamefont {Tran}}, \bibinfo {author}
  {\bibfnamefont {M.}~\bibnamefont {Kolarczik}}, \bibinfo {author}
  {\bibfnamefont {S.}~\bibnamefont {Helmrich}}, \bibinfo {author}
  {\bibfnamefont {A.~W.}\ \bibnamefont {Achtstein}}, \bibinfo {author}
  {\bibfnamefont {U.}~\bibnamefont {Woggon}}, \bibinfo {author} {\bibfnamefont
  {N.}~\bibnamefont {Owschimikow}}, \bibinfo {author} {\bibfnamefont
  {F.}~\bibnamefont {Jahnke}}, \emph {et~al.},\ }\bibfield  {title} {\bibinfo
  {title} {Biexciton fine structure in monolayer transition metal
  dichalcogenides},\ }\href {https://doi.org/10.1038/s41567-018-0282-x}
  {\bibfield  {journal} {\bibinfo  {journal} {Nature Physics}\ }\textbf
  {\bibinfo {volume} {14}},\ \bibinfo {pages} {1199} (\bibinfo {year}
  {2018})}\BibitemShut {NoStop}%
\bibitem [{\citenamefont {Conway}\ \emph {et~al.}(2022)\citenamefont {Conway},
  \citenamefont {Muir}, \citenamefont {Earl}, \citenamefont {Wurdack},
  \citenamefont {Mishra}, \citenamefont {Tollerud},\ and\ \citenamefont
  {Davis}}]{conway2022direct}%
  \BibitemOpen
  \bibfield  {author} {\bibinfo {author} {\bibfnamefont {M.}~\bibnamefont
  {Conway}}, \bibinfo {author} {\bibfnamefont {J.}~\bibnamefont {Muir}},
  \bibinfo {author} {\bibfnamefont {S.}~\bibnamefont {Earl}}, \bibinfo {author}
  {\bibfnamefont {M.}~\bibnamefont {Wurdack}}, \bibinfo {author} {\bibfnamefont
  {R.}~\bibnamefont {Mishra}}, \bibinfo {author} {\bibfnamefont
  {J.}~\bibnamefont {Tollerud}},\ and\ \bibinfo {author} {\bibfnamefont
  {J.}~\bibnamefont {Davis}},\ }\bibfield  {title} {\bibinfo {title} {Direct
  measurement of biexcitons in monolayer \ce{WS2}},\ }\href
  {https://doi.org/10.1088/2053-1583/ac4779} {\bibfield  {journal} {\bibinfo
  {journal} {2D Materials}\ }\textbf {\bibinfo {volume} {9}},\ \bibinfo {pages}
  {021001} (\bibinfo {year} {2022})}\BibitemShut {NoStop}%
\bibitem [{\citenamefont {Sharma}\ \emph {et~al.}(2022)\citenamefont {Sharma},
  \citenamefont {Zhu}, \citenamefont {Halbich}, \citenamefont {Sun},
  \citenamefont {Zhang}, \citenamefont {Wang},\ and\ \citenamefont
  {Lu}}]{sharma2022engineering}%
  \BibitemOpen
  \bibfield  {author} {\bibinfo {author} {\bibfnamefont {A.}~\bibnamefont
  {Sharma}}, \bibinfo {author} {\bibfnamefont {Y.}~\bibnamefont {Zhu}},
  \bibinfo {author} {\bibfnamefont {R.}~\bibnamefont {Halbich}}, \bibinfo
  {author} {\bibfnamefont {X.}~\bibnamefont {Sun}}, \bibinfo {author}
  {\bibfnamefont {L.}~\bibnamefont {Zhang}}, \bibinfo {author} {\bibfnamefont
  {B.}~\bibnamefont {Wang}},\ and\ \bibinfo {author} {\bibfnamefont
  {Y.}~\bibnamefont {Lu}},\ }\bibfield  {title} {\bibinfo {title} {Engineering
  the dynamics and transport of excitons, trions, and biexcitons in monolayer
  \ce{WS2}},\ }\href {https://doi.org/10.1021/acsami.2c08199} {\bibfield
  {journal} {\bibinfo  {journal} {ACS Applied Materials \& Interfaces}\
  }\textbf {\bibinfo {volume} {14}},\ \bibinfo {pages} {41165} (\bibinfo {year}
  {2022})}\BibitemShut {NoStop}%
\bibitem [{\citenamefont {Kunneman}\ \emph {et~al.}(2014)\citenamefont
  {Kunneman}, \citenamefont {Schins}, \citenamefont {Pedetti}, \citenamefont
  {Heuclin}, \citenamefont {Grozema}, \citenamefont {Houtepen}, \citenamefont
  {Dubertret},\ and\ \citenamefont {Siebbeles}}]{kunneman2014nature}%
  \BibitemOpen
  \bibfield  {author} {\bibinfo {author} {\bibfnamefont {L.~T.}\ \bibnamefont
  {Kunneman}}, \bibinfo {author} {\bibfnamefont {J.~M.}\ \bibnamefont
  {Schins}}, \bibinfo {author} {\bibfnamefont {S.}~\bibnamefont {Pedetti}},
  \bibinfo {author} {\bibfnamefont {H.}~\bibnamefont {Heuclin}}, \bibinfo
  {author} {\bibfnamefont {F.~C.}\ \bibnamefont {Grozema}}, \bibinfo {author}
  {\bibfnamefont {A.~J.}\ \bibnamefont {Houtepen}}, \bibinfo {author}
  {\bibfnamefont {B.}~\bibnamefont {Dubertret}},\ and\ \bibinfo {author}
  {\bibfnamefont {L.~D.}\ \bibnamefont {Siebbeles}},\ }\bibfield  {title}
  {\bibinfo {title} {Nature and decay pathways of photoexcited states in
  \ce{CdSe} and \ce{CdSe}/\ce{CdS} nanoplatelets},\ }\href
  {https://doi.org/10.1021/nl503406a} {\bibfield  {journal} {\bibinfo
  {journal} {Nano Letters}\ }\textbf {\bibinfo {volume} {14}},\ \bibinfo
  {pages} {7039} (\bibinfo {year} {2014})}\BibitemShut {NoStop}%
\bibitem [{\citenamefont {Peng}\ \emph {et~al.}(2021)\citenamefont {Peng},
  \citenamefont {Cho}, \citenamefont {Zhang}, \citenamefont {Talapin},\ and\
  \citenamefont {Ma}}]{peng2021observation}%
  \BibitemOpen
  \bibfield  {author} {\bibinfo {author} {\bibfnamefont {L.}~\bibnamefont
  {Peng}}, \bibinfo {author} {\bibfnamefont {W.}~\bibnamefont {Cho}}, \bibinfo
  {author} {\bibfnamefont {X.}~\bibnamefont {Zhang}}, \bibinfo {author}
  {\bibfnamefont {D.}~\bibnamefont {Talapin}},\ and\ \bibinfo {author}
  {\bibfnamefont {X.}~\bibnamefont {Ma}},\ }\bibfield  {title} {\bibinfo
  {title} {Observation of biexciton emission from single semiconductor
  nanoplatelets},\ }\href {https://doi.org/10.1103/PhysRevMaterials.5.L051601}
  {\bibfield  {journal} {\bibinfo  {journal} {Physical Review Materials}\
  }\textbf {\bibinfo {volume} {5}},\ \bibinfo {pages} {L051601} (\bibinfo
  {year} {2021})}\BibitemShut {NoStop}%
\bibitem [{\citenamefont {Thouin}\ \emph {et~al.}(2018)\citenamefont {Thouin},
  \citenamefont {Neutzner}, \citenamefont {Cortecchia}, \citenamefont
  {Dragomir}, \citenamefont {Soci}, \citenamefont {Salim}, \citenamefont {Lam},
  \citenamefont {Leonelli}, \citenamefont {Petrozza}, \citenamefont {Kandada}
  \emph {et~al.}}]{thouin2018stable}%
  \BibitemOpen
  \bibfield  {author} {\bibinfo {author} {\bibfnamefont {F.}~\bibnamefont
  {Thouin}}, \bibinfo {author} {\bibfnamefont {S.}~\bibnamefont {Neutzner}},
  \bibinfo {author} {\bibfnamefont {D.}~\bibnamefont {Cortecchia}}, \bibinfo
  {author} {\bibfnamefont {V.~A.}\ \bibnamefont {Dragomir}}, \bibinfo {author}
  {\bibfnamefont {C.}~\bibnamefont {Soci}}, \bibinfo {author} {\bibfnamefont
  {T.}~\bibnamefont {Salim}}, \bibinfo {author} {\bibfnamefont {Y.~M.}\
  \bibnamefont {Lam}}, \bibinfo {author} {\bibfnamefont {R.}~\bibnamefont
  {Leonelli}}, \bibinfo {author} {\bibfnamefont {A.}~\bibnamefont {Petrozza}},
  \bibinfo {author} {\bibfnamefont {A.~R.~S.}\ \bibnamefont {Kandada}}, \emph
  {et~al.},\ }\bibfield  {title} {\bibinfo {title} {Stable biexcitons in
  two-dimensional metal-halide perovskites with strong dynamic lattice
  disorder},\ }\href {https://doi.org/10.1103/PhysRevMaterials.2.034001}
  {\bibfield  {journal} {\bibinfo  {journal} {Physical Review Materials}\
  }\textbf {\bibinfo {volume} {2}},\ \bibinfo {pages} {034001} (\bibinfo {year}
  {2018})}\BibitemShut {NoStop}%
\bibitem [{\citenamefont {Cho}\ \emph {et~al.}(2024)\citenamefont {Cho},
  \citenamefont {Sato}, \citenamefont {Yamada}, \citenamefont {Sato},
  \citenamefont {Saruyama}, \citenamefont {Teranishi}, \citenamefont
  {Suzuura},\ and\ \citenamefont {Kanemitsu}}]{cho2024size}%
  \BibitemOpen
  \bibfield  {author} {\bibinfo {author} {\bibfnamefont {K.}~\bibnamefont
  {Cho}}, \bibinfo {author} {\bibfnamefont {T.}~\bibnamefont {Sato}}, \bibinfo
  {author} {\bibfnamefont {T.}~\bibnamefont {Yamada}}, \bibinfo {author}
  {\bibfnamefont {R.}~\bibnamefont {Sato}}, \bibinfo {author} {\bibfnamefont
  {M.}~\bibnamefont {Saruyama}}, \bibinfo {author} {\bibfnamefont
  {T.}~\bibnamefont {Teranishi}}, \bibinfo {author} {\bibfnamefont
  {H.}~\bibnamefont {Suzuura}},\ and\ \bibinfo {author} {\bibfnamefont
  {Y.}~\bibnamefont {Kanemitsu}},\ }\bibfield  {title} {\bibinfo {title} {Size
  dependence of trion and biexciton binding energies in lead halide perovskite
  nanocrystals},\ }\href {https://doi.org/10.1021/acsnano.3c11842} {\bibfield
  {journal} {\bibinfo  {journal} {ACS Nano}\ }\textbf {\bibinfo {volume}
  {18}},\ \bibinfo {pages} {5723} (\bibinfo {year} {2024})}\BibitemShut
  {NoStop}%
\bibitem [{\citenamefont {Inoue}\ \emph {et~al.}(1998)\citenamefont {Inoue},
  \citenamefont {Brandes},\ and\ \citenamefont {Shimizu}}]{Inoue_1998}%
  \BibitemOpen
  \bibfield  {author} {\bibinfo {author} {\bibfnamefont {J.-I.}\ \bibnamefont
  {Inoue}}, \bibinfo {author} {\bibfnamefont {T.}~\bibnamefont {Brandes}},\
  and\ \bibinfo {author} {\bibfnamefont {A.}~\bibnamefont {Shimizu}},\
  }\bibfield  {title} {\bibinfo {title} {Effective {H}amiltonian for excitons
  with spin degrees of freedom},\ }\href {https://doi.org/10.1143/jpsj.67.3384}
  {\bibfield  {journal} {\bibinfo  {journal} {Journal of the Physical Society
  of Japan}\ }\textbf {\bibinfo {volume} {67}},\ \bibinfo {pages} {3384}
  (\bibinfo {year} {1998})}\BibitemShut {NoStop}%
\bibitem [{\citenamefont {Elistratov}\ and\ \citenamefont
  {Lozovik}(2016)}]{Elistratov_2016}%
  \BibitemOpen
  \bibfield  {author} {\bibinfo {author} {\bibfnamefont {A.~A.}\ \bibnamefont
  {Elistratov}}\ and\ \bibinfo {author} {\bibfnamefont {Y.~E.}\ \bibnamefont
  {Lozovik}},\ }\bibfield  {title} {\bibinfo {title} {Coupled exciton-photon
  {B}ose condensate in path integral formalism},\ }\href
  {https://doi.org/10.1103/PhysRevB.93.104530} {\bibfield  {journal} {\bibinfo
  {journal} {Physical Review B}\ }\textbf {\bibinfo {volume} {93}},\ \bibinfo
  {pages} {104530} (\bibinfo {year} {2016})}\BibitemShut {NoStop}%
\bibitem [{\citenamefont {Berman}\ and\ \citenamefont
  {Kezerashvili}(2017)}]{Berman_2017}%
  \BibitemOpen
  \bibfield  {author} {\bibinfo {author} {\bibfnamefont {O.~L.}\ \bibnamefont
  {Berman}}\ and\ \bibinfo {author} {\bibfnamefont {R.~Y.}\ \bibnamefont
  {Kezerashvili}},\ }\bibfield  {title} {\bibinfo {title} {Superfluidity of
  dipolar excitons in a transition metal dichalcogenide double layer},\ }\href
  {https://doi.org/10.1103/PhysRevB.96.094502} {\bibfield  {journal} {\bibinfo
  {journal} {Physical Review B}\ }\textbf {\bibinfo {volume} {96}},\ \bibinfo
  {pages} {094502} (\bibinfo {year} {2017})}\BibitemShut {NoStop}%
\bibitem [{\citenamefont {Ben-Tabou~de Leon}\ and\ \citenamefont
  {Laikhtman}(2001)}]{deLeon_2001}%
  \BibitemOpen
  \bibfield  {author} {\bibinfo {author} {\bibfnamefont {S.}~\bibnamefont
  {Ben-Tabou~de Leon}}\ and\ \bibinfo {author} {\bibfnamefont {B.}~\bibnamefont
  {Laikhtman}},\ }\bibfield  {title} {\bibinfo {title} {Exciton-exciton
  interactions in quantum wells: {O}ptical properties and energy and spin
  relaxation},\ }\href {https://doi.org/10.1103/PhysRevB.63.125306} {\bibfield
  {journal} {\bibinfo  {journal} {Physical Review B}\ }\textbf {\bibinfo
  {volume} {63}},\ \bibinfo {pages} {125306} (\bibinfo {year}
  {2001})}\BibitemShut {NoStop}%
\bibitem [{\citenamefont {Thilagam}(2001)}]{Thilagam_2001}%
  \BibitemOpen
  \bibfield  {author} {\bibinfo {author} {\bibfnamefont {A.}~\bibnamefont
  {Thilagam}},\ }\bibfield  {title} {\bibinfo {title} {Exciton-exciton
  interaction in semiconductor quantum wells},\ }\href
  {https://doi.org/10.1103/PhysRevB.63.045321} {\bibfield  {journal} {\bibinfo
  {journal} {Physical Review B}\ }\textbf {\bibinfo {volume} {63}},\ \bibinfo
  {pages} {045321} (\bibinfo {year} {2001})}\BibitemShut {NoStop}%
\bibitem [{\citenamefont {Shahnazaryan}\ \emph {et~al.}(2017)\citenamefont
  {Shahnazaryan}, \citenamefont {Iorsh}, \citenamefont {Shelykh},\ and\
  \citenamefont {Kyriienko}}]{Shahnazaryan_2017}%
  \BibitemOpen
  \bibfield  {author} {\bibinfo {author} {\bibfnamefont {V.}~\bibnamefont
  {Shahnazaryan}}, \bibinfo {author} {\bibfnamefont {I.}~\bibnamefont {Iorsh}},
  \bibinfo {author} {\bibfnamefont {I.~A.}\ \bibnamefont {Shelykh}},\ and\
  \bibinfo {author} {\bibfnamefont {O.}~\bibnamefont {Kyriienko}},\ }\bibfield
  {title} {\bibinfo {title} {Exciton-exciton interaction in transition-metal
  dichalcogenide monolayers},\ }\href
  {https://doi.org/10.1103/PhysRevB.96.115409} {\bibfield  {journal} {\bibinfo
  {journal} {Physical Review B}\ }\textbf {\bibinfo {volume} {96}},\ \bibinfo
  {pages} {115409} (\bibinfo {year} {2017})}\BibitemShut {NoStop}%
\bibitem [{\citenamefont {Katsch}\ \emph {et~al.}(2018)\citenamefont {Katsch},
  \citenamefont {Selig}, \citenamefont {Carmele},\ and\ \citenamefont
  {Knorr}}]{Katsch_2018}%
  \BibitemOpen
  \bibfield  {author} {\bibinfo {author} {\bibfnamefont {F.}~\bibnamefont
  {Katsch}}, \bibinfo {author} {\bibfnamefont {M.}~\bibnamefont {Selig}},
  \bibinfo {author} {\bibfnamefont {A.}~\bibnamefont {Carmele}},\ and\ \bibinfo
  {author} {\bibfnamefont {A.}~\bibnamefont {Knorr}},\ }\bibfield  {title}
  {\bibinfo {title} {Theory of exciton-exciton interactions in monolayer
  transition metal dichalcogenides},\ }\href
  {https://doi.org/10.1002/pssb.201800185} {\bibfield  {journal} {\bibinfo
  {journal} {Physica Status Solidi (b)}\ }\textbf {\bibinfo {volume} {255}},\
  \bibinfo {pages} {1800185} (\bibinfo {year} {2018})}\BibitemShut {NoStop}%
\bibitem [{\citenamefont {Gribakin}\ \emph {et~al.}(2021)\citenamefont
  {Gribakin}, \citenamefont {Khramtsov}, \citenamefont {Trifonov},\ and\
  \citenamefont {Ignatiev}}]{Gribakin_2021}%
  \BibitemOpen
  \bibfield  {author} {\bibinfo {author} {\bibfnamefont {B.~F.}\ \bibnamefont
  {Gribakin}}, \bibinfo {author} {\bibfnamefont {E.~S.}\ \bibnamefont
  {Khramtsov}}, \bibinfo {author} {\bibfnamefont {A.~V.}\ \bibnamefont
  {Trifonov}},\ and\ \bibinfo {author} {\bibfnamefont {I.~V.}\ \bibnamefont
  {Ignatiev}},\ }\bibfield  {title} {\bibinfo {title} {Exciton-exciton and
  exciton--charge carrier interaction and exciton collisional broadening in
  {G}a{A}s/{A}l{G}a{A}s quantum wells},\ }\href
  {https://doi.org/10.1103/PhysRevB.104.205302} {\bibfield  {journal} {\bibinfo
   {journal} {Physical Review B}\ }\textbf {\bibinfo {volume} {104}},\ \bibinfo
  {pages} {205302} (\bibinfo {year} {2021})}\BibitemShut {NoStop}%
\bibitem [{\citenamefont {Heitler}\ and\ \citenamefont
  {London}(1927)}]{Heitler_1927}%
  \BibitemOpen
  \bibfield  {author} {\bibinfo {author} {\bibfnamefont {W.}~\bibnamefont
  {Heitler}}\ and\ \bibinfo {author} {\bibfnamefont {F.}~\bibnamefont
  {London}},\ }\bibfield  {title} {\bibinfo {title} {Wechselwirkung neutraler
  {A}tome und hom{\"o}opolare {B}indung nach der {Q}uantenmechanik},\ }\href
  {https://doi.org/10.1007/BF01397394} {\bibfield  {journal} {\bibinfo
  {journal} {Zeitschrift f{\"u}r Physik}\ }\textbf {\bibinfo {volume} {44}},\
  \bibinfo {pages} {455} (\bibinfo {year} {1927})}\BibitemShut {NoStop}%
\bibitem [{\citenamefont {Zimmermann}\ and\ \citenamefont
  {Schindler}(2007)}]{Zimmermann_2007}%
  \BibitemOpen
  \bibfield  {author} {\bibinfo {author} {\bibfnamefont {R.}~\bibnamefont
  {Zimmermann}}\ and\ \bibinfo {author} {\bibfnamefont {C.}~\bibnamefont
  {Schindler}},\ }\bibfield  {title} {\bibinfo {title} {Exciton-exciton
  interaction in coupled quantum wells},\ }\href
  {https://www.sciencedirect.com/science/article/pii/S0038109807005832}
  {\bibfield  {journal} {\bibinfo  {journal} {Solid State Communications}\
  }\textbf {\bibinfo {volume} {144}},\ \bibinfo {pages} {395} (\bibinfo {year}
  {2007})}\BibitemShut {NoStop}%
\bibitem [{\citenamefont {Schindler}\ and\ \citenamefont
  {Zimmermann}(2008)}]{Zimmermann_2008}%
  \BibitemOpen
  \bibfield  {author} {\bibinfo {author} {\bibfnamefont {C.}~\bibnamefont
  {Schindler}}\ and\ \bibinfo {author} {\bibfnamefont {R.}~\bibnamefont
  {Zimmermann}},\ }\bibfield  {title} {\bibinfo {title} {Analysis of the
  exciton-exciton interaction in semiconductor quantum wells},\ }\href
  {https://doi.org/10.1103/PhysRevB.78.045313} {\bibfield  {journal} {\bibinfo
  {journal} {Physical Review B}\ }\textbf {\bibinfo {volume} {78}},\ \bibinfo
  {pages} {045313} (\bibinfo {year} {2008})}\BibitemShut {NoStop}%
\bibitem [{\citenamefont {Hanamura}(1974)}]{Hanamura_1974}%
  \BibitemOpen
  \bibfield  {author} {\bibinfo {author} {\bibfnamefont {E.}~\bibnamefont
  {Hanamura}},\ }\bibfield  {title} {\bibinfo {title} {Theory of many {W}annier
  excitons. {I}},\ }\href {https://doi.org/10.1143/JPSJ.37.1545} {\bibfield
  {journal} {\bibinfo  {journal} {Journal of the Physical Society of Japan}\
  }\textbf {\bibinfo {volume} {37}},\ \bibinfo {pages} {1545} (\bibinfo {year}
  {1974})}\BibitemShut {NoStop}%
\bibitem [{\citenamefont {Haug}\ and\ \citenamefont
  {Schmitt-Rink}(1984)}]{Haug_1984}%
  \BibitemOpen
  \bibfield  {author} {\bibinfo {author} {\bibfnamefont {H.}~\bibnamefont
  {Haug}}\ and\ \bibinfo {author} {\bibfnamefont {S.}~\bibnamefont
  {Schmitt-Rink}},\ }\bibfield  {title} {\bibinfo {title} {Electron theory of
  the optical properties of laser-excited semiconductors},\ }\href
  {https://www.sciencedirect.com/science/article/pii/0079672784900260}
  {\bibfield  {journal} {\bibinfo  {journal} {Progress in Quantum Electronics}\
  }\textbf {\bibinfo {volume} {9}},\ \bibinfo {pages} {3} (\bibinfo {year}
  {1984})}\BibitemShut {NoStop}%
\bibitem [{\citenamefont {Rochat}\ \emph {et~al.}(2000)\citenamefont {Rochat},
  \citenamefont {Ciuti}, \citenamefont {Savona}, \citenamefont {Piermarocchi},
  \citenamefont {Quattropani},\ and\ \citenamefont
  {Schwendimann}}]{Rochat_2001}%
  \BibitemOpen
  \bibfield  {author} {\bibinfo {author} {\bibfnamefont {G.}~\bibnamefont
  {Rochat}}, \bibinfo {author} {\bibfnamefont {C.}~\bibnamefont {Ciuti}},
  \bibinfo {author} {\bibfnamefont {V.}~\bibnamefont {Savona}}, \bibinfo
  {author} {\bibfnamefont {C.}~\bibnamefont {Piermarocchi}}, \bibinfo {author}
  {\bibfnamefont {A.}~\bibnamefont {Quattropani}},\ and\ \bibinfo {author}
  {\bibfnamefont {P.}~\bibnamefont {Schwendimann}},\ }\bibfield  {title}
  {\bibinfo {title} {Excitonic {B}loch equations for a two-dimensional system
  of interacting excitons},\ }\href {https://doi.org/10.1103/PhysRevB.61.13856}
  {\bibfield  {journal} {\bibinfo  {journal} {Physical Review B}\ }\textbf
  {\bibinfo {volume} {61}},\ \bibinfo {pages} {13856} (\bibinfo {year}
  {2000})}\BibitemShut {NoStop}%
\bibitem [{\citenamefont {Okumura}\ and\ \citenamefont
  {Ogawa}(2001)}]{Okumura_2001}%
  \BibitemOpen
  \bibfield  {author} {\bibinfo {author} {\bibfnamefont {S.}~\bibnamefont
  {Okumura}}\ and\ \bibinfo {author} {\bibfnamefont {T.}~\bibnamefont
  {Ogawa}},\ }\bibfield  {title} {\bibinfo {title} {Boson representation of
  two-exciton correlations: {A}n exact treatment of composite-particle
  effects},\ }\href {https://doi.org/10.1103/PhysRevB.65.035105} {\bibfield
  {journal} {\bibinfo  {journal} {Physical Review B}\ }\textbf {\bibinfo
  {volume} {65}},\ \bibinfo {pages} {035105} (\bibinfo {year}
  {2001})}\BibitemShut {NoStop}%
\bibitem [{\citenamefont {Combescot}\ \emph {et~al.}(2007)\citenamefont
  {Combescot}, \citenamefont {Betbeder-Matibet},\ and\ \citenamefont
  {Combescot}}]{Combescot_2007}%
  \BibitemOpen
  \bibfield  {author} {\bibinfo {author} {\bibfnamefont {M.}~\bibnamefont
  {Combescot}}, \bibinfo {author} {\bibfnamefont {O.}~\bibnamefont
  {Betbeder-Matibet}},\ and\ \bibinfo {author} {\bibfnamefont {R.}~\bibnamefont
  {Combescot}},\ }\bibfield  {title} {\bibinfo {title} {Exciton-exciton
  scattering: {C}omposite boson versus elementary boson},\ }\href
  {https://doi.org/10.1103/PhysRevB.75.174305} {\bibfield  {journal} {\bibinfo
  {journal} {Physical Review B}\ }\textbf {\bibinfo {volume} {75}},\ \bibinfo
  {pages} {174305} (\bibinfo {year} {2007})}\BibitemShut {NoStop}%
\bibitem [{\citenamefont {Keldysh}\ and\ \citenamefont
  {Kozlov}(1968)}]{Keldysh_1968}%
  \BibitemOpen
  \bibfield  {author} {\bibinfo {author} {\bibfnamefont {L.~V.}\ \bibnamefont
  {Keldysh}}\ and\ \bibinfo {author} {\bibfnamefont {A.}~\bibnamefont
  {Kozlov}},\ }\bibfield  {title} {\bibinfo {title} {Collective properties of
  excitons in semiconductors},\ }\href
  {http://jetp.ras.ru/cgi-bin/e/index/e/27/3/p521?a=list} {\bibfield  {journal}
  {\bibinfo  {journal} {Journal of Experimental and Theoretical Physics}\
  }\textbf {\bibinfo {volume} {27}},\ \bibinfo {pages} {521} (\bibinfo {year}
  {1968})}\BibitemShut {NoStop}%
\bibitem [{\citenamefont {Maisel~Licer{\'a}n}\ and\ \citenamefont
  {Stoof}(2025)}]{liceran2025unconventional}%
  \BibitemOpen
  \bibfield  {author} {\bibinfo {author} {\bibfnamefont {L.}~\bibnamefont
  {Maisel~Licer{\'a}n}}\ and\ \bibinfo {author} {\bibfnamefont {H.~T.~C.}\
  \bibnamefont {Stoof}},\ }\bibfield  {title} {\bibinfo {title} {Unconventional
  excitonic insulators in two-dimensional topological materials},\ }\href
  {https://doi.org/10.1103/PhysRevB.111.245102} {\bibfield  {journal} {\bibinfo
   {journal} {Physical Review B}\ }\textbf {\bibinfo {volume} {111}},\ \bibinfo
  {pages} {245102} (\bibinfo {year} {2025})}\BibitemShut {NoStop}%
\bibitem [{\citenamefont {Wu}\ \emph {et~al.}(2015)\citenamefont {Wu},
  \citenamefont {Qu},\ and\ \citenamefont {MacDonald}}]{Wu_2015}%
  \BibitemOpen
  \bibfield  {author} {\bibinfo {author} {\bibfnamefont {F.}~\bibnamefont
  {Wu}}, \bibinfo {author} {\bibfnamefont {F.}~\bibnamefont {Qu}},\ and\
  \bibinfo {author} {\bibfnamefont {A.~H.}\ \bibnamefont {MacDonald}},\
  }\bibfield  {title} {\bibinfo {title} {Exciton band structure of monolayer
  \ce{MoS2}},\ }\href {https://doi.org/10.1103/PhysRevB.91.075310} {\bibfield
  {journal} {\bibinfo  {journal} {Physical Review B}\ }\textbf {\bibinfo
  {volume} {91}},\ \bibinfo {pages} {075310} (\bibinfo {year}
  {2015})}\BibitemShut {NoStop}%
\bibitem [{\citenamefont {Elliott}(1957)}]{elliott1957intensity}%
  \BibitemOpen
  \bibfield  {author} {\bibinfo {author} {\bibfnamefont {R.}~\bibnamefont
  {Elliott}},\ }\bibfield  {title} {\bibinfo {title} {Intensity of optical
  absorption by excitons},\ }\href {https://doi.org/10.1103/PhysRev.108.1384}
  {\bibfield  {journal} {\bibinfo  {journal} {Physical Review}\ }\textbf
  {\bibinfo {volume} {108}},\ \bibinfo {pages} {1384} (\bibinfo {year}
  {1957})}\BibitemShut {NoStop}%
\bibitem [{\citenamefont {Michel}(2008)}]{michel2008direct}%
  \BibitemOpen
  \bibfield  {author} {\bibinfo {author} {\bibfnamefont {N.}~\bibnamefont
  {Michel}},\ }\bibfield  {title} {\bibinfo {title} {Direct demonstration of
  the completeness of the eigenstates of the {S}chr{\"o}dinger equation with
  local and nonlocal potentials bearing a {C}oulomb tail},\ }\href
  {https://doi.org/10.1063/1.2830976} {\bibfield  {journal} {\bibinfo
  {journal} {Journal of Mathematical Physics}\ }\textbf {\bibinfo {volume}
  {49}} (\bibinfo {year} {2008})}\BibitemShut {NoStop}%
\bibitem [{\citenamefont {Mukhamedzhanov}\ and\ \citenamefont
  {Akin}(2008)}]{mukhamedzhanov2008completeness}%
  \BibitemOpen
  \bibfield  {author} {\bibinfo {author} {\bibfnamefont {A.~M.}\ \bibnamefont
  {Mukhamedzhanov}}\ and\ \bibinfo {author} {\bibfnamefont {M.}~\bibnamefont
  {Akin}},\ }\bibfield  {title} {\bibinfo {title} {Completeness of the
  {C}oulomb scattering wave functions},\ }\href
  {https://doi.org/10.1140/epja/i2007-10613-1} {\bibfield  {journal} {\bibinfo
  {journal} {The European Physical Journal A}\ }\textbf {\bibinfo {volume}
  {37}},\ \bibinfo {pages} {185} (\bibinfo {year} {2008})}\BibitemShut
  {NoStop}%
\bibitem [{\citenamefont {Parlett}(1998)}]{parlett1998symmetric}%
  \BibitemOpen
  \bibfield  {author} {\bibinfo {author} {\bibfnamefont {B.~N.}\ \bibnamefont
  {Parlett}},\ }\href@noop {} {\emph {\bibinfo {title} {The symmetric
  eigenvalue problem}}}\ (\bibinfo  {publisher} {SIAM},\ \bibinfo {year}
  {1998})\BibitemShut {NoStop}%
\bibitem [{\citenamefont {Hanamura}\ and\ \citenamefont
  {Haug}(1977)}]{Hanamura_1977}%
  \BibitemOpen
  \bibfield  {author} {\bibinfo {author} {\bibfnamefont {E.}~\bibnamefont
  {Hanamura}}\ and\ \bibinfo {author} {\bibfnamefont {H.}~\bibnamefont
  {Haug}},\ }\bibfield  {title} {\bibinfo {title} {Condensation effects of
  excitons},\ }\href
  {https://www.sciencedirect.com/science/article/pii/0370157377900126}
  {\bibfield  {journal} {\bibinfo  {journal} {Physics Reports}\ }\textbf
  {\bibinfo {volume} {33}},\ \bibinfo {pages} {209} (\bibinfo {year}
  {1977})}\BibitemShut {NoStop}%
\bibitem [{\citenamefont {Weiner}\ \emph {et~al.}(1999)\citenamefont {Weiner},
  \citenamefont {Bagnato}, \citenamefont {Zilio},\ and\ \citenamefont
  {Julienne}}]{weiner1999experiments}%
  \BibitemOpen
  \bibfield  {author} {\bibinfo {author} {\bibfnamefont {J.}~\bibnamefont
  {Weiner}}, \bibinfo {author} {\bibfnamefont {V.~S.}\ \bibnamefont {Bagnato}},
  \bibinfo {author} {\bibfnamefont {S.}~\bibnamefont {Zilio}},\ and\ \bibinfo
  {author} {\bibfnamefont {P.~S.}\ \bibnamefont {Julienne}},\ }\bibfield
  {title} {\bibinfo {title} {Experiments and theory in cold and ultracold
  collisions},\ }\href {https://doi.org/10.1103/RevModPhys.71.1} {\bibfield
  {journal} {\bibinfo  {journal} {Reviews of Modern Physics}\ }\textbf
  {\bibinfo {volume} {71}},\ \bibinfo {pages} {1} (\bibinfo {year}
  {1999})}\BibitemShut {NoStop}%
\bibitem [{\citenamefont {K{\"o}hler}\ \emph {et~al.}(2006)\citenamefont
  {K{\"o}hler}, \citenamefont {G{\'o}ral},\ and\ \citenamefont
  {Julienne}}]{kohler2006production}%
  \BibitemOpen
  \bibfield  {author} {\bibinfo {author} {\bibfnamefont {T.}~\bibnamefont
  {K{\"o}hler}}, \bibinfo {author} {\bibfnamefont {K.}~\bibnamefont
  {G{\'o}ral}},\ and\ \bibinfo {author} {\bibfnamefont {P.~S.}\ \bibnamefont
  {Julienne}},\ }\bibfield  {title} {\bibinfo {title} {Production of cold
  molecules via magnetically tunable {F}eshbach resonances},\ }\href
  {https://doi.org/10.1103/RevModPhys.78.1311} {\bibfield  {journal} {\bibinfo
  {journal} {Reviews of Modern Physics}\ }\textbf {\bibinfo {volume} {78}},\
  \bibinfo {pages} {1311} (\bibinfo {year} {2006})}\BibitemShut {NoStop}%
\bibitem [{\citenamefont {Chin}\ \emph {et~al.}(2010)\citenamefont {Chin},
  \citenamefont {Grimm}, \citenamefont {Julienne},\ and\ \citenamefont
  {Tiesinga}}]{chin2010feshbach}%
  \BibitemOpen
  \bibfield  {author} {\bibinfo {author} {\bibfnamefont {C.}~\bibnamefont
  {Chin}}, \bibinfo {author} {\bibfnamefont {R.}~\bibnamefont {Grimm}},
  \bibinfo {author} {\bibfnamefont {P.}~\bibnamefont {Julienne}},\ and\
  \bibinfo {author} {\bibfnamefont {E.}~\bibnamefont {Tiesinga}},\ }\bibfield
  {title} {\bibinfo {title} {Feshbach resonances in ultracold gases},\ }\href
  {https://doi.org/10.1103/RevModPhys.82.1225} {\bibfield  {journal} {\bibinfo
  {journal} {Reviews of Modern Physics}\ }\textbf {\bibinfo {volume} {82}},\
  \bibinfo {pages} {1225} (\bibinfo {year} {2010})}\BibitemShut {NoStop}%
\bibitem [{\citenamefont {Chao}\ and\ \citenamefont
  {Chuang}(1991)}]{chao1991analytical}%
  \BibitemOpen
  \bibfield  {author} {\bibinfo {author} {\bibfnamefont {C.~Y.-P.}\
  \bibnamefont {Chao}}\ and\ \bibinfo {author} {\bibfnamefont {S.~L.}\
  \bibnamefont {Chuang}},\ }\bibfield  {title} {\bibinfo {title} {Analytical
  and numerical solutions for a two-dimensional exciton in momentum space},\
  }\href {https://doi.org/10.1103/PhysRevB.43.6530} {\bibfield  {journal}
  {\bibinfo  {journal} {Physical Review B}\ }\textbf {\bibinfo {volume} {43}},\
  \bibinfo {pages} {6530} (\bibinfo {year} {1991})}\BibitemShut {NoStop}%
\bibitem [{\citenamefont {Yang}\ \emph {et~al.}(1991)\citenamefont {Yang},
  \citenamefont {Guo}, \citenamefont {Chan}, \citenamefont {Wong},\ and\
  \citenamefont {Ching}}]{yang1991analytic}%
  \BibitemOpen
  \bibfield  {author} {\bibinfo {author} {\bibfnamefont {X.}~\bibnamefont
  {Yang}}, \bibinfo {author} {\bibfnamefont {S.}~\bibnamefont {Guo}}, \bibinfo
  {author} {\bibfnamefont {F.}~\bibnamefont {Chan}}, \bibinfo {author}
  {\bibfnamefont {K.}~\bibnamefont {Wong}},\ and\ \bibinfo {author}
  {\bibfnamefont {W.}~\bibnamefont {Ching}},\ }\bibfield  {title} {\bibinfo
  {title} {Analytic solution of a two-dimensional hydrogen atom. {I}.
  {N}onrelativistic theory},\ }\href {https://doi.org/10.1103/PhysRevA.43.1186}
  {\bibfield  {journal} {\bibinfo  {journal} {Physical Review A}\ }\textbf
  {\bibinfo {volume} {43}},\ \bibinfo {pages} {1186} (\bibinfo {year}
  {1991})}\BibitemShut {NoStop}%
\bibitem [{\citenamefont {Parfitt}\ and\ \citenamefont
  {Portnoi}(2002)}]{Parfitt_2002}%
  \BibitemOpen
  \bibfield  {author} {\bibinfo {author} {\bibfnamefont {D.~G.~W.}\
  \bibnamefont {Parfitt}}\ and\ \bibinfo {author} {\bibfnamefont {M.~E.}\
  \bibnamefont {Portnoi}},\ }\bibfield  {title} {\bibinfo {title} {The
  two-dimensional hydrogen atom revisited},\ }\href
  {https://doi.org/10.1063/1.1503868} {\bibfield  {journal} {\bibinfo
  {journal} {Journal of Mathematical Physics}\ }\textbf {\bibinfo {volume}
  {43}},\ \bibinfo {pages} {4681} (\bibinfo {year} {2002})}\BibitemShut
  {NoStop}%
\bibitem [{\citenamefont {Efimkin}\ \emph {et~al.}(2021)\citenamefont
  {Efimkin}, \citenamefont {Laird}, \citenamefont {Levinsen}, \citenamefont
  {Parish},\ and\ \citenamefont {MacDonald}}]{efimkin2021electron}%
  \BibitemOpen
  \bibfield  {author} {\bibinfo {author} {\bibfnamefont {D.~K.}\ \bibnamefont
  {Efimkin}}, \bibinfo {author} {\bibfnamefont {E.~K.}\ \bibnamefont {Laird}},
  \bibinfo {author} {\bibfnamefont {J.}~\bibnamefont {Levinsen}}, \bibinfo
  {author} {\bibfnamefont {M.~M.}\ \bibnamefont {Parish}},\ and\ \bibinfo
  {author} {\bibfnamefont {A.~H.}\ \bibnamefont {MacDonald}},\ }\bibfield
  {title} {\bibinfo {title} {Electron-exciton interactions in the
  exciton-polaron problem},\ }\href
  {https://doi.org/10.1103/PhysRevB.103.075417} {\bibfield  {journal} {\bibinfo
   {journal} {Physical Review B}\ }\textbf {\bibinfo {volume} {103}},\ \bibinfo
  {pages} {075417} (\bibinfo {year} {2021})}\BibitemShut {NoStop}%
\bibitem [{Note1()}]{Note1}%
  \BibitemOpen
  \bibinfo {note} {Note that we could have also eliminated $S_{\protect \mathrm
  {c}}$ in favor of $S_{\protect \mathrm {v}}$ in Eq.\ \protect \eqref {eq:
  exciton--exciton potential effective} and performed the heavy-electron limit
  instead of the heavy-hole limit, which would result in Eq.\ \protect \eqref
  {eq: Heitler--London potential real space} depending only on $S_{\protect
  \mathrm {v}}$.}\BibitemShut {Stop}%
\bibitem [{\citenamefont {Cam}\ \emph {et~al.}(2022)\citenamefont {Cam},
  \citenamefont {Phuc},\ and\ \citenamefont {Osipov}}]{cam2022symmetry}%
  \BibitemOpen
  \bibfield  {author} {\bibinfo {author} {\bibfnamefont {H.~N.}\ \bibnamefont
  {Cam}}, \bibinfo {author} {\bibfnamefont {N.~T.}\ \bibnamefont {Phuc}},\ and\
  \bibinfo {author} {\bibfnamefont {V.~A.}\ \bibnamefont {Osipov}},\ }\bibfield
   {title} {\bibinfo {title} {Symmetry-dependent exciton-exciton interaction
  and intervalley biexciton in monolayer transition metal dichalcogenides},\
  }\href {https://doi.org/10.1038/s41699-022-00290-z} {\bibfield  {journal}
  {\bibinfo  {journal} {npj 2D Materials and Applications}\ }\textbf {\bibinfo
  {volume} {6}},\ \bibinfo {pages} {22} (\bibinfo {year} {2022})}\BibitemShut
  {NoStop}%
\bibitem [{\citenamefont {Stoof}\ \emph {et~al.}(2008)\citenamefont {Stoof},
  \citenamefont {Gubbels},\ and\ \citenamefont {Dickerscheid}}]{Stoof_2008}%
  \BibitemOpen
  \bibfield  {author} {\bibinfo {author} {\bibfnamefont {H.~T.~C.}\
  \bibnamefont {Stoof}}, \bibinfo {author} {\bibfnamefont {K.~B.}\ \bibnamefont
  {Gubbels}},\ and\ \bibinfo {author} {\bibfnamefont {D.~B.~M.}\ \bibnamefont
  {Dickerscheid}},\ }\href {https://doi.org/10.1007/978-1-4020-8763-9} {\emph
  {\bibinfo {title} {Ultracold Quantum Fields}}},\ Vol.~\bibinfo {volume} {1}\
  (\bibinfo  {publisher} {Springer},\ \bibinfo {year} {2008})\BibitemShut
  {NoStop}%
\bibitem [{\citenamefont {Steinhoff}\ \emph {et~al.}(2017)\citenamefont
  {Steinhoff}, \citenamefont {Florian}, \citenamefont {R{\"o}sner},
  \citenamefont {Sch{\"o}nhoff}, \citenamefont {Wehling},\ and\ \citenamefont
  {Jahnke}}]{Steinhoff_2017}%
  \BibitemOpen
  \bibfield  {author} {\bibinfo {author} {\bibfnamefont {A.}~\bibnamefont
  {Steinhoff}}, \bibinfo {author} {\bibfnamefont {M.}~\bibnamefont {Florian}},
  \bibinfo {author} {\bibfnamefont {M.}~\bibnamefont {R{\"o}sner}}, \bibinfo
  {author} {\bibfnamefont {G.}~\bibnamefont {Sch{\"o}nhoff}}, \bibinfo {author}
  {\bibfnamefont {T.~O.}\ \bibnamefont {Wehling}},\ and\ \bibinfo {author}
  {\bibfnamefont {F.}~\bibnamefont {Jahnke}},\ }\bibfield  {title} {\bibinfo
  {title} {Exciton fission in monolayer transition metal dichalcogenide
  semiconductors},\ }\href {https://doi.org/10.1038/s41467-017-01298-6}
  {\bibfield  {journal} {\bibinfo  {journal} {Nature Communications}\ }\textbf
  {\bibinfo {volume} {8}},\ \bibinfo {pages} {1166} (\bibinfo {year}
  {2017})}\BibitemShut {NoStop}%
\bibitem [{\citenamefont {Ma}\ \emph {et~al.}(2021)\citenamefont {Ma},
  \citenamefont {Nguyen}, \citenamefont {Wang}, \citenamefont {Zeng},
  \citenamefont {Watanabe}, \citenamefont {Taniguchi}, \citenamefont
  {MacDonald}, \citenamefont {Mak},\ and\ \citenamefont {Shan}}]{Ma_2021}%
  \BibitemOpen
  \bibfield  {author} {\bibinfo {author} {\bibfnamefont {L.}~\bibnamefont
  {Ma}}, \bibinfo {author} {\bibfnamefont {P.~X.}\ \bibnamefont {Nguyen}},
  \bibinfo {author} {\bibfnamefont {Z.}~\bibnamefont {Wang}}, \bibinfo {author}
  {\bibfnamefont {Y.}~\bibnamefont {Zeng}}, \bibinfo {author} {\bibfnamefont
  {K.}~\bibnamefont {Watanabe}}, \bibinfo {author} {\bibfnamefont
  {T.}~\bibnamefont {Taniguchi}}, \bibinfo {author} {\bibfnamefont {A.~H.}\
  \bibnamefont {MacDonald}}, \bibinfo {author} {\bibfnamefont {K.~F.}\
  \bibnamefont {Mak}},\ and\ \bibinfo {author} {\bibfnamefont {J.}~\bibnamefont
  {Shan}},\ }\bibfield  {title} {\bibinfo {title} {Strongly correlated
  excitonic insulator in atomic double layers},\ }\href
  {https://doi.org/10.1038/s41586-021-03947-9} {\bibfield  {journal} {\bibinfo
  {journal} {Nature}\ }\textbf {\bibinfo {volume} {598}},\ \bibinfo {pages}
  {585} (\bibinfo {year} {2021})}\BibitemShut {NoStop}%
\bibitem [{sup()}]{supp}%
  \BibitemOpen
  \href@noop {} {}\bibinfo {note} {See Supplemental Material at
  \textcolor{blue}{URL-will-be-inserted-by-publisher} for numerical details and
  field-theoretical derivations.}\BibitemShut {Stop}%
\bibitem [{Note2()}]{Note2}%
  \BibitemOpen
  \bibinfo {note} {In principle, because of the Hubbard--Stratonovich
  transformation, every polarization field always comes ``attached'' to a
  single-particle interaction. Essentially, this is an interaction between the
  conduction and valence electrons of the same exciton. This kind of
  interaction can always be rewritten into energy terms via the
  temperature-dependent BSE. Hence, this single-particle interaction is no
  longer explicitly present, neither in the upcoming Feynman diagrams nor in
  the mathematical expressions. Therefore, when there is a mention of an
  exciton--exciton interaction being of the order $\protect \mathcal {O}(V)$,
  this will refer to a single-particle interaction that occurs between the
  electrons of different excitons.}\BibitemShut {Stop}%
\bibitem [{\citenamefont {Perali}\ \emph {et~al.}(2013)\citenamefont {Perali},
  \citenamefont {Neilson},\ and\ \citenamefont {Hamilton}}]{Perali_2013}%
  \BibitemOpen
  \bibfield  {author} {\bibinfo {author} {\bibfnamefont {A.}~\bibnamefont
  {Perali}}, \bibinfo {author} {\bibfnamefont {D.}~\bibnamefont {Neilson}},\
  and\ \bibinfo {author} {\bibfnamefont {A.~R.}\ \bibnamefont {Hamilton}},\
  }\bibfield  {title} {\bibinfo {title} {High-temperature superfluidity in
  double-bilayer graphene},\ }\href
  {https://doi.org/10.1103/PhysRevLett.110.146803} {\bibfield  {journal}
  {\bibinfo  {journal} {Physical Review Letters}\ }\textbf {\bibinfo {volume}
  {110}},\ \bibinfo {pages} {146803} (\bibinfo {year} {2013})}\BibitemShut
  {NoStop}%
\bibitem [{\citenamefont {Haque}\ \emph {et~al.}(2024)\citenamefont {Haque},
  \citenamefont {Michael}, \citenamefont {Zhu}, \citenamefont {Zhang},
  \citenamefont {Windgätter}, \citenamefont {Latini}, \citenamefont
  {Wakefield}, \citenamefont {Zhang}, \citenamefont {Zhang}, \citenamefont
  {Rubio}, \citenamefont {Checkelsky}, \citenamefont {Demler},\ and\
  \citenamefont {Averitt}}]{Haque_2024}%
  \BibitemOpen
  \bibfield  {author} {\bibinfo {author} {\bibfnamefont {S.~R.~U.}\
  \bibnamefont {Haque}}, \bibinfo {author} {\bibfnamefont {M.~H.}\ \bibnamefont
  {Michael}}, \bibinfo {author} {\bibfnamefont {J.}~\bibnamefont {Zhu}},
  \bibinfo {author} {\bibfnamefont {Y.}~\bibnamefont {Zhang}}, \bibinfo
  {author} {\bibfnamefont {L.}~\bibnamefont {Windgätter}}, \bibinfo {author}
  {\bibfnamefont {S.}~\bibnamefont {Latini}}, \bibinfo {author} {\bibfnamefont
  {J.~P.}\ \bibnamefont {Wakefield}}, \bibinfo {author} {\bibfnamefont {G.-F.}\
  \bibnamefont {Zhang}}, \bibinfo {author} {\bibfnamefont {J.}~\bibnamefont
  {Zhang}}, \bibinfo {author} {\bibfnamefont {A.}~\bibnamefont {Rubio}},
  \bibinfo {author} {\bibfnamefont {J.~G.}\ \bibnamefont {Checkelsky}},
  \bibinfo {author} {\bibfnamefont {E.}~\bibnamefont {Demler}},\ and\ \bibinfo
  {author} {\bibfnamefont {R.~D.}\ \bibnamefont {Averitt}},\ }\bibfield
  {title} {\bibinfo {title} {Terahertz parametric amplification as a reporter
  of exciton condensate dynamics},\ }\href
  {https://doi.org/10.1038/s41563-023-01755-2} {\bibfield  {journal} {\bibinfo
  {journal} {Nature Materials}\ }\textbf {\bibinfo {volume} {23}},\ \bibinfo
  {pages} {796} (\bibinfo {year} {2024})}\BibitemShut {NoStop}%
\bibitem [{\citenamefont {Berman}\ and\ \citenamefont
  {Kezerashvili}(2016)}]{Berman_2016}%
  \BibitemOpen
  \bibfield  {author} {\bibinfo {author} {\bibfnamefont {O.~L.}\ \bibnamefont
  {Berman}}\ and\ \bibinfo {author} {\bibfnamefont {R.~Y.}\ \bibnamefont
  {Kezerashvili}},\ }\bibfield  {title} {\bibinfo {title} {High-temperature
  superfluidity of the two-component {B}ose gas in a transition metal
  dichalcogenide bilayer},\ }\href {https://doi.org/10.1103/PhysRevB.93.245410}
  {\bibfield  {journal} {\bibinfo  {journal} {Physical Review B}\ }\textbf
  {\bibinfo {volume} {93}},\ \bibinfo {pages} {245410} (\bibinfo {year}
  {2016})}\BibitemShut {NoStop}%
\bibitem [{\citenamefont {Lin}\ and\ \citenamefont
  {Jin}(2010)}]{lin2010charge}%
  \BibitemOpen
  \bibfield  {author} {\bibinfo {author} {\bibfnamefont {H.-C.}\ \bibnamefont
  {Lin}}\ and\ \bibinfo {author} {\bibfnamefont {B.-Y.}\ \bibnamefont {Jin}},\
  }\bibfield  {title} {\bibinfo {title} {Charge-transfer interactions in
  organic functional materials},\ }\href {https://doi.org/10.3390/ma3084214}
  {\bibfield  {journal} {\bibinfo  {journal} {Materials}\ }\textbf {\bibinfo
  {volume} {3}},\ \bibinfo {pages} {4214} (\bibinfo {year} {2010})}\BibitemShut
  {NoStop}%
\bibitem [{\citenamefont {Bardeen}(2014)}]{bardeen2014structure}%
  \BibitemOpen
  \bibfield  {author} {\bibinfo {author} {\bibfnamefont {C.~J.}\ \bibnamefont
  {Bardeen}},\ }\bibfield  {title} {\bibinfo {title} {The structure and
  dynamics of molecular excitons},\ }\href
  {https://doi.org/10.1146/annurev-physchem-040513-103654} {\bibfield
  {journal} {\bibinfo  {journal} {Annual Review of Physical Chemistry}\
  }\textbf {\bibinfo {volume} {65}},\ \bibinfo {pages} {127} (\bibinfo {year}
  {2014})}\BibitemShut {NoStop}%
\bibitem [{\citenamefont {Nematiaram}\ \emph {et~al.}(2021)\citenamefont
  {Nematiaram}, \citenamefont {Padula},\ and\ \citenamefont
  {Troisi}}]{nematiaram2021bright}%
  \BibitemOpen
  \bibfield  {author} {\bibinfo {author} {\bibfnamefont {T.}~\bibnamefont
  {Nematiaram}}, \bibinfo {author} {\bibfnamefont {D.}~\bibnamefont {Padula}},\
  and\ \bibinfo {author} {\bibfnamefont {A.}~\bibnamefont {Troisi}},\
  }\bibfield  {title} {\bibinfo {title} {Bright {F}renkel excitons in molecular
  crystals: a survey},\ }\href {https://doi.org/10.1021/acs.chemmater.1c00645}
  {\bibfield  {journal} {\bibinfo  {journal} {Chemistry of Materials}\ }\textbf
  {\bibinfo {volume} {33}},\ \bibinfo {pages} {3368} (\bibinfo {year}
  {2021})}\BibitemShut {NoStop}%
\bibitem [{\citenamefont {G{\"o}tting}\ \emph {et~al.}(2022)\citenamefont
  {G{\"o}tting}, \citenamefont {Lohof},\ and\ \citenamefont
  {Gies}}]{gotting2022moire}%
  \BibitemOpen
  \bibfield  {author} {\bibinfo {author} {\bibfnamefont {N.}~\bibnamefont
  {G{\"o}tting}}, \bibinfo {author} {\bibfnamefont {F.}~\bibnamefont {Lohof}},\
  and\ \bibinfo {author} {\bibfnamefont {C.}~\bibnamefont {Gies}},\ }\bibfield
  {title} {\bibinfo {title} {Moir{\'e}-{B}ose-{H}ubbard model for interlayer
  excitons in twisted transition metal dichalcogenide heterostructures},\
  }\href {https://doi.org/10.1103/PhysRevB.105.165419} {\bibfield  {journal}
  {\bibinfo  {journal} {Physical Review B}\ }\textbf {\bibinfo {volume}
  {105}},\ \bibinfo {pages} {165419} (\bibinfo {year} {2022})}\BibitemShut
  {NoStop}%
\bibitem [{\citenamefont {Haber}\ \emph {et~al.}(2023)\citenamefont {Haber},
  \citenamefont {Qiu}, \citenamefont {da~Jornada},\ and\ \citenamefont
  {Neaton}}]{haber2023maximally}%
  \BibitemOpen
  \bibfield  {author} {\bibinfo {author} {\bibfnamefont {J.~B.}\ \bibnamefont
  {Haber}}, \bibinfo {author} {\bibfnamefont {D.~Y.}\ \bibnamefont {Qiu}},
  \bibinfo {author} {\bibfnamefont {F.~H.}\ \bibnamefont {da~Jornada}},\ and\
  \bibinfo {author} {\bibfnamefont {J.~B.}\ \bibnamefont {Neaton}},\ }\bibfield
   {title} {\bibinfo {title} {Maximally localized exciton {W}annier functions
  for solids},\ }\href {https://doi.org/10.1103/PhysRevB.108.125118} {\bibfield
   {journal} {\bibinfo  {journal} {Physical Review B}\ }\textbf {\bibinfo
  {volume} {108}},\ \bibinfo {pages} {125118} (\bibinfo {year}
  {2023})}\BibitemShut {NoStop}%
\bibitem [{\citenamefont {Chang}(2023)}]{chang2023continuum}%
  \BibitemOpen
  \bibfield  {author} {\bibinfo {author} {\bibfnamefont {Y.-W.}\ \bibnamefont
  {Chang}},\ }\bibfield  {title} {\bibinfo {title} {Continuum model study of
  optical absorption by hybridized moir{\'e} excitons in transition metal
  dichalcogenide heterobilayers},\ }\href
  {https://doi.org/10.1103/PhysRevB.108.155424} {\bibfield  {journal} {\bibinfo
   {journal} {Physical Review B}\ }\textbf {\bibinfo {volume} {108}},\ \bibinfo
  {pages} {155424} (\bibinfo {year} {2023})}\BibitemShut {NoStop}%
\bibitem [{\citenamefont {Peng}\ \emph {et~al.}(2025)\citenamefont {Peng},
  \citenamefont {Vignale},\ and\ \citenamefont {Adam}}]{peng2025many}%
  \BibitemOpen
  \bibfield  {author} {\bibinfo {author} {\bibfnamefont {L.}~\bibnamefont
  {Peng}}, \bibinfo {author} {\bibfnamefont {G.}~\bibnamefont {Vignale}},\ and\
  \bibinfo {author} {\bibfnamefont {S.}~\bibnamefont {Adam}},\ }\bibfield
  {title} {\bibinfo {title} {Many-body perturbation theory for moir{\'e}
  systems},\ }\href {https://doi.org/10.1103/5qws-l9ny} {\bibfield  {journal}
  {\bibinfo  {journal} {Physical Review B}\ }\textbf {\bibinfo {volume}
  {112}},\ \bibinfo {pages} {075146} (\bibinfo {year} {2025})}\BibitemShut
  {NoStop}%
\bibitem [{Note3()}]{Note3}%
  \BibitemOpen
  \bibinfo {note} {In fact, $\protect \mathcal {U}$ corresponds to the
  exciton--exciton interaction that results from setting the overlap integrals
  $\protect \mathcal {K}^{\protect \mathrm {c}}$ and $\protect \mathcal
  {K}^{\protect \mathrm {v}}$ to zero in the generalized eigenvalue problem of
  Eq.\ \protect \eqref {eq: biexciton eigenvalue equation}. For this reason, it
  only partially includes the effects of the composite nature of the excitons,
  but we nevertheless expect it to be a better approximation for this problem
  than the use of a simple potential between point charges. Additional exchange
  effects can be included by adding also the exchange part of the kinetic term
  of Eq.\ \protect \eqref {eq: Hamiltonian exciton--exciton}, if
  desired.}\BibitemShut {Stop}%
\bibitem [{\citenamefont {Singh}(1994)}]{Singh_1994}%
  \BibitemOpen
  \bibfield  {author} {\bibinfo {author} {\bibfnamefont {J.}~\bibnamefont
  {Singh}},\ }\href {https://doi.org/10.1007/978-1-4899-0996-1} {\emph
  {\bibinfo {title} {Excitation Energy Transfer Processes in Condensed
  Matter}}}\ (\bibinfo  {publisher} {Springer New York, NY},\ \bibinfo {year}
  {1994})\BibitemShut {NoStop}%
\bibitem [{\citenamefont {Young}(2024)}]{Young_2024}%
  \BibitemOpen
  \bibfield  {author} {\bibinfo {author} {\bibfnamefont {D.}~\bibnamefont
  {Young}},\ }\href
  {https://www.mathworks.com/matlabcentral/fileexchange/31012-2-d-convolution-using-the-fft}
  {\bibinfo {title} {2-{D} convolution using the {FFT}}} (\bibinfo {year}
  {2024}),\ \bibinfo {note} {{MATLAB Central File Exchange. Retrieved December
  6, 2024}}\BibitemShut {NoStop}%
\end{thebibliography}%









 

\end{document}